\documentclass[%
reprint,
 amsmath,amssymb,
]{revtex4-1}

\usepackage{graphicx}
\usepackage{dcolumn}
\usepackage{bm}
\usepackage{multirow}

\newcommand{\zero}{|0 \rangle}
\newcommand{\one}{|1 \rangle}

\begin{document}


\title{Trapped-Ion Quantum Computing: Progress and Challenges}
\author{Colin D. Bruzewicz}
\email{colin.bruzewicz@ll.mit.edu}
\affiliation{Lincoln Laboratory, Massachusetts Institute of Technology, Lexington, MA, 02420}
\author{John Chiaverini}
\email{john.chiaverini@ll.mit.edu}
\affiliation{Lincoln Laboratory, Massachusetts Institute of Technology, Lexington, MA, 02420}
\author{Robert McConnell}
\email{robert.mcconnell@ll.mit.edu}
\affiliation{Lincoln Laboratory, Massachusetts Institute of Technology, Lexington, MA, 02420}
\author{Jeremy M. Sage}
\email{jsage@ll.mit.edu}
\affiliation{Lincoln Laboratory, Massachusetts Institute of Technology, Lexington, MA, 02420}

\date{\today}

\begin{abstract}
Trapped ions are among the most promising systems for practical quantum computing (QC).  The basic requirements for universal QC have all been demonstrated with ions and quantum algorithms using few-ion-qubit systems have been implemented.  We review the state of the field, covering the basics of how trapped ions are used for QC and their strengths and limitations as qubits.  In addition, we discuss what is being done, and what may be required, to increase the scale of trapped ion quantum computers while mitigating decoherence and control errors.  Finally, we explore the outlook for trapped-ion QC.  In particular, we discuss near-term applications, considerations impacting the design of future systems of trapped ions, and experiments and demonstrations that may further inform these considerations.

\end{abstract}

\pacs{Valid PACS appear here}

\maketitle
\tableofcontents
\section{Introduction}
\subsection{Trapped Ions for Quantum Computing}

Soon after Shor developed the factoring algorithm that bears his name \cite{ShorAlgorithm}, demonstrating that a large-scale quantum computer could efficiently solve useful tasks that were classically intractable, Cirac and Zoller proposed an implementation of such a device using individual atomic ions~\cite{CiracZollerGate}.  In this scheme, ions confined in radiofrequency (RF) traps serve as quantum bits, with entanglement achieved by using the shared ion motional modes as a quantum bus. RF Paul traps had been used to confine single ions since 1980 \cite{NeuhauserTrappedIon1980} and appeared to be a promising platform due to the ions' robust trap lifetimes, long internal-state coherence, strong ion-ion interactions, and the existence of cycling transitions between internal states of ions for measurement and laser cooling. Controlled-NOT (CNOT) gates entangling one ion's internal state with its motional state were rapidly demonstrated \cite{MonroeCNOT1995} and multi-ion entangled states were demonstrated soon afterwards~\cite{TurchetteEntanglement1998, Sackett4IonEntanglement2000}.

Since then, trapped ions have remained one of the leading technology platforms for large-scale QC. Using trapped ions, single-qubit gates \cite{NIST:HifiMicrogate:12}, two-qubit gates \cite{BenhelmMSGate2008}, and qubit state preparation and readout \cite{MyersonReadoutIons2008} have all been performed with fidelity exceeding that required for fault-tolerant QC using high-threshold quantum error correction codes \cite{RaussendorfSurfaceCode2007}.  However, despite the promise shown by trapped ions, there are still many challenges that must be addressed in order to realize a practically useful quantum computer.  Chief among these is increasing the number of simultaneously trapped ions while maintaining the ability to control and measure them individually with high fidelity.

\subsection{Scope of this Review}

The goal of this paper is to review recent progress in QC with trapped ions, with a particular emphasis on the challenges inherent in going from high-fidelity demonstrations using a few ions, where the field is today, to demonstrations using hundreds or many more. Several excellent review papers exist which treat various aspects of trapped-ion physics in detail \cite{Wineland1998, leibfried2003quantum, Blatt:Wine:Nat08, HaffnerIonReview2008, WinelandIonReview2009, ozeri_tutorial_2011, SchindlerIonReview2013}. As a result, in this paper, we will not present a detailed review of the mechanics of ion trapping or of the equations governing the interaction between ions and electromagnetic control fields. For these topics the interested reader is referred to the aforementioned reviews.

After a brief introduction to trapped ions as a qubit technology, we will discuss methods of controlling trapped-ion qubits. We will review experiments demonstrating single- and two-qubit gates using ion qubits, the achieved fidelities, other important aspects of qubit control including loading and detection, and key outstanding limitations to the scalable implementation of the demonstrated methods. We will next look at recent efforts to increase the number of simultaneously-trapped ions and to develop technologies and methods for robustly controlling large numbers of ions. Finally, we will discuss near-term experiments which might hope to achieve interesting results with traps of 50 to 100 ions and without quantum error correction, and preview the long-term outlook for trapped-ion QC.

This paper primarily addresses gate-based approaches to QC, which includes digital quantum simulation \cite{UniversalQSimLloyd1996}. Recently, a different approach known as quantum annealing has gathered widespread interest \cite{JohnsonQuantumAnneal2011, BoixoQuantumAnneal100Qubits2014}. However, it has still not been shown even theoretically whether quantum annealing can yield a speedup over the best classical approaches to a problem. It remains to be seen whether this highly interesting avenue of research will yield useful results, but we will not discuss it further in this article.

\subsection{DiVincenzo Criteria}
\label{sec:DiV_criteria}

In 2000, DiVincenzo outlined five key criteria for a quantum information processor \cite{DiVincenzoCriteria2000}. These criteria have been used as one basis for assessing the viability of different possible physical implementations of a quantum computer. DiVincenzo's five criteria include: 1) a physical system containing well-defined two-level quantum systems, or qubits (whose computational basis states are usually written as $\zero$ and $\one$), which can be isolated from the environment; 2) the ability to initialize the system into a well-defined and determinate initial state; 3) qubit decoherence times much longer than the gate times; 4) a set of universal quantum gates which can be applied to each qubit (or pair of qubits, in the case of two-qubit gates); and 5) the ability to read out the qubit state with high accuracy. Trapped ions represent one of only a few qubit technologies which have yet fulfilled all of DiVincenzo's original criteria with high fidelity.

For trapped ions, internal electronic states of the ion are used for the qubit states $\zero$ and $\one$. Trapped-ion qubits can generally be considered as one of four types: \emph{hyperfine} qubits, where the qubit states are hyperfine states of the ion separated by an energy splitting of order gigahertz; \emph{Zeeman} qubits, where the qubit states are magnetic sublevels split by an applied field and typically have tens of megahertz frequencies; \emph{fine structure} qubits, where the qubit states reside in the fine structure levels and are separated by typically tens of terahertz; and \emph{optical} qubits, where the qubit states are separated by an optical transition (typically hundreds of terahertz). Each type of qubit comes with its own particular benefits and drawbacks, as will be described later (see Section \ref{subInternal}).

\begin{figure*}[t b h]
\includegraphics[width=2.0\columnwidth]{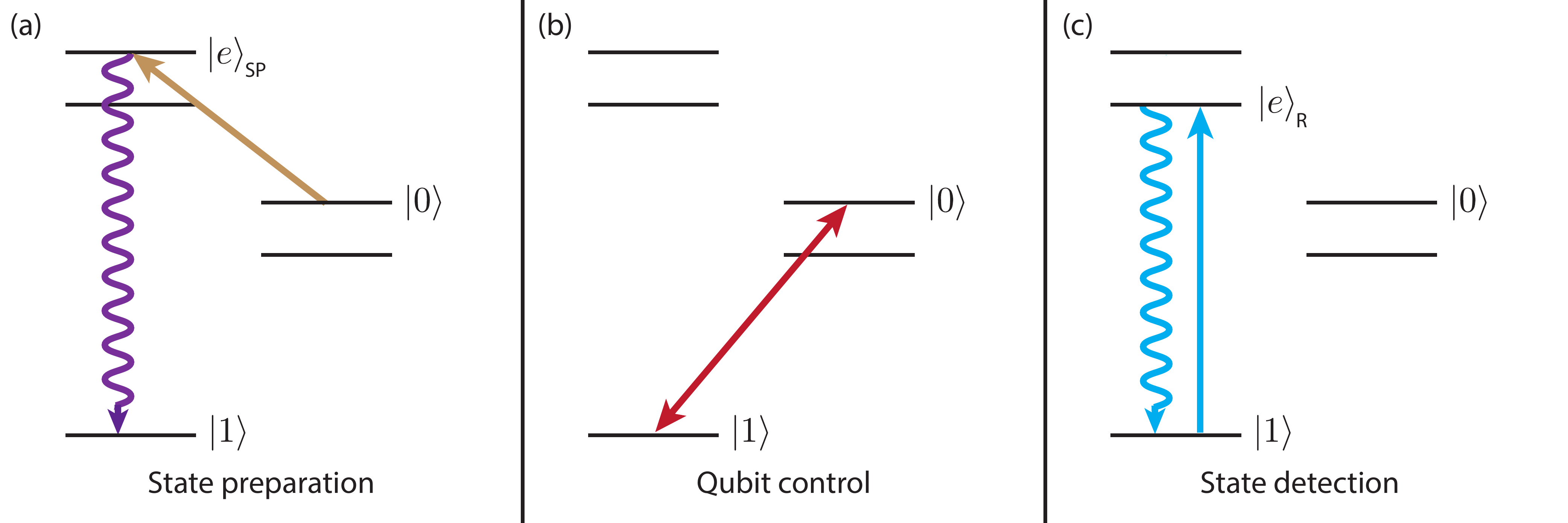}
\caption{Simplified depiction of state preparation, control, and detection in trapped-ion optical qubits. (a) The ion can be quickly optically pumped to the $\one$ state by coupling the long-lived $\zero$ state to an auxiliary state $|e\rangle_{\mathrm{SP}}$ that rapidly decays. (b) Qubit control is achieved by directly coupling the $\zero$ and $\one$ states using a narrow electric quadrupole transition. (c) Readout is achieved by shining light resonant on the broad transition $\one \rightarrow | e \rangle_{\mathrm{R}}$, and collecting the resulting scattered fluorescence photons. There is no similar transition $\zero \rightarrow | e \rangle_{\mathrm{R}}$ so the $\zero$ state appears dark.}
\label{fig:stateprep}
\end{figure*}

Initialization and readout in trapped ions are both performed by laser manipulation of the ion internal and motional states. These operations are shown schematically in Fig.~\ref{fig:stateprep} for an optical qubit. Initialization is performed via optical pumping into the $\one$ state, often accompanied by cooling of the ions' quantized motion to the trap harmonic oscillator ground state.  State readout is likewise very simple: a resonant laser couples the $\one$ state to a cycling transition which scatters many photons that can be collected by a detector, while no similar transition exists for the $\zero$ state which therefore remains dark. High-fidelity state preparation and readout have both been performed in less than 1 ms \cite{MyersonReadoutIons2008, HartyHighFidelityIons2014, CrainSNSPDdetect2019} (see Section \ref{Detection} for more details).

Trapped-ion qubits have also allowed for a demonstration of a universal, high-fidelity set of quantum gates. Laser or microwave drives applied to the ions allow arbitrary and high-fidelity single-qubit rotations to be performed. In addition, a two-qubit entangling gate is required, which is typically chosen to be the CNOT gate \cite{BarencoElemGates1995}. Trapped-ion entangling gates utilize the shared motional modes of two or more ions as a bus for transferring quantum information among ions, with a few single-qubit rotations required to transform such an operation on two qubits into a CNOT. Several schemes to perform these two-qubit gates have been proposed \cite{CiracZollerGate, MolmerSorensenGate, LeibfriedDidiGate2003} and demonstrated with high fidelity for both hyperfine qubits \cite{Ballance2QubitHyperfineGate2016} and optical qubits \cite{BenhelmMSGate2008}. The demonstrated single- and two-qubit gates combine to achieve a universal gate set for quantum computation. Typical single-qubit gate times are on the order of a few microseconds, with two-qubit gate times typically $10$--$100$~$\mu$s (though some have been performed faster). The achieved gate fidelities are sufficient to be compatible with error-correction schemes such as the surface code \cite{RaussendorfSurfaceCode2007}. Meanwhile, ion coherence times are much longer than gate times, with achieved values---depending upon qubit type---ranging from 0.2~s in optical qubits~\cite{BermudezAssessing2017} to up to 600 s for hyperfine qubits~\cite{bollinger_IEEE_550s_ramsey,Wang10MinuteCoherence2017}. The combination of long coherence times and a universal set of quantum gates thus fulfills the remaining two of DiVincenzo's criteria.

DiVincenzo's original paper also specified two additional criteria for quantum communications purposes: the ability to interconvert between stationary and so-called ``flying'' qubits (which would likely be photons with quantum information encoded in polarization, frequency, or phase), and the ability to transmit these flying qubits from one location to another with high fidelity. These criteria are not essential if the goal is to build a stationary large-scale quantum computer, but would be necessary for some other applications including quantum networks. Furthermore, some proposals for realizing a quantum processor rely on photonic interconnects between medium-scale modules of trapped ions \cite{MonroeModularArch2014}. Ions themselves---although they may be ``shuttled'' around the surface of a microfabricated trap---are unlikely to themselves be the flying qubits used for long-distance quantum communication or quantum networks, but high-fidelity entanglement between ions and photons has been demonstrated \cite{BlinovIonPhotonEntangle2004}.

In summary, ions satisfy the five main DiVincenzo criteria for QC and the ability to transfer their quantum information to flying qubits has also been achieved. In fact, all of these criteria for trapped-ion qubits have essentially been satisfied since 2004 \cite{LeibfriedDidiGate2003,BlinovIonPhotonEntangle2004}, yet the largest fully-controlled quantum register of trapped ions has contained only 20 ions \cite{Friis20QubitEntanglement2018}. As with other qubit technologies, it has become clear that---in any practical sense---there are other criteria which must be fulfilled to make trapped-ion quantum computers scalable; these additional criteria are discussed in Section \ref{Sec:Practical} below.

\subsection{Pros and Cons of Trapped Ions as Qubits}
\label{ProsandCons}

Trapped ions are recognized as having several advantages over competing qubit modalities. One of these is their coherence times, which can be exceptionally long for all four types of qubits enumerated above. Hyperfine qubit coherence times as high as 50~s have been achieved without using spin-echo or other dynamical decoupling techniques \cite{PhysRevLett.113.220501} and, as mentioned in Sec.~\ref{sec:DiV_criteria}, such coherence times were extended up to 600~s with the aid of dynamical decoupling \cite{bollinger_IEEE_550s_ramsey,Wang10MinuteCoherence2017}. These coherence times are effectively $T_2$ times, limited by technical sources of dephasing rather than by the fundamental state lifetime. With two-qubit gate times of typically $1$ to $100$~$\mu$s, even the achieved coherence times represent ratios of coherence time to gate time of ${\sim}10^6$, a far higher ratio than has been achieved for superconducting qubits (${\sim}1000$) \cite{BarendsSCQubitgate2014} or for Rydberg atom qubits (${\sim}200$) \cite{LevineRydbergGate2018}.

Another advantage is that both single and two-qubit gates can be implemented with very high fidelity using trapped ions. Single-qubit rotations, with fidelities as high as $99.9999 \%$ have been achieved \cite{HartyHighFidelityIons2014}, which surpasses the performance of any other modality. In addition, two-qubit entangling gates have been demonstrated with fidelities as high as $99.9 \%$ for hyperfine qubits \cite{Ballance2QubitHyperfineGate2016, nist_gate_2016} and $99.6 \%$ for optical qubits \cite{ErhardBlattCycleBench2019}, with only superconducting qubits achieving comparable performance.

State preparation and readout are also straightforward for trapped ions. The use of lasers for measurement enabled readout fidelity greater than $99.99 \%$ in less than 200 $\mu$s detection time \cite{MyersonReadoutIons2008} and $99.93\%$ in 11~$\mu$s \cite{CrainSNSPDdetect2019}.  Additionally, combined laser-based state preparation and readout with $99.93 \%$ fidelity was demonstrated \cite{HartyHighFidelityIons2014}, from which state preparation errors of $2 \times 10^{-4}$ were inferred. The achieved initialization and readout fidelities are better than those demonstrated in any other qubit technology.

Trapped ions also benefit from the fact that all ions of a given species and isotope are fundamentally identical. Thus, the microwave or laser frequency required to address each ion in the system will be the same and each ion will have the same coherence time.  This improves the reproducibility of the qubits and limits the number of calibration steps which are required at the beginning of the computation when compared with technologies such as superconducting qubits.  This is because the superconducting qubit frequencies and coherence times are defined and affected by fabrication and so will vary slightly from qubit to qubit due to fabrication process variability; these properties in superconducting qubits have also been observed to vary with thermal cycling \cite{KilmovGoogleSCqubitvary2018}.  At the same time, taking advantage of the benefit of the identical nature of ions requires that spatially-varying external perturbations to the trapped-ion qubit (such as magnetic field inhomogeneities, Stark shifts, or decoherence-inducing noise) be minimized or trapped-ion qubits at different locations will \emph{de facto} have different frequencies or coherence times.

While any ion contains additional internal states beyond the simplistic structure shown in Fig \ref{fig:stateprep}, the number of additional levels that must be accounted for in performing quantum operations is small when compared with the continuum of additional states that exist in solid-state qubits. While this additional ion internal structure must be accounted for when performing a quantum computation, the existence of some additional states---such as a short-lifetime state which can be used for readout---is a useful feature. At the same time, off-resonant light shifts and photon scattering can degrade quantum operations, and there is often the possibility that the ion becomes trapped in an undesirable internal state, one other than the cycling transition or $\zero$ and $\one$ states (i.e. a \emph{leakage} error occurs). Additional repumping lasers are needed to reinitialize the ion into the $\zero$ state, which add to the complexity of the system.

As mentioned previously, an ion can be trapped for many hours, or in some cases up to months for heavier ion species in deep traps, without being lost. While these lifetimes are long, they are not infinite and, as a result, the need to reload lost ions and to correct for computation errors due to their loss is a complication when compared with some modalities. However, some other promising QC modalities, including Rydberg atoms in optical lattices, suffer from much shorter lifetimes.

While trapped ions have demonstrated the highest ratio of coherence time to gate operation time for any qubit technology, their absolute gate speeds are much slower than those of some other types of qubits. High-fidelity two-qubit gates for trapped ions have been demonstrated as fast as $1.6$~$\mu$s \cite{SchaferFastIonGates2018}, but two-qubit gates in superconducting qubits have been performed in tens of nanoseconds. Depending on the number of operations required, a trapped-ion based quantum computation may take a considerable amount of time even if it is ultimately successful. One recent estimate put the time to factor a 1024-bit and 2048-bit number using a trapped-ion based quantum computer, with optimistic but achievable gate and readout parameters, at ${\sim}10$~days and ${\sim}100$~days, respectively \cite{LekitscheMicrowaveBlueprint2017}. Long gate times may also pose a challenge for trapped-ion quantum processors to perform meaningful quantum simulations or calculations in the near term. Achieving ``quantum supremacy'' \cite{HarrowQuantumSupremacy}, where a quantum processor can outperform the best classical processor for a task, may be difficult if the gate speed in a classical computer (${\sim}10$ GHz) greatly exceeds that in a trapped-ion quantum processor (${\sim}1$ MHz). One promising avenue of research is to perform entangling gates using sequences of ultrafast pulses \cite{WongCamposUltrafast2017} or shaped pulses of continuous-wave light~\cite{SchaferFastIonGates2018}, but so far fidelities for sub-microsecond gates have not exceeded $76 \%$.

Finally, while it is in principle easy to trap larger and larger numbers of ions in linear chains \cite{PaganoCryoChains2018} or two-dimensional arrays \cite{BruzewiczArrayLoading2016}, in practice the scaling to larger numbers of trapped ions has been slow. Arrays of up to thousands of superconducting qubits---such as the D-Wave 2000Q machine \cite{DWave}---have been fabricated with elementary control over each qubit, although these large arrays have limited connectivity, typically very short coherence times, and have not been used to demonstrate entanglement even between two qubits. While clouds of many thousands of ions can easily be trapped in deep macroscopic RF traps, such large clouds typically afford little meaningful control over individual ions and lack ion-specific readout. The largest systems of trapped ions with meaningful control and readout include 300-ion crystals in Penning traps \cite{BohnetSpinDynamics2016} and linear chains of ${\sim}100$ ions in RF traps \cite{PaganoCryoChains2018}; neither of these systems has yet demonstrated entanglement between arbitrary ions in the system. The difficulties of implementing the necessary optical and electronic control have slowed progress towards larger numbers of trapped ions as compared with other technologies where analogous control elements are cofabricated into the qubit chip itself. At the same time, trapped ions have made greater strides in performing high-fidelity operations \cite{HartyHighFidelityIons2014, Ballance2QubitHyperfineGate2016} and quantum algorithms on small numbers of qubits \cite{LanyonDigSim2011, LinkeArchCompare2017}. The winning technological modality for large-scale quantum computation is still far from certain.

\subsection{Considerations for Scaling a Trapped-Ion Quantum Computer}
\label{Sec:Practical}

A \emph{scalable} computer is one where the number of basic computational elements can be increased on demand without suffering a loss in performance and without an incommensurate increase in cost, energy usage, or footprint. While this increase will of course not be possible without bound, it is imperative that it allows for a marked improvement in functionality for some practical task.  Classical computers achieved scalability in that, for a period of many decades, the empirical rule of thumb known as Moore's Law was followed: the number of transistors that could be placed on a single chip doubled roughly every 18 months. Achieving scalability in a QC technology would mean that the number of available qubits could similarly be increased rapidly, over at least several orders of magnitude, while maintaining full quantum control of the system, achieving high-fidelity gates, and retaining long coherence times. No QC technology currently achieves scalability in this sense.

There are a number of approaches and capabilities which will likely be required to achieve a scalable quantum computer.  The first approach is that of modularity, in which a larger system is built through the combination of smaller subsystems.  In such a modular system, each subsystem can be built and tested independently, has a particular and well-defined functionality, and is compatible with the other subsystems. Modularity not only provides a means to predict and assess full system performance via tests and measurements on the individual components, but also allows the manufacturing process for one component to be tailored to achieve desired functionality with minimal impact on the others.  It is likely that modularity will be required to increase the scale of quantum computers, as it has played such an important role in large-scale classical technologies.  However, we note that the need to generate and maintain entangled states that span multiple modules may introduce challenges unique to quantum technology; these challenges would then need to be addressed to exploit the full benefits of a modular approach.

Another approach that may be necessary to achieve scalability is monolithic integration.  Monolithic integration is the technique of combining functions into a single component such as a microfabricated chip, as has been realized for classical computers.  Monolithic integration and modularity are complementary approaches.  For instance, on-chip control components for ion systems (such as waveguides for light delivery or on-chip detectors) can be considered modular to the degree that their fabrication and functionality can be made independent of other monolithically integrated components or other subsystems of the overall ion-trapping system. Such chip-integrated elements represent an important path towards scalability and we discuss them further in Sec.~\ref{ScalableHardware}. At the same time, integrated components introduce additional challenges: they require more complex fabrication techniques and better process reliability than simpler ion traps. Hence, some aspects of a scalable technology will likely still need to be made up of independent components. Ultimately, a modularity hierarchy may be required, with some elements monolithically integrated, in much the same way that monolithic microprocessor cores are placed together as modules in today's highest-performance classical computers.

A key capability needed for scaling---mentioned in DiVincenzo's original paper---is a mechanism for error correction. The first quantum error correcting codes were introduced in the mid-nineties \cite{PhysRevA.52.R2493,CalderbankErrorCorrection1996,SteaneErrorCorrection1996}, whereas more recent error-correcting codes have improved on these by reducing the necessary requirements for gate fidelity \cite{RaussendorfSurfaceCode2007}. Most codes work by encoding information in a logical qubit which is made up of multiple physical qubits, and thus introduce significant overhead, in terms of the number of qubits required to perform a given calculation, as well as in gate count. A physical arrangement of qubits that is compatible with an error-correcting code, and which can accommodate enough qubits to deal with the necessary overhead, is thus necessary to achieve scalable QC. Furthermore, gate errors must be reduced below the threshold for fault tolerance \cite{GottesmanFTQC1998}. At present, the highest thresholds, which are typically calculated assuming only depolarizing error channels, are on the order of $1 \%$ error \cite{RaussendorfSurfaceCode2007}.  This gives a rough idea of the gate fidelities that are required, though the depolarizing error model likely leads to overestimates of the true threshold of a realistic system that has additional coherent errors which can arise, for example, from a miscalibration of gates. It is important to note that the amount of overhead increases dramatically as the error rate approaches this threshold. In a practical sense, all gate errors must be reduced to significantly below this threshold for error correction to become feasible.

For this reason, an architecture which allows robust and low-error operations on many qubits is also a necessity for QC. However, this architecture must inherently be able to accommodate large numbers of trapped ions as qubits while allowing for high-fidelity gates, readout, and other key operations to be performed on any ion. Furthermore, the architecture must allow all of these necessary operations to be performed on the qubits without fidelity being degraded by crosstalk or other effects of scaling.

Some means of ensuring a sufficient degree of connectivity within the architecture will likewise be necessary as entanglement will need to be generated among qubits throughout the quantum computer.  In principle nearest-neighbor connectivity is sufficient, but higher degrees of connectivity may be beneficial as well. Higher levels of connectivity may require the ability to move individual ions within the architecture, so that two-qubit gates between different pairs of ions can be implemented. It may instead be possible to achieve high connectivity among ions in a large linear chain, though entangling gates suffer from slower speed and/or reduced fidelity due to the presence of many motional modes.  Techniques have been developed to mitigate this concern, which utilize temporal variation of the amplitude \cite{ChoiMultimodeIonControl2014}, frequency \cite{LeungFreqMod2QGate2018}, or phase \cite{milne2018phase} of the optical fields that couple to the multiple collective modes of motion in the ion chain.  However, entangling operations with these methods have not yet been demonstrated for chains of more than 5 ion qubits.

Physically maintaining a large array of qubits for the duration of a computation will also be required.  While this can be taken for granted in many systems, it is not necessarily straightforward with trapped ions since ions are sometimes lost from the trap due to collisions with background gas molecules or other experimental imperfections. From a QC perspective, ion loss can be seen as an amplitude-damping error which can be corrected by suitable codes as long as the loss can be detected in a state-insensitive way and the lost ion can be reliably reloaded \cite{ValaAtomLossCorrection2005}. Even for very long ion lifetimes, e.g. $> 24$~h, in large arrays with tens of thousands (or more) ions, one ion would be lost every few seconds (or faster). Thus a method of rapidly reloading ions without disturbing the coherence of other ions involved in the computation \cite{Bruzewicz2016} seems necessary for large-scale systems.

Methods of scalably addressing and measuring a large array of ion qubits will also be needed. Nearly all trapped-ion experiments currently make use of bulk optics to route and focus laser beams needed for state manipulation of ions, as well as to collect fluorescence from ions to measure them.  Likewise, nearly all make use of external voltage supplies to control the DC and RF voltages required for robust ion trapping. The challenge of working with the sheer number of bulk optics or external supplies required to control a large-scale quantum processor seems likely to become intractable unless some methods to improve the scalability of control are introduced. One option is monolithic integration of photonics and electronics into ion traps to interface with ions~\cite{MehtaIntegrated2016, StuartDACTRap2018}.

In this review, we will focus our discussion of different trapped-ion QC methodologies and technologies on what currently known challenges must be overcome to reach any reasonable level of scalability with a particular approach. We emphasize that there are many outstanding questions in the field of trapped-ion QC and it is hard to be sure of which approaches will ultimately bear fruit; as trapped-ion systems move from the few-qubit scale to hundreds or thousands of qubits, new challenges will certainly appear.

Specifically, in Secs. \ref{sec_trapped_ion_qubits} and \ref{sec_ion_control}, we will discuss the basic elements required for trapped-ion QC, namely the ion qubits themselves and the general methods for their control. An understanding of these basic elements is necessary to determine what methodologies and technologies are likely to help enable scalabilty, and we will discuss these methodologies in Sec. \ref{PracticalQC} and these technologies in Sec \ref{ScalableHardware}.  In Sec. \ref{Outlook} we will explore the near-term outlook for trapped-ion systems that utilize these methodologies and technologies and discuss the impact particular choices will have on prospects for scalability.  In addition, we will highlight experiments that might be performed in the near future to help understand this impact even further.

\section{Trapped Ions as Qubits}
\label{sec_trapped_ion_qubits}

Individual atomic ions were first suggested for use as quantum bits in a quantum computer more than twenty years ago~\cite{CiracZollerGate}.  The proposal for their use in this manner grew out of the development of single-ion atomic clocks.  Both applications benefit from the isolation from the environment and the resulting long coherence times available in the electronic states of trapped ions.  Additional benefits of trapped ions for QC are the combination of short- and long-lived electronic levels, shared vibrational states in the trapping potential, and the ability to couple the electronic and motional states using electromagnetic radiation.  In this section, we discuss ion trapping methods, such that individual ions can be maintained for long periods of time in a very small volume.  We also describe the states, internal and external, used for trapped-ion QC, as well as the fundamental and technological limitations to their quantum coherence properties.

\subsection{Trapping Individual Ions}

One of the chief advantages of trapped ions for QC is the straightforward methodology for localizing individual atomic ions for long periods of time.  While trapping of charged particles in three dimensions is not possible with static electric fields alone, a time-dependent electric field or a combination of static electric and magnetic fields can allow for localization, such that an effective average potential that can confine charged particles is created~\cite{DEHMELT196853,RevModPhys.62.531}.

\subsubsection{Types of Ion Traps}

Ions are typically maintained in space using either Penning or Paul traps; in the former, a static electric field provides confinement in one, axial dimension, while a parallel static magnetic field allows for confinement in the two perpendicular, radial directions.  In the latter, an oscillating electric field sets up a ponderomotive confining pseudopotential in two or three dimensions.  In the case of cylindrically symmetric trapping due to this oscillating field, an additional static field can be applied for trapping in the third, axial dimension.  In the presence of ultra-high vacuum conditions, and with careful consideration of trap parameters to satisfy effective potential stability requirements, charged particles including atomic ions can be held in these types of traps for hours, days, and even months in some cases~\cite{PhysRevLett.65.1317}.

Penning traps provide the ability to maintain large, two-dimensional ion crystals if the trap frequency in the direction parallel to the magnetic field is made much higher than the frequencies in the perpendicular directions,   Due to the radial component of the electric field in combination with the magnetic field, however, these crystals rotate at constant angular velocity, and are stationary only in a frame rotating with respect to the laboratory at that rate.  Stroboscopic methods of address can be used to control ions in such a system, but most work to date has effected uniform excitation.  Recent work using these systems has resulted in the creation of many-body entanglement in large 2D ion crystals of hundreds of ions~\cite{Britton2012} with application to quantum simulation of critical systems and non-equilibrium dynamics in general~\cite{Garttner2017,PhysRevLett.121.040503}, as well as enhanced quantum sensing~\cite{PhysRevLett.118.263602}.  However, since it is generally more straightforward to individually manipulate ions that are part of stationary arrays, Paul traps, with oscillating electric fields in the RF range, are the main focus for researchers in QC.  Significant literature concerned with solving the equations of motion of ions in an RF Paul trap exists, e.g.~\cite{RevModPhys.62.531,Wineland1998,leibfried2003quantum}, and hence we only summarize it here.

\subsubsection{Paul Traps for QC}

RF trapping relies on time variation of a potential that, at any instant in time, is anti-confining in at least one dimension.  Confinement using this time-variation is enabled due to the inertia of a massive charged particle.  It is thus clear that the stability of an ion in such a ponderomotive trapping potential created in this manner would depend on the parameters of both the RF potential, as well as of the ion itself.  In fact, the motion of an ion in an RF Paul trap satisfies the Mathieu equation~\cite{Wineland1998} and depends in detail on these parameters, i.e. the ion's charge-to-mass ratio, the RF frequency, the RF amplitude, and the curvature of the potential.  Solutions to the Mathieu equation result in so-called ``secular'' harmonic bounded motion at a frequency typically  somewhat less than half the RF drive frequency.  Upon the secular motion is superimposed a higher frequency motion, at the RF drive frequency, known as ``micromotion,'' and its amplitude is in general time-dependent.  The stability of RF Paul traps used for containment of singly-ionized atoms requires the voltage amplitude and frequency of the applied RF to fall in a certain range.  Traps being explored for QC applications have ion-electrode distances in the range of 30~$\mu$m to 1~mm, leading to RF voltage amplitudes of 10-1000~V at 10-100~MHz, depending on the exact trap size and atomic species.

\begin{figure*}[t b h]
\includegraphics[width=2.0\columnwidth]{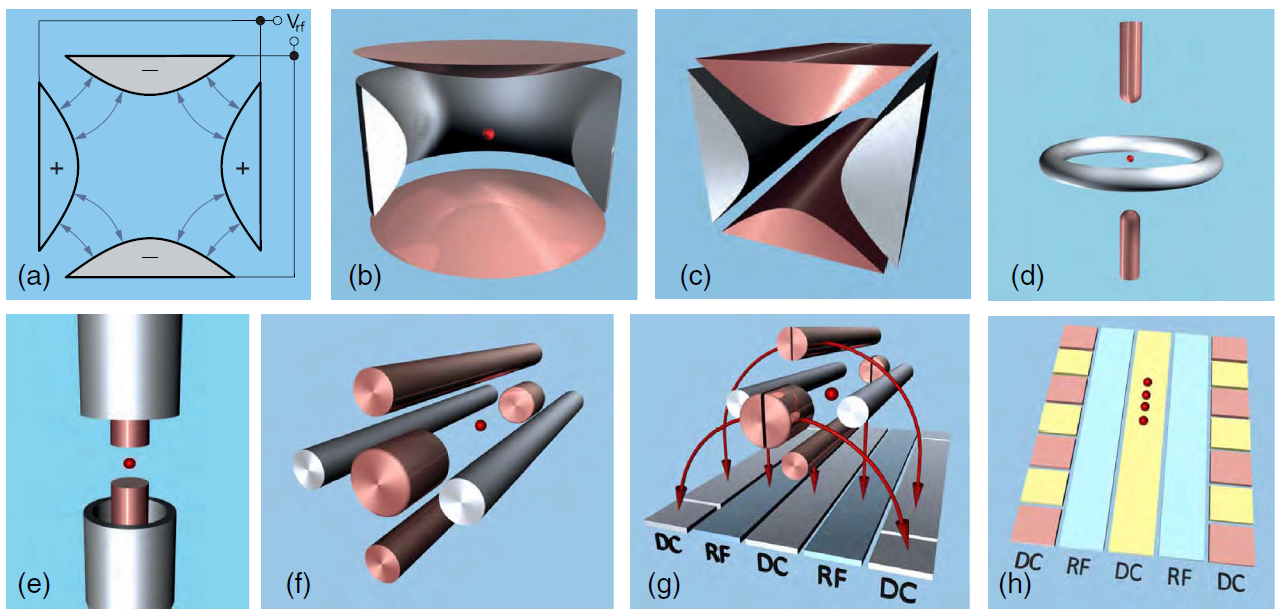}
\caption{(Reproduced from~\cite{brownnutt_2015}.) RF Paul trap geometries. (a) The basic concept of RF trapping, where quadrupolar fields oscillating at an RF frequency are produced using a set of (parabolic) electrodes.  (b) The simplest cylindrically symmetric version of the basic RF trap.  This is of the ``ring and endcap'' point-trap geometry.  (c) The simplest translationally symmetric version of the basic RF trap.  This will form a quadrupole mass filter and can be used to make a linear trap.  (d,e) Topologically equivalent deformations of the geometry shown in (b).  (f) Topologically equivalent deformations of the geometry shown in (c) with additional endcap electrodes added to form a four-rod, linear trap.  (g) The four-rod trap in (f) may be deformed such that all electrodes reside in a single plane, forming a linear ``surface-electrode trap.''  (h) A subset of the electrodes in a linear trap [a surface-electrode trap is depicted here, but segmentation may be applied to other linear trap geometries, such as that shown in (f)] may be segmented to allow trapping in multiple zones, along the axial direction.}
\label{fig:rf_trap_types}
\end{figure*}

The two main configurations of Paul trap that are used in QC are quadrupolar electrode layouts that lead to RF trapping in all three dimensions, known as point traps, and those which have two-dimensional RF trapping plus static electric-field trapping in the third dimension, known as linear traps (See Fig.~\ref{fig:rf_trap_types}).  In a point trap there is only one point, known as the RF null, where the RF field is zero.  Therefore, when more than a single ion is held in a point trap, the ions will in general suffer excess micromotion, motion at the RF frequency whose amplitude is proportional to the distance between an ion and the RF null.  Micromotion can in some cases lead to RF heating of ions~\cite{Wineland1998}, reducing quantum logic fidelities.  Linear traps, on the other hand, have zero RF field along a line, in general.  This means ions can be held in a 1D crystal along this line without suffering excess micromotion.  Moreover, through concerted variation of the static field that is responsible for trapping along the axial direction, ions can be moved along the RF null in this direction such that ion crystals may be separated into constituent ions, or vice versa, and ions can be independently transported between zones of an array~\cite{NIST:ion_transport:2002}.  This capability is a key component of some proposed architectures for large-scale trapped-ion QC, as will be discussed in Sec. \ref{PracticalQC}.

Traditional RF traps for trapped-ion QC are fashioned from metallic electrodes geometrically arranged to create the largest fields for given voltages (see Fig.~\ref{fig:rf_trap_types}a, b, and c).  The optical access required to deliver and collect light to and from ions, as well as ease of fabrication, are beneficial features of such traps.  While the optimal shape for electrodes would match the (hyperbolic) equipotential surfaces of a quadrupolar field, in practice, much simpler shapes are used.  Point traps can be formed using a ``ring and endcap'' geometry (Fig.~~\ref{fig:rf_trap_types}d and e), in which an RF potential is applied between a ring and two cylindrical electrodes placed symmetrically above and below the ring along its line of cylindrical symmetry.  This forms a three-dimensional quadrupolar field with the RF null at the center of the ring.  Linear traps can be formed using four parallel rods placed at the corners of a square, much like a quadrupole mass filter, such that an RF potential is applied between pairs of diametrically opposed rods (Fig.~~\ref{fig:rf_trap_types}f).  This forms an RF null along the line of symmetry between and parallel to the rods.  Trapping along this axial direction can be accomplished either via two endcap electrodes placed along the RF null at opposite ends of the rods, or via segments of the rods at either end, to which static electric voltages are applied to create a harmonic potential along the axial direction.

In terms of QC, where large numbers of ions will be needed to surpass the capabilities of classical computers, desiderata include traps which can contain many ions that may be individually addressable, thus forming what has been termed in the field an ``ion register.''  Putting more than one ion into a point trap leads to undesired micromotion as described above, but one possible architecture consists of arrays of point traps, each containing a single ion.  In a linear trap, however, multiple ions may be trapped along the RF null in a linear array for a sufficiently strong radial potential compared to the axial potential; this produces a linear ion register or ion chain.  For a harmonic potential in the axial direction, the ions will in general not be spaced equally; their positions are set by the equalization of the harmonic trap forces and the nonlinear Coulomb repulsion of co-trapped ions~\cite{James1998}.  A consequence of this is that the ion spacing is independent of the mass for ions of the same charge, so multispecies ion crystals in a linear trap will be spaced identically regardless of composition.  This is not true for confinement in a point trap or for radial confinement in a linear trap since the RF pseudopotential is mass dependent.  Non-harmonic potentials may be applied along the axis of a linear trap in order to obtain equal spacing, but as a practical matter, this generally requires much larger voltages on a subset of the electrodes~\cite{PaganoCryoChains2018}.

\subsubsection{Miniature, Microfabricated, and Surface-Electrode Traps}
\label{MicroTraps}

Both point and linear Paul traps were first (and in some cases continue to be) constructed of macroscopic, conventionally machined metal pieces, but beginning approximately two decades ago, miniature traps made from laser-etched insulating substrates, selectively coated with patterned metal electrodes, were created in the hopes of obtaining smaller, more precisely defined structures~\cite{NIST:ion_transport:2002}.  While these goals were partially achieved, and these devices are still in use for many experiments, the fact that the substrates were held together with conventional mechanical means, such as bolts and alignment rods, limited the attainable precision and level of complexity.  Subsequently, microfabrication techniques were utilized to create trapping structures with micron-scale (or better) precision and alignment accuracy, as well as access to increased complexity enabled by this accuracy in combination with the parallel pattern definition afforded by photolithographic methods.

There have been a few notable demonstrations of complex non-microfabricated linear traps with multiple, non-co-linear segments to allow movement and reordering of ions along multiple paths and through junctions~\cite{HensingerTJunction2006,blakestad_junction_2009}, some still in use.  But the move toward lithographic techniques in conjunction with multi-layer pattern alignment through microfabrication~\cite{Stick2005,NIST:SET:PRL:06} has ushered in the current era of more complex trap design, including examples of multi-linear-segment array structures with hundreds of separate electrode segments~\cite{amini_racetrack}, segmented circular rings~\cite{PhysRevApplied.4.031001}, multi-site point trap arrays~\cite{Sterling2014,BruzewiczArrayLoading2016,kumph_array,Mielenz2016}, and traps with electrodes with switchable or variable RF amplitudes, or of varying geometry across a linear, segmented region ~\cite{KimIntFiberTrap2011,kumph_array,PhysRevLett.120.023201,PhysRevA.97.020302_2018}.

Some of these advanced designs are based on the ``surface-electrode'' architecture for ion traps~\cite{NIST:SET:QIC:05}.  In contrast to the three-dimensional nature of the electrode geometry for the point and linear Paul traps described above, surface-electrode traps contain all the electrodes in a single plane.  They are essentially a deformation of the three-dimensional geometries onto a surface, with trapping potential minima (the RF null, either a point or a line) formed above the surface of the plane.  This can be accomplished for a point trap by, e.g., taking a ring-and-endcap trap and allowing the bottom endcap to become a region in the center of a plane, transforming the ring RF electrode into an annular region surrounding the planar endcap, and deforming the top endcap to be the entirety of the plane outside the ring annulus~\cite{PhysRevA.78.063410}.  Similarly, for a linear trap, the four rods can be deformed into four or five long, parallel electrodes in the plane, with RF electrodes alternating with DC ones (Fig.~~\ref{fig:rf_trap_types}g); a subset of them can be segmented along their length for application of static fields for axial confinement~\cite{NIST:SET:QIC:05} (Fig.~~\ref{fig:rf_trap_types}h).  The surface-electrode paradigm has the advantages of substantial optical access to the ions, more straightforward design and simulation~\cite{PhysRevA.78.033402,PhysRevA.78.063410,PhysRevLett.102.233002_2009,Hong2016}, and straightforward 2D microfabrication, while also allowing for integration of additional control components beneath the electrodes, making it very amenable to combination with, e.g., CMOS-based technologies~\cite{mehta_cmos_2014}.  The drawbacks include lower trap frequencies and potential depths for the same applied voltage, but these effects are not severe, and the benefits of this platform have enabled significant progress in trap functionality and integration~\cite{VanDevenderFiberTrap2010,sandia_junction_2011,KimIntFiberTrap2011,Allcock2012,oxford_muwave_trap_2013,MehtaIntegrated2016,kim_group_mirror_2016,Willitsch_group_junctions_2017,Ghadimi2017}, much of which is described in more detail in later sections of this review.

We note that the quadrupolar-field generating electrode structure of a Penning trap may also be unfolded into a plane, such that charged particles may be trapped above such a trap in the presence of a magnetic field oriented perpendicular (and in some cases parallel~\cite{verdu_2011}) to the surface.  Such surface-electrode Penning traps have been explored for QC-based experiments~\cite{Stahl2005,planar_penning_designs_2010,PhysRevA.81.052335_2010}, but they have not seen wide use for ion-based QC as of yet.

\subsubsection{Loading Ions into Traps}
All trapped ion experiments begin by loading one or more ions into the trap. This process involves the ionization of a neutral precursor and subsequent confinement of the charged species. Due to the comparatively deep (${\sim}0.1$ to 1~eV) ion trap depths, and subsequently long trapping lifetimes, many experiments can be carried out following successful loading of the trap. As experiments continue to become more complex, comprising large arrays of many ions, it is likely to become necessary to be able to reload the ion register quickly even for single-ion trap lifetimes of many hours~\cite{BruzewiczArrayLoading2016}.

In many of the earliest experiments~\cite{NeuhauserTrappedIon1980,raizen1992ionic}, ion traps were loaded from a hot, neutral atomic vapor subject to electron bombardment. The electron bombardment technique is non-resonant and can therefore be readily applied to different atomic species. However, this general purpose loading scheme lacks isotopic selectivity, often giving rise to ion registers with defects consisting of unwanted isotopes present in the neutral precursor. Due to isotope frequency shifts, registers with such defects cannot easily be controlled with high fidelity, making them impractical for scalable quantum processing. The electrons used for bombardment can also cause charging of exposed dielectrics near to or part of the trap, which can affect trapping potentials and stability.

Defect loading can be reduced by orders of magnitude by using an alternate scheme based on resonance-enhanced photoionization~\cite{kjaergaard2000isotope,GuldePhotoIonize2001}. This technique exploits isotope frequency shifts to excite only the desired isotope with high probability to an ionizing state. Due to the relatively large ionization energies of the atoms generally used as trapped ion qubits, the excitation is often done in at least two steps using photons of different energies, at least one of which is typically in or near the UV part of the spectrum (notable exceptions are Be$^{+}$ and Mg$^{+}$, typically formed via single-wavelength, two-step photoionization~\cite{Wolf2018,kjaergaard2000isotope}). The first step is generally resonant with a strong bound-to-bound optical transition and can often be saturated with modest laser intensity. At this modest first-step laser intensity, the detuned excitation probability for other isotopes is greatly reduced. The second step, which must be executed before the atom spontaneously decays or leaves the trapping volume, need not be resonant, as the atom is excited either to the free-electron continuum or, as in the case of Sr, to a broad auto-ionizing state. This second step is generally not saturated and is therefore often driven with higher laser intensity in order to achieve high photoionization rates. Unfortunately, high laser intensities, especially for laser beams in the UV, have been shown to cause charging in microfabricated ion traps~\cite{harlander2010trapped,wang2011laser}.  Alternate photoionization pathways that use a larger number of lower energy photons have been explored and may be useful in applications that are particularly sensitive to stray fields due to charging~\cite{zhang2017realizing}.

Trap performance can also be degraded following the deposition of the neutral precursor atoms onto the electrode surface. This contamination is especially dangerous when using microfabricated surface-electrode traps, as the precursor metal can cause shorting between the electrodes if the inter-electrode dielectric is not undercut. The technique of backside loading, which uses an atomic beam that propagates through a hole in the trap chip, is widely used \cite{NIST:Bksideloading:APL:09,stick2010demonstration,merrill2011demonstration}. This approach becomes more difficult as the number of ions is increased, since it will require either more apertures (with the concomitant risks of charging of the hole edges and perturbation of the trapped ions), or a ``loading zone'' located far from the computation regions of the trap. More recently, alternate approaches that employ laser cooling of the the neutral atoms have been reported~\cite{cetina2007bright,sage2012loading,BruzewiczArrayLoading2016}. Lowering the temperature of the atomic vapor compresses the Boltzmann velocity distribution such that a larger fraction of the incident flux can potentially be trapped, permitting high loading rates with much reduced deposition. Further, laser cooling can provide additional levels of isotopic selectivity. For example, recent experiments have studied loading from remotely-located 2D and 3D magneto-optical traps (MOTs) of neutral strontium and calcium~\cite{sage2012loading,BruzewiczArrayLoading2016,BruzewiczQLAR2017}. The transitions used for laser cooling and subsequent acceleration of the pre-cooled atoms to the ion trap for ionization are all subject to isotope frequency shifts, and the probability of loading the wrong isotope is greatly reduced due to the multiple stages of resonant laser excitation. The demonstration via this method of site-selective loading in an ion-array trap \cite{BruzewiczArrayLoading2016} also showed that the coherence of an ion at one array location could be maintained while loading in different array sites, which will become increasingly important as the number of simultaneously-trapped ions increases.

\subsection{Internal States:  Qubit Levels}
\label{subInternal}

\begin{figure*}[t b h]
\includegraphics[width=2.0\columnwidth]{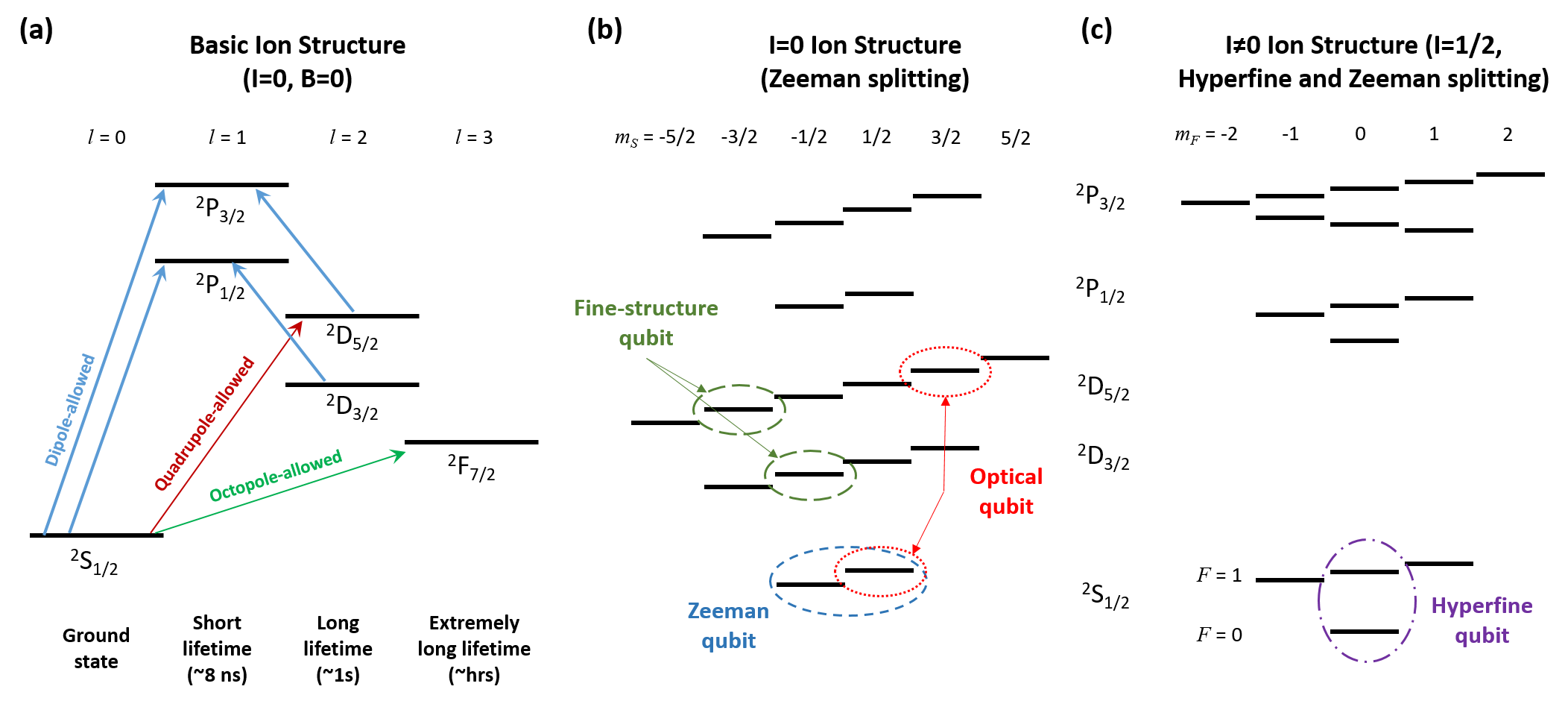}
\caption{Level structure and pairs of levels utilized for qubits in monovalent ions (energy splittings not to scale).  (a) Basic electronic structure of ions used in QC.  All have $S$ and $P$ levels.  If present, low-lying $D$ (as in Ca$^{+}$, etc.) and $F$ (as in Yb$^{+}$) levels require quadrupole or octupole transitions, respectively, from the ground state.  (b) Structure of zero-nuclear-spin ($I=0$, typically even isotopes) ions in a small magnetic field.  Example Zeeman, optical, and fine-structure qubit level choices are depicted.  (c) Structure of non-zero-nuclear-spin ($I\neq0$, odd isotopes) ions in a small magnetic field.  An $I=1/2$ level structure is depicted and $D$ and $F$ levels are omitted for clarity.  Hyperfine levels are labeled in the ground state only.  A ``clock''-type hyperfine qubit is depicted, but all four qubit types can be implemented in $I\neq0$ ions (if $D$ or $F$ levels are present).  Typical (order of magnitude) level splittings for the various qubit types are:  Zeeman qubits, 1--10~MHz; Optical qubits, 100--1000~THz; fine-structure qubits; 1--10~THz; hyperfine qubits, 1--10~GHz.  Levels are labeled using spectroscopic notation with the principle quantum number omitted, i.e. as $^{2s+1}L_{j}$, where $s$ is the total spin quantum number ($1/2$ in the case of a single valence electron), $L$ refers to the orbital momentum quantum number written as $S, P, D, F, \ldots$, and $j$ is the total angular momentum quantum number.}
\label{fig:qubit_types}
\end{figure*}

The multitude of states of the valence electrons in the mostly Group-II or Group-II-like atomic ions used for QC experiments allows for many choices of qubit.  Pairs of states employed for qubit levels can come from any combination of long-lived levels in the ground or metastable manifolds of ions with or without nuclear spin or low-lying $D$~levels.  Non-zero nuclear spin, as is present in odd-mass isotopes of ions of interest (and in even-mass isotopes with net nuclear spin), generates hyperfine levels due to interaction of the nuclear spin with the valence electron.  The ground-state hyperfine levels are some of the most long-lived states available, with spontaneous-emission-limited lifetimes approaching the age of the universe.  Low-lying $D$~levels, which are present in several ions of interest, form metastable manifolds with lifetimes in the range of seconds.  Low-lying $F$~levels, as exist in, e.g., Yb$^{+}$, can also be used; these have even longer lifetimes than the $D$ states, but their extremely narrow linewidth means that significant optical power is required to drive transitions to these levels (for an equivalent gate time).  Furthermore, the laser linewidth needs to be especially narrow in order to take advantage of the extended coherence that can result from the longer lifetime. The addition of a non-zero magnetic field splits the Zeeman sublevels in the ground and metastable manifolds, creating many well-defined and addressable levels.  Figure~\ref{fig:qubit_types} shows  a basic level structure diagram of species of interest for QC; this figure also depicts level choices for the various types of qubit described below.

The states used for qubits almost always include one from the ground state manifold (for an exception, see Sec.~\ref{subsubFine-Structure} below).  The other state can be another Zeeman sublevel or another hyperfine level in the same manifold; in these cases, we will refer to these qubits as Zeeman or hyperfine qubits, respectively.  If the other state is instead a level in a $D$~state manifold, we will refer to these qubits as optical qubits.  Energy splittings of these types of qubits are typically in the megahertz range for Zeeman qubits, the gigahertz range for hyperfine qubits, and the hundreds-of-terahertz range for the optical qubits.  Each has advantages and drawbacks for QC as will be described below, but all have been used in recent experiments and demonstrations in the field.

\subsubsection{Zeeman Qubits}
\label{subsubZeeman}

Zeeman qubits, consisting of a pair of states in the same electronic orbital and hyperfine level, and separated by megahertz frequencies by means of a small magnetic field, offer essentially infinite qubit lifetimes while allowing one to take advantage of the simpler level structure of the even-isotope ions.  These species have straightforward methods for state preparation, Doppler and sideband cooling and optical pumping, and the small splitting between neighboring Zeeman levels affords addressing with a minimal set of laser frequencies.  Single and two-qubit logic operations are typically performed using two-photon stimulated-Raman transitions, with two beams derived from the same laser that is tuned near resonance with one of the $P$ levels in the ion.  These operations can in principle be performed using a direct RF drive near the qubit frequency, a few megahertz, but it is difficult to spatially focus radiation at this frequency, presenting a challenge to low cross-talk operation, and the long RF wavelength leads to the requirement of its use in combination with a higher-gradient magnetic-field to enact two-qubit logic.

State discrimination for Zeeman qubit measurement requires an auxiliary operation before resonant photon scattering.  This can be accomplished via shelving of one of the qubit levels in a metastable $D$ level, leading to a requirement that these levels are available.  This shelving must be done via an electric-quadrupole-allowed transition using resonant light in a small magnetic field.  This requires an additional laser that is narrow in linewidth and with appreciable intensity to transfer population from one of the qubit states to a sublevel in the $D$ manifold.  The purity of the state transfer in this case can be improved via multiple pulses to separate $D$ sublevels~\cite{KeselmanZeemanQubit2011}.

Zeeman qubits, almost by definition, generally have high sensitivity to magnetic-field variations.  Field fluctuations lead to varying rates of phase accrual in the qubit, and this looks like dephasing when averaged over multiple uncorrelated experimental instantiations.  Great care must be taken to shield the ions from magnetic field variation to achieve long coherence times.  Nonetheless, coherence times of 300~ms (and 2.1~s with dynamical decoupling pulses) have been achieved through use of mu-metal magnetic shielding of the ion vacuum chamber in combination with the use of permanent magnets for bias-field production~\cite{Ruster2016}.  The coherence is limited at this level by residual thermal fluctuations affecting both the shielding properties of the mu-metal and the magnetic moment of the permanent magnets, so significant improvement may require better temperature control and/or new materials with better magnetic properties (assuming other sources of magnetic technical noise do not begin to limit coherence).

\subsubsection{Hyperfine Qubits}

Hyperfine qubits, consisting of a pair of states in the ground-state hyperfine manifold, can offer the long lifetimes afforded to Zeeman qubits, while also allowing for a high degree of magnetic-field-fluctuation insensitivity, easing many of the challenges associated with obtaining long coherence times.  In addition, state detection is more straightforward than with a Zeeman qubit, since there is a significant qubit splitting.  The price paid for these advantages is a more complicated level structure as is present in the odd-isotope ions that possess hyperfine levels, leading to more lasers, or laser frequency components, to address all the electronic levels for state preparation and measurement.

Hyperfine qubits based on pairs of so-called ``stretched'' states, i.e. the highest (or lowest) $z$-projection Zeeman levels in each hyperfine level, allow for straightforward state preparation and detection using circularly-polarized light.  The qubit state in the higher hypefine level $F'$, with $m_{F}=F'$, can be prepared (detected) via excitation to the $m_{F}=F'+1$ sublevel in the $P_{3/2}$ manifold in the presence (absence) of a repumping light component that couples the lower $F=F$, $m_{F}=F$ level to the $F'$ level through an upper state.  Qubits of this type are susceptible to magnetic-field noise due to their stretched-state composition.  The so called ``clock'' states, however, with $m_{F},m_{F'}=0$, provide a qubit that is first-order insensitive to magnetic field at $B=0$.  In practice, working at zero field is not convenient, due to the frequency selectivity provided by a small quantizing field.  Furthermore, laser cooling and readout of ions are inhibited at zero field due to the creation of dark states that prevent cycling transitions from being maintained using static laser polarizations \cite{BerkelandDarkStates2002}. Therefore, clock state qubits are typically operated in a regime with a reduced, but not zero, first-order sensitivity to magnetic field.  This nonetheless leads to increased coherence times when compared with stretched-state qubits.  Most experiments utilizing $^{171}$Yb$^{+}$ are based on its clock-state qubit in a small magnetic bias field, with demonstrated coherence times in the range of seconds~\cite{PhysRevA.76.052314}, or even as high as 600~s with the use of dynamical decoupling \cite{Wang10MinuteCoherence2017}.  Other experimenters using this ion employ a variation on this theme, where ``dressed states,'' superpositions of the ground-state hyperfine levels created by means of application of multiple RF coupling fields, are used to obtain similar coherence times~\cite{Timoney2011,PhysRevLett.111.140501}.  The potential advantage of the dressed-state hyperfine qubits is one of addressability; the qubit can be tuned using a magnetic field such that different frequencies may be used to control different ions in a magnetic field gradient.  This comes at a cost of experimental complexity and potential challenges with RF crosstalk and magnetic field gradient fluctuations.

At intermediate magnetic fields there exist other pairs of hyperfine sublevels whose difference is insensitive to magnetic field to first order.  These finite-field, clock-type qubits, which we will refer to as first-order field insensitive (FOFI) qubits, provide the practical utility of operating at a nonzero field while also possessing extremely low sensitivity to field fluctuations.  FOFI qubits have been demonstrated to have coherence times of minutes~\cite{bollinger_IEEE_550s_ramsey,HartyHighFidelityIons2014}.  These coherence times can be obtained in standard Ramsey-type measurements without dynamical decoupling or refocusing~\cite{PhysRev.80.580,ViolaDynamicDecoup1998}, meaning no algorithmic reconfiguration is required to use them in long experiments.  Current limitations to the coherence times obtained with FOFI qubits are technical in nature~\cite{langer2005long,HartyHighFidelityIons2014} and include residual magnetic field drift, fluctuations in trap RF voltage amplitude which lead to fluctuating AC Zeeman shifts, and even instability of the local oscillator used to make the measurements.

\subsubsection{Optical Qubits}

Optical qubits, consisting of a state in the ground-state manifold and a state in a metastable $D$~level, can benefit from the straightforward level structure of the zero-nuclear-spin ions while also utilizing quantum logic control wavelengths in the visible to near-IR region of the spectrum.  One drawback is the fact that the lasers used for control of optical qubits must be made narrow, around 1~Hz, to fully take advantage of the second-scale lifetimes available.  Moreover, since the laser is essentially the local oscillator for the optical qubit, phase fluctuations in the laser lead directly to decoherence in the qubit; if magnetic fields are controlled well, the laser is often the limiting factor in optical qubit coherence time. With work in the last two decades toward stabilizing visible and near-IR lasers \cite{ZHAOSubHertz2010}, however, it is relatively routine to get optical sources narrower than 100~Hz (commercial lasers are even available with hertz-scale linewidths~\footnote{Stable Laser Systems, Boulder, CO, USA}), with the best lasers at the sub-hertz level~\cite{Kessler2012}.  Experimenters have achieved upwards of 0.2~s coherence times for optical qubits in zero-nuclear-spin ions with careful control of the laser linewidth and optical component vibration~\cite{BermudezAssessing2017}.  The ultimate limit in coherence time for optical qubits will however be set by the upper-state decay time, typically approximately one to tens of seconds.

With one quantum state of the qubit being an optically-separated, metastable state, optical qubits allow for very high detection efficiency based on electron shelving~\cite{dehmelt_shelving}.  This technique allows for near-unit detection efficiency based on resonance fluorescence; application of light resonant with a transition from the ion ground state to an auxiliary rapidly decaying $P$ level will produce light upon decay from the auxiliary level if the qubit is projected to the ground state.  In contrast, the metastable upper level of the optical qubit is far off-resonant with this light, and so the ion will remain dark if the qubit is projected to the upper state.  Up to the decay time of the upper state, quantum non-demolition measurement can continue, as the measurement process will not further change the state after projection, and therefore a high signal-to-noise ratio is attainable, even for small fluorescence collection efficiency.  As mentioned in Sec.~\ref{subsubZeeman}, Zeeman qubits are typically measured in this manner with transfer of one qubit state to a metastable level after which measurement proceeds as for an optical qubit.

Perhaps most important for scalability, the lasers needed for direct optical qubit excitation are in the red to near IR for many ion species of interest.  Integrated technologies such as optical waveguides for on-chip routing and grating-based waveguide-to-free-space couplers, as will be discussed in more detail in Sec.~\ref{IntPhot}, are much more challenging to fabricate for blue and UV wavelengths as feature size scales roughly with wavelength; fixed fabrication and design tolerances hence lead to bigger errors for smaller wavelengths.  Moreover, scattering loss in the waveguide (due to surface roughness) increases at lower wavelengths.  Even for near-term experiments, where free-space and fiber optics will be utilized predominantly, the optical quality and consistency of components made for use in the red and IR is far superior to those for use in the blue and UV.  The qubit-control beams also have the highest intensity requirements of all the wavelengths needed, independent of qubit type, for ion QC, suggesting that they be at the most friendly wavelengths possible.  All these scalability arguments highlight the favorability of optical qubits as systems are scaled up.

We note that FOFI-type optical qubits exist, in non-zero nuclear-spin ions, where one of the qubit states is in the ground state manifold and the other is in the $D$~state manifold.  Due to the rather small hyperfine splitting in the metastable state, these transitions can be at conveniently low magnetic fields; on the other hand, this also means that the level splittings can be in the tens of megahertz range, potentially giving rise to large AC-Zeeman shifts in the case of imperfect trap potentials~\cite{PhysRevA.75.032506}.

\subsubsection{Fine-Structure Qubits}
\label{subsubFine-Structure}

It is also possible to use a pair of states in the $D$ manifold, one from each of the fine-structure split levels $D_{3/2}$ and $D_{5/2}$, to form a qubit with energy splittings in the terahertz range~\cite{PhysRevA.81.032322}.  Like the optical qubits, lifetimes (due to leakage, not relaxation to the other qubit state, in this case) are typically in the second range.  Quantum logic can proceed either via Raman transitions using two IR laser beams tuned near the $P$ levels, or potentially directly at terahertz frequencies, though both generating narrow-linewidth terahertz radiation at arbitrary frequencies and addressing individual ions using this radiation represent challenges. The Raman method is similar to that used with the Zeeman and hyperfine qubits, although in this case, the two Raman fields are much farther apart in frequency, typically requiring two separate phase-locked lasers, with the degree of relative phase stability as a potential limit to coherence times.  These lasers are in the IR, however, and are therefore more straightforwardly scalable, via integrated photonics technologies, than the blue and UV lasers needed for Raman transitions in the Zeeman and hyperfine qubits.  Like Zeeman qubits, detection requires transfer from one of the qubit levels to another manifold.  In this case it is relatively straightforward, as this transfer is accomplished using the laser that is typically applied to repump from the $D_{3/2}$ level during detection of an optical qubit, so the same techniques, with the afforded high detection efficiency, are available.

\subsection{Motional States}

A powerful aspect of trapped ions is their combination of long-lived internal states and external, shared vibrational states in a system that allows for their independent or coupled manipulation.  For ions of interest to QC, these vibrational states of the harmonic trapping potential typically have frequencies in the megahertz range, set by the potentials applied to trap electrodes as described above.  The ladder of harmonic oscillator states is set by this splitting.  With multiple ions in the trap, the vibrational levels are shared, as they correspond to normal modes of motion of the coupled ion harmonic oscillators.  For $N$ ions, there are $3N$ of these normal modes of vibration, essentially phonon modes of the ion crystal, and each mode $i$ can be in a superposition of its harmonic oscillator levels $|n\rangle_{i}$, where $n=0, 1, 2, \ldots$; since many ions participate in each mode, the modes act like a quantum bus.  And since lasers can be used to excite the internal electronic levels dependent upon the ions' vibrational states, the motional bus allows coupling of the internal electronic levels of separate ions.

The strength of the coupling between the internal electronic states and the motional levels $|n\rangle$ of a particular mode is set by the red and blue ``sideband'' Rabi frequencies, $\Omega_{r}=\eta \Omega_{0} \sqrt{n}$ and  $\Omega_{b}=\eta \Omega_{0} \sqrt{n+1}$ respectively, where $\Omega_{0}$ is the Rabi frequency for the corresponding electronic transition that does not couple to the motion (the so-called ``carrier'' Rabi frequency), and $\eta$ is the Lamb-Dicke parameter which characterizes the strength with which an electromagnetic field couples to the ion motion.  The Lamb-Dicke parameter is given by $\eta=k z_{0}\cos\theta$ for an optical field with wavevector $k$ oriented at an angle $\theta$ with respect to the direction of the motional mode, and a trapped-ion of mass $m$ whose ground-state wavefunction has a width $z_{0}=\sqrt{\hbar/(2 m \omega)}$, set primarily by the mode oscillation frequency $\omega$.  Experiments are often performed in the so-called Lamb-Dicke limit, where $\eta\sqrt{n+1}\ll 1$, due to the tractable dynamics and high fidelity afforded by an effectively reduced set of transitions involving the motion.  Here the transitions on the red and blue sidebands of a mode correspond to terms in an effective system Hamiltonian in which the excitation of the internal electronic state is accompanied by the decrement or increment, respectively, of the phononic mode excitation by a single vibrational quantum, equivalent in energy to the Planck constant times the mode frequency.  These sideband transitions are the basic components of multiqubit quantum logic in ion systems, and their use for this purpose will be highlighted in Sec.~\ref{Multiqubitgates}.

The controlled excitation of motional states and their coupling to ion internal states has been described in detail elsewhere~\cite{Wineland1998,leibfried2003quantum,ozeri_tutorial_2011}, so here we will focus on decoherence of motional states and a primary cause of that decoherence, anomalous motional heating.  This is a current practical limit to multi-qubit gate fidelity, and it will be a hindrance to miniaturization of trap structures for higher-frequency quantum logic.

\subsubsection{Motional State Decoherence}

The motional states of trapped ions are influenced by the local electric field environment; electric-field noise can heat the system, changing the motional state (a $T_{1}$-type process), but fluctuations will in general also lead to decoherence of motional-state superpositions (a $T_{2}$-type process).  While heating is primarily due to noise near resonant with the ion's secular mode frequencies~\cite{brownnutt_2015} (and in some cases near the trap RF drive frequency~\cite{blakestad_junction_2009,sedlacek_tnoise_2018}) due to the high quality factor of ion oscillation in an electromagnetic trap, lower frequency noise, up to the secular frequency~\cite{PhysRevA.93.043415_2016}, can lead to motional state decoherence without heating.  For instance, slow trap-frequency fluctuations, on the time-scale of experiments, alter the mode frequency, changing the superposition phase evolution, effectively leading to motional decoherence over many experiments.  Ramsey experiments using superpositions of Fock states of a vibrational mode (with the ion in the same internal state in both cases) can be used to measure this decoherence rate~\cite{PhysRevA.62.053807_2000,innsbruck_decoherence_2003,oxford_decoherence_2007,PhysRevA.93.043415_2016}.  These measurements generally find rough agreement between the motional decoherence rate and the heating rate from the ground state to the first excited state.  Superpositions of larger states, however, decay faster, as is typically seen in quantum mechanical settings~\cite{PhysRevA.62.053807_2000,brownnutt_2015}.

Since motional heating is the primary motional decoherence mechanism in most cases, we discuss it further in the next section.  However, recent work exploring trap-frequency fluctuations highlights the importance of this low-frequency noise source for scalable trapped-ion QC~\cite{nist_gate_2016}.  As many of the relevant ion wavelengths are in the UV part of the spectrum, time-dependent trap frequencies can be due to charging and discharging of photo-electrons onto and off insulators that are part of the trap or support apparatus.  Environmental temperature fluctuations can also bring about drift in power supplies used to generate the voltages applied to electrodes.   Methods for quantum-enhanced frequency measurement including Fock state interferometry have recently been employed to measure typical fluctuations and drifts in trap frequencies~\cite{schmidt_et_al_fock_state_measurement_2018,nist_fock_interferometry_2018}.  The results show fractional trap-frequency fluctuations at the $10^{-6}$--$10^{-5}$ level on the tens to hundreds of seconds timescale.  Keeping this stability level across a large array of traps, or improving it as will likely be required to reach fault-tolerant two-qubit gate fidelities, is an engineering challenge to large-scale ion QC.

\subsubsection{Anomalous Motional Heating}
\label{AnomHeating}

As first discovered a couple of decades ago~\cite{turchette_2000}, electric-field noise near the secular trap frequency that causes the ion motional mode occupation to increase incoherently is widely observed, and the resultant heating rates are much larger than would be expected from known sources, such as Johnson noise from the electrode metal, blackbody radiation, or background gas collisions~\cite{brownnutt_2015}.  Due to its unknown source, this heating is termed ``anomalous.''

Motional heating leads directly to an error in multi-qubit logic gates \cite{MolmerSorensenGate} based on the Coulomb interaction since quantum-bus mode decoherence is a source of gate infidelity.  In cases where this error is a significant contribution to the overall infidelity, its mitigation is paramount. Thus, the existence of anomalous motional heating (AMH) has  implications for scalability.  Due to the strong observed scaling of AMH with ion-electrode distance $d$, approximately $d^{-4}$~\cite{PhysRevLett.97.103007_2006,hite_mckay_kotler_leibfried_wineland_pappas_2017,PhysRevLett.120.023201,PhysRevA.97.020302_2018},  miniaturization of ion trap arrays is not straightforward.  Most methodologies for multi-qubit logic gates in trapped-ion systems, and all techniques that have been demonstrated with high fidelity, are limited in speed by the trap frequency, assuming the required control field intensity is available.  The trap frequency can be increased with larger applied potentials or smaller ion-electrode distances; applied voltage is limited, however, by dielectric breakdown (in vacuum or along surfaces), and in this case the achievable frequency will generally scale as $d^{-1/2}$ (due to the requirement of maintaining RF-trap stability while scaling trap size down~\cite{NIST:SET:QIC:05}).  AMH is therefore a potential roadblock to high-speed, high-fidelity quantum logic due to these scalings with $d$. On the other hand, using a large ion-electrode separation to minimize ion heating, and as a result operating more slowly, leads to requirements of large voltages, RF currents, and overall physical processor sizes; maintaining stability over an extended area is a challenge due to the deleterious effects of temperature and magnetic field gradients that will exist in any real system.  Vibration sensitivity will grow with system size as well.

Recent experiments have shed some light on ion heating, although its origins are not generally understood, except in a few experiments where technical noise has been found to predominate \cite{brownnutt_2015} or one experiment where an ion trap was specifically designed to have atypically high thermal (Johnson) noise \cite{innsbruck_first_time_2018}.  It has been demonstrated that  traps show a reduction in AMH of approximately two orders of magnitude upon cooling the electrodes from room temperature to approximately 4~K~\cite{PhysRevLett.100.013001_2008_1,PhysRevA.91.041402_2015}, independent of material~\cite{PhysRevA.89.012318_2014}.  This suggests cryogenic operation to achieve the lowest electric-field noise.  Related to this finding, a state of superconductivity of the electrode material does not appear to alter AMH levels~\cite{Chuang_SC_trap_2010,PhysRevA.89.012318_2014}, giving weight to the hypothesis that AMH is not a bulk effect but is dominated by surface effects.  Along these lines, it has been shown that surface treatment of the electrodes can lead to lower levels of AMH at room temperature.  In particular, pulsed-laser treatment~\cite{laser_cleaning_2011}, plasma treatment~\cite{PhysRevA.92.020302_2015}, and energetic-ion milling~\cite{PhysRevLett.109.103001_2012,PhysRevB.89.245435_2014,NIST_Ne_2014,sedlacek_milling_2018} have all been shown to reduce electric-field noise that causes AMH for room-temperature traps.  The removal of surface contaminants and/or the alteration of surface morphology is therefore implicated in surface-generated electric-field reduction.  Moreover, it appears that after ion milling of the surface, material-dependent behavior is uncovered, with different trap materials exhibiting different AMH amplitudes as a function of temperature~\cite{sedlacek_milling_2018}.  This suggests that making systems more scalable will include determining which trap-electrode materials or surface passivation techniques provide sufficient mitigation of AMH.

Carbon contaminants have been implicated as a contributor to AMH~\cite{PhysRevA.95.033407_2017}, but it is not clear which carbon-containing compounds are the most deleterious, and how much coverage is required to cause significant heating (e.g. some readsorption of carbonaceous contaminants has been shown not to increase heating rates~\cite{PhysRevB.89.245435_2014}).  Moreover, the carbon-based contamination present on ion trap electrodes most likely varies in composition depending on the fabrication steps, cleaning methods, and local environments of fabrication and testing facilities.  There is also the complication of high-temperature baking that is often required to obtain UHV pressures.  It is expected that additional carbonaceous contaminants accrue during the baking process~\cite{hite_colombe_wilson_allcock_leibfried_wineland_pappas_2013,PhysRevB.89.245435_2014}, and there is some evidence that unbaked systems have lower initial heating rates~\cite{PhysRevA.89.012318_2014,PhysRevA.91.041402_2015}, but the role of high-temperature baking in AMH has not been systematically investigated.  We note that there is evidence that the targeted removal, via a local chip bake, of water (and presumably other low-boiling-point solvents) remaining on the surface of electrodes in unbaked systems does not lead to a reduction in AMH~\cite{PhysRevA.91.041402_2015}, suggesting that amounts of water beyond the molecular level are not major contributors to ion heating.

For the smallest ion-electrode distances typically in use, $d=40$--$80$~$\mu$m, heating rates of atomic-ion species of interest can be made sufficiently low for fault-tolerant QC with high threshold codes~\cite{PhysRevA.80.052312_2009} with the use of \textit{in situ} ion milling treatment or cryogenic operation below 10~K or so.  However, in order to go to smaller structures to improve the likelihood of successful scalability of the trapped-ion platform, a more detailed understanding of AMH will be required, such that heating rates closer to the Johnson-noise limit are achieved.  The most straightforward route to further mitigation of AMH may be through study of various electrode materials and fabrication methods in conjunction with surface treatment techniques that can prepare a more ideal surface.  In addition, a more ideal surface is easier to model, potentially allowing for more effective prediction of electric-field noise properties of a particular surface.  Effort in this direction potentially includes electrode-film annealing~\cite{PhysRevLett.100.013001_2008_1} or high-temperature treatment~\cite{UCB_heater_heating_rates_2018} of the trap electrodes to heal defects or remove contaminants.  On the other hand, traps may be fabricated from more ideal materials, through e.g. epitaxial metal growth or the addition of self-assembled monolayers for passivation of the surface.

\section{Trapped Ion Qubit Control}
\label{sec_ion_control}

Any QC modality requires precise control in order to initialize the quantum state of the system, perform gate operations, and read out the final state. Trapped ions benefit from robust and high-fidelity methods of performing these key control operations. In this section, we will discuss methods for trapped-ion quantum control, the experimental performance achieved so far, and the implications of different methods for future scalability.  We also survey the key quantum computing experimental demonstrations preformed using these techniques.

\subsection{State Preparation}
Once loaded into the trap, the ion register must be prepared in the desired initial state before quantum operations can proceed. Unlike trap loading, however, high-fidelity initial state preparation must be repeated after each experimental realization. Certain operations, such as Doppler cooling and state-dependent fluorescence detection, can transfer ions to internal states outside of the subspace spanned by the $|0\rangle$ and  $|1\rangle$ qubit states. Even state measurement itself can project a superposition of $\zero$ and $\one$ into the long-lived $\zero$ state, that then must be quenched to avoid prohibitive delays in the experimental cycle time. Hence, it is necessary to optically pump the ions into either the desired initial state or into some intermediate state that can be coupled to the initial state with high fidelity. Optical pumping schemes can take a number of different forms, but they generally take advantage of photon absorption and emission selection rules to sequester quantum state amplitude in a single state with high probability after repeated absorption and emission cycles. Limitations to the state preparation fidelity include off-resonant excitation during photon absorption and residual branching to undesired levels. However, errors on the scale of $10^{-4}$ can be achieved \cite{HartyHighFidelityIons2014}.

In addition to internal state preparation, it is often necessary to control the ion register's motional state as well. Laser-based Doppler cooling is very useful for rapidly reducing the effective ion temperature to the milliKelvin scale, but for the trapping frequencies ($\sim$1~MHz) generally used in quantum processing experiments, this leaves the ions in a thermal distribution spread over several motional states.  When addressing small numbers of ions or controlling a small number of motional modes, resolved sideband cooling can be efficiently used to further lower the motional state occupation of the ion register \cite{DiedrichZeroPoint1989,MonroeCNOT1995}. Absorption of a photon tuned to a narrow red sideband transition associated with a particular motional mode reduces the state occupation of that mode. Subsequent state quenching and spontanteous decay, both of which favor keeping the motional state unchanged in the Lamb-Dicke regime, return the ion to the ground electronic state, permitting the cooling cycle to begin again.  As this technique cools only a single mode at a time and requires repeated resonant addressing of weak transitions, it can become prohibitively slow for large ion chains with many motional modes.

Alternatively, cooling can be achieved by altering the light absorption profile of the ion register using electromagnetic induced transparency (EIT). Typical Doppler cooling uses atomic transitions with natural linewidths outside the resolved sideband regime ($\Gamma\sim$ tens of megahertz), where red and blue sideband transitions are driven with approximately equal probability.  With a judicious choice of laser frequencies and polarizations, however, a $\Lambda$-level scheme can be used to inhibit photon scattering on blue sideband transitions and thereby preferentially reduce the motional state occupation. This technique was first applied to single ions \cite{PhysRevLett.85.4458,roos2000experimental} but has recently been extended to cool longer chains and simultaneously address multiple motional modes \cite{lechner2016electromagnetically}. Chains of up to 18 ions have been efficiently cooled using EIT to motional phonon occupations of around $0.01$--$0.02$, where performance has been limited by the purity of laser polarizations. The EIT technique has also been successfully applied to ion chains consisting of different atomic ion species~\cite{lin2013sympathetic}.

\subsection{Qubit Logic}

A quantum logic gate is simply any transformation which takes some qubits as inputs and transforms their state to an output state in a deterministic and reversible way. Coherent operations on the qubit can generally be characterized as gates, although irreversible operations on ion qubits, such as optical pumping and state measurement, are not characterized as gates. Any actual gate includes imperfections which effectively introduce random fluctuations into its performance, degrading the gate quality. In some cases these imperfections may themselves be reversible (for example, calibration errors in the amplitude or frequency of the control field), thus allowing them to be characterized as small-magnitude undesired gates. Other errors, including decay from the $\zero$ state in optical qubits, are stochastic and thus considered more as a leakage channel than an undesired gate. This section will discuss methods of performing quantum gates on ions, their fidelity, and ways of characterizing gate performance.

A quantum computer generally needs to perform arbitrary gates on arbitrarily large qubit registers in order to perform any significant computation. Fortunately, just as in a classical computer, it is generally possible to decompose an arbitrary quantum gate into a product of gates chosen from a much smaller gate set. A set of gates which can be combined to achieve any arbitrary gate is referred to as a \emph{universal} gate set. In classical computing, the NAND gate is universal: any other gate on a classical computer can be decomposed into a product of NANDs on various sets of the classical bits.

In a universal gate set for QC, at least one gate which generates entanglement between two qubits is required, such as the CNOT gate (which inverts the state of one qubit conditioned on the state of a second qubit). In fact, almost any two-qubit entangling gate can achieve universality if a few additional single-qubit rotations are also included in the gateset \cite{LloydAlmostAny1995, DiVincenzoUniversal2Qgate1995}. In practice, high-accuracy single qubit rotations of arbitrary angles and phases are comparatively easy to perform in ions, so universality is achieved by demonstrating a single two-qubit entangling gate.


    \subsubsection{Types of Gates: Optical, Raman, Microwave}
\label{Gatetypes}
As described in Sec.~\ref{subInternal}, trapped-ion qubits come in different types. Hyperfine qubits use two hyperfine internal states of the ion, typically separated by GHz frequencies, as $\zero$ and $\one$ states. For these ions single-qubit gates are implemented with microwaves or Raman transitions. Optical qubits use a metastable excited state as $\zero$ with a transition frequency in the optical range ($> 100$ THz). For these qubits, single-qubit gates can be performed with a single resonant laser.

Optical qubits typically use two internal states separated by an electric quadrupole $S \rightarrow D$ transition. With lifetimes of ${\sim}1$~s, lasers with linewidths approaching the hertz level are required to address and drive gates in these qubits. While longer-lifetime octupole qubits exist (see Sec. \ref{subInternal}), achievable coherence times are also limited by $T_2$ constraints due to laser phase noise and fluctuations in the optical path length, and very-long-lifetime optical transitions also require higher optical power to address. Thus these transitions have not been used as frequently. They remain promising for future QC efforts, particularly since millihertz-class lasers~\cite{Kessler2012} have recently been demonstrated for optical clocks.

\begin{figure}
    \centering
    \includegraphics[width=0.8\columnwidth]{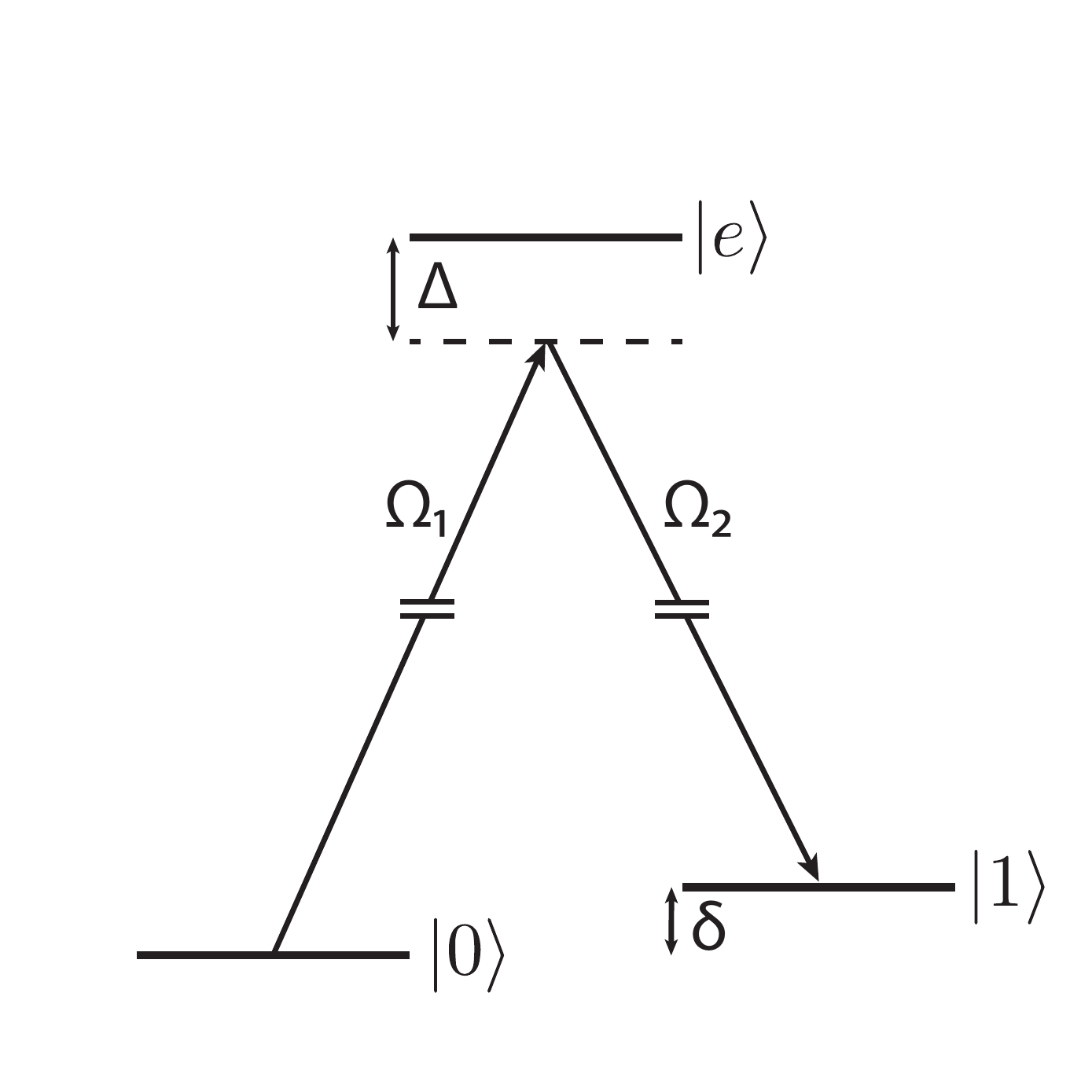}
    \caption{Schematic of level structure used in a stimulated Raman transition. The $\zero$ and $\one$ states are coupled by two lasers separated in frequency by the qubit splitting $\delta$. For sufficiently large Rabi frequencies $\Omega_{1,2}$ and Raman detuning $\Delta$, qubit transitions can be efficiently driven with negligible population of the lossy, short-lived state $|e \rangle$.}
    \label{fig:ramangate}
\end{figure}

For hyperfine qubits, laser-based gates can be implemented using stimulated Raman coupling, as shown in Fig.~\ref{fig:ramangate}. In this scheme, two laser beams detuned from a dipole-allowed transition and detuned from each other by the splitting of the $|0\rangle$ and  $|1\rangle$ states drive the desired qubit gates. Laser linewidth requirements can be reduced when using Raman gates, as the difference frequency between the two Raman beams, which drives the coherent operations, can be controlled very precisely using acousto-optical modulators powered by high-quality commercial RF and microwave synthesizers \cite{leibfried2003quantum}. Gates of this type are also frequently used for Zeeman qubits.

Gates on hyperfine qubits can also be implemented with a direct microwave drive at the GHz-scale transition frequency, which couples to the magnetic dipole of the ion. While a microwave horn can easily couple radiation to drive transitions in a single ion, the centimeter-scale wavelengths of microwaves make individual addressability very difficult when compared with gates driven with focused laser beams. A different approach is to use current-carrying wires integrated into a microfabricated chip trap to drive near-field microwave transitions \cite{OspelkausOscillating2008}. The use of near-field microwaves means that crosstalk to other zones on the same chip will be limited, but all ions trapped in the same zone will experience the gate. Controllably moving ions in and out of such microwave zones allows such gates to be implemented on one ion at a time in a multi-ion system.

One way to get around the requirement of moving ions in order to individually address them with microwaves is to utilize a magnetic-field gradient. For qubits with a first-order Zeeman shift, slightly changing the effective magnetic field seen by different ions can allow for spectroscopic addressing due to the resulting spatially-dependent Zeeman shifts \cite{MintertMicrowaveScheme2001}. In Ref. \cite{MintertMicrowaveScheme2001}, it was also shown that a magnetic gradient allows microwave drives to couple the internal and motional states of ions, enabling two-qubit gates to be performed (discussed further below). Magnetic gradients have been used to demonstrate single-ion addressing with a global RF drive \cite{JohanningRFAddress2009}, although crosstalk between ions (due to the finite magnetic gradient and Zeeman shifts) remains an issue.

    \subsubsection{Single Qubit Gates}
The dynamics of single-qubit gates in trapped-ion systems have been reviewed elsewhere~\cite{leibfried2003quantum}. Instead we focus on the recent leading gate times and fidelities of various qubit control protocols. A selection of state-of-the-art gate performances for a number of different schemes is given in Table~I.

\begin{table*}
\caption{Selected state-of-the-art gate demonstrations.}
\renewcommand{\arraystretch}{1.15}
\begin{tabular}{l c c c c c}
\hline
\hline
Gate & Gate & Fidelity & Gate Time & Ion & Ref. \\
 Type & Method & & ($\mu$s) & Species \\
\hline

Single-Qubit & & &\\
 & Optical & 0.99995 & 5 &$^{40}\mathrm{Ca}^{+}$ & \cite{BermudezAssessing2017}\\
 & Raman & 0.99993  & 7.5  & $^{43}\mathrm{Ca}^{+}$ &\cite{Ballance2QubitHyperfineGate2016}\\
 & Raman & 0.99996  & 2  & $^{9}\mathrm{Be}^{+}$ & \cite{nist_gate_2016} \\
 & Raman & 0.99  & 0.00005  & $^{171}\mathrm{Yb}^{+}$ & \cite{CampbellUltrafast2010} \\
 & Raman & 0.999  & 8  & $^{88}\mathrm{Sr}^{+}$ & \cite{KeselmanZeemanQubit2011} \\
 & Microwave & 0.999999 & 12& $^{43}\mathrm{Ca}^{+}$ &\cite{HartyHighFidelityIons2014}\\
 & Microwave &  & 0.0186& $^{25}\mathrm{Mg}^{+}$& \cite{ospelkaus2011microwave}\\
Two-Qubit & & &\\
(1 species) \\
 & Optical & 0.996 & --& $^{40}\mathrm{Ca}^{+}$ & \cite{ErhardBlattCycleBench2019}\\
 & Optical & 0.993 & 50& $^{40}\mathrm{Ca}^{+}$ & \cite{BenhelmMSGate2008}\\
 & Raman & 0.9991(6) & 30 &  $^{9}\mathrm{Be}^{+}$ & \cite{nist_gate_2016} \\
 & Raman & 0.999 & 100 &  $^{43}\mathrm{Ca}^{+}$ &  \cite{Ballance2QubitHyperfineGate2016}\\
 & Raman & 0.998 & 1.6 &  $^{43}\mathrm{Ca}^{+}$ & \cite{SchaferFastIonGates2018}\\
 & Raman & 0.60 & 0.5 &  $^{43}\mathrm{Ca}^{+}$ & \cite{SchaferFastIonGates2018}\\
 & Microwave & 0.997 & 3250 &  $^{43}\mathrm{Ca}^{+}$ & \cite{HartyNearFieldMicrowaves2016}\\
& (AC B-field gradient) & \\
 & Microwave & 0.985 & 2700 & $^{171}\mathrm{Yb}^{+}$ &\cite{WeidtMicrowaveGates2017}\\
 & (DC B-field gradient) &\\
Two-Qubit & & &\\
(2 species) \\
& Raman/Raman & 0.998(6) & 27.4 &  $^{40}\mathrm{Ca}^{+}/^{43}\mathrm{Ca}^{+}$ & \cite{BallanceHybridLogic2015} \\
& Raman/Raman & 0.979(1) & 35 &  $^{9}\mathrm{Be}^{+}/^{25}\mathrm{Mg}^{+}$ & \cite{TanMultiElement2015}\\
\hline
\hline
\end{tabular}
\label{table:gateperformance}
\end{table*}

For optical qubits, single-qubit gates have been performed in a few microseconds and have achieved fidelities up to $99.995 \%$ \cite{AkermanUniversal2015, BermudezAssessing2017}. The ease of ion-selective single-qubit gates with low crosstalk is a significant strength of optical qubits, however, the ultimate fidelity of these gates will be limited by the $T_1$ time of the excited state, which is on the order of 1 s for the commonly-used quadrupole transitions.

Microwave gates driven by an on-chip microwave antenna have achieved single-ion fidelities of $99.9999\%$ in 12 $\mu$s \cite{HartyHighFidelityIons2014}. Fidelities with Raman beams have been limited to about $99.993 \%$, with gate times of $7.5$~$\mu$s, due to off-resonant scattering \cite{Ballance2QubitHyperfineGate2016}. Optimization of single-qubit hyperfine gates has generally shown a tradeoff between gate speed and fidelity: in Ref.~\cite{Ballance2QubitHyperfineGate2016}, gates in less than $2$~$\mu$s were demonstrated but with errors of ${\sim}2 \times 10^{-4}$.

Zeeman single-qubit gates have been implemented via RF drive with $99.9 \%$ fidelity in $8$~$\mu$s \cite{KeselmanZeemanQubit2011} via an RF signal applied to a wire near the trap, although this demonstration did not include any means of site-selectively addressing one ion without crosstalk. Zeeman-qubit gates via Raman beams detuned from one another by the Zeeman resonance, which can allow ion selectivity by tightly focusing the beam, were also demonstrated \cite{PoschingerZeeman2009, Ruster2016} but at lower fidelities of $96 \%$ to $99 \%$.

Some effort has been made to increase the speed of single-qubit ion gates, albeit typically at the expense of accuracy. Far-detuned ultrafast pulses were used to drive Raman $\pi$-pulses in the $^{171}$Yb$^+$ hyperfine qubit in less than 50 ps with $99 \%$ fidelity \cite{CampbellUltrafast2010}. On-chip microwaves were used to drive gates in $^{25}$Mg$^+$ hyperfine qubits in less than 20 ns \cite{ospelkaus2011microwave}, although fidelity was not indicated in that publication.  The fidelity of single-qubit gates can generally be improved through the reduction of control-field noise or through the use of composite pulse sequences, as will be discussed further in Sec.~\ref{ErrorReducandMit}.

    \subsubsection{Multi Qubit Gates}
    \label{Multiqubitgates}
Multi-qubit gates entangle the internal and motional states of trapped ions by means of the Coulomb interaction \cite{James2QGatesReview2000}.  The first proposal of an entangling gate between two ions was made by Cirac and Zoller \cite{CiracZollerGate}, and launched the field of ion-trap QC. While many different gate schemes have been proposed since then, with various advantages over the Cirac-Zoller (CZ) gate, all multi-qubit trapped-ion gates proposed so far share the essential feature introduced by Cirac and Zoller: using the shared motional modes of ions as a bus to transfer quantum information between them. The CZ gate is a controlled phase gate, but a single-qubit rotation can transform it into the CNOT gate.

\begin{figure}
\includegraphics[width=\columnwidth]{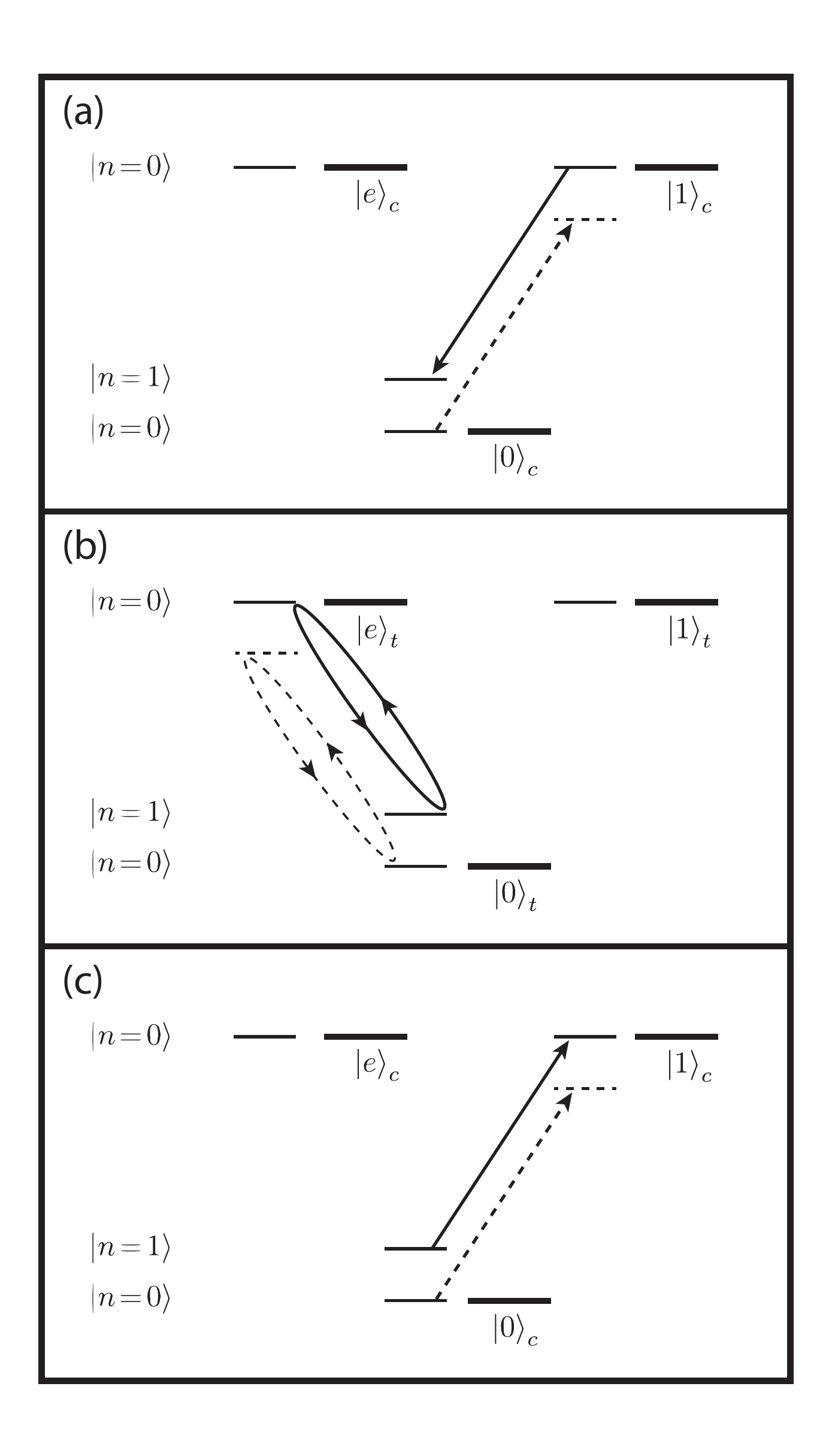}
\caption{Schematic representation of the action of the CZ gate.  a) A red sideband $\pi$-pulse on the control ion transfers amplitude from the $|1\rangle_{c}$  state to the ground electronic state $|0\rangle_{c}$ in the first excited collective motional state $|n=1\rangle$. b) A red sideband $2\pi$-pulse on the target ion through an auxiliary excited state $|e\rangle_{t}|n=0\rangle$ proceeds, conditioned on population of the $|n=1\rangle$ motional state in the first step. c) A final red sideband $\pi$-pulse on the control ion returns it to its initial state. Dashed lines denote forbidden transitions to nonexistent motional states.}
\label{CZFigure}
\end{figure}

The CZ gate itself requires the ions to be cooled to the ground state of their collective motion. For an explanation of this gate, we will denote the internal states of ion $j$ as $\zero_j$, $\one_j$ (with $j \in c,t$ denoting the control and target ions, respectively) and the shared motional state as $| n = 0 \rangle$, $| n = 1 \rangle$, etc. An initial $\pi$-pulse applied to the control ion and detuned to the red motional sideband excites the system to $|n = 1 \rangle$ if the control ion is in $\one$ but leaves the system unchanged if the control ion is in $\zero$ (since there is no lower-$n$ state for the motion to reach). A $2 \pi$-pulse, also tuned to the red motional sideband, is then applied to the target ion. This $2 \pi$ pulse does not excite to the state $\one$ but instead to an ``auxiliary'' excited state, distinguished from $\one$ by polarization or, in some later demonstrations, frequency (see Fig. \ref{CZFigure}). The target ion can be rotated through the auxiliary state and picks up an overall negative sign if it is in state $\zero$ and if the ions' collective motion is in $|n = 1 \rangle$. The target ion cannot be rotated if it is in $\one$ because this laser's polarization (or frequency) does not allow it to couple to the ground state, and it cannot be rotated in $\zero |n = 0 \rangle$ because the drive is red-detuned. A final $\pi$-pulse on the control ion will return the control ion to its initial state. The resulting state transformation looks like:
\begin{eqnarray}
\zero_c \zero_t \rightarrow \zero_c \zero_t \\
\zero_c \one_t \rightarrow \zero_c \one_t \\
\one_c \zero_t \rightarrow \one_c \zero_t \\
\one_c \one_t \rightarrow - \one_c \one_t
\end{eqnarray}
The gate thus inverts the phase of only the $\one \one$ state, realizing an entangling controlled-phase interaction. Besides cooling to the motional ground state, the CZ gate requires individual addressing of each ion and multiple polarizations for the drive laser. Despite these limitations, a modified CZ interaction was demonstrated the same year it was proposed~\cite{MonroeCNOT1995}, entangling the internal state and motional state of a single $^9$Be$^+$ ion. In 1998, a two-ion entangling gate with fidelity of 0.7 was demonstrated between two Be$^+$ ions with gate time of $\sim 10 \, \mu$s \cite{TurchetteEntanglement1998}, while a Cirac-Zoller gate and single-qubit rotations were used to implement the CNOT operations on two trapped $^{40}$Ca$^+$ ions with $71 \%$ fidelity in $600$~$\mu$s \cite{SchmidtKalerCNOT2003}. A Cirac-Zoller gate was later implemented with $77 \%$ fidelity on the $1.82$ THz transition separating the $D_{3/2}$ and $D_{5/2}$ states in $^{40}$Ca$^+$ in 400~$\mu$s~\cite{PhysRevA.81.032322}.

The requirement that the ions remain in the motional ground state is a significant limitation on the original Cirac-Zoller proposal. As discussed in Sec.~\ref{AnomHeating}, even when the ions have been cooled to the motional ground state, they can be subsequently heated by electric-field noise. In 1999, M{\o}lmer and S{\o}rensen introduced a controlled-phase gate which could be implemented without the need to be in the motional ground state \cite{MolmerSorensenGate}. The M{\o}lmer-S{\o}rensen (MS) gate generates a state-dependent force with bichromatic laser fields tuned near first-order sideband transitions. The motional-state wavepacket executes a closed trajectory in phase space, giving rise to a state-dependent geometric phase. At the conclusion of the gate, internal and motional states are disentangled for all values of $n$. Hence, the MS gate can be used for ions that are not cooled to the motional ground state. An additional feature of the MS interaction is that entanglement among multiple ions can be generated using only global control lasers (that is, it does not require lasers independently focused on each ion). The MS entangling gate was first demonstrated for chains of 2 and 4 $\mathrm{Be}^{+}$ ions in 2000 \cite{Sackett4IonEntanglement2000}. To date, the highest-achieved fidelities in both optical and hyperfine two-qubit gates have been achieved using the MS interaction. For optical qubits, a fidelity of $99.6 \%$ was obtained \cite{ErhardBlattCycleBench2019} and, while the gate time was not reported in this work, a similar fidelity ($99.3 \%$) was achieved in a gate time of $50$~$\mu$s \cite{BenhelmMSGate2008}.  For hyperfine qubits, an MS two-qubit gate was demonstrated with $99.91 \%$ fidelity in $30$~$\mu$s \cite{nist_gate_2016}.

\begin{figure}
    \centering
    \includegraphics[width=\columnwidth]{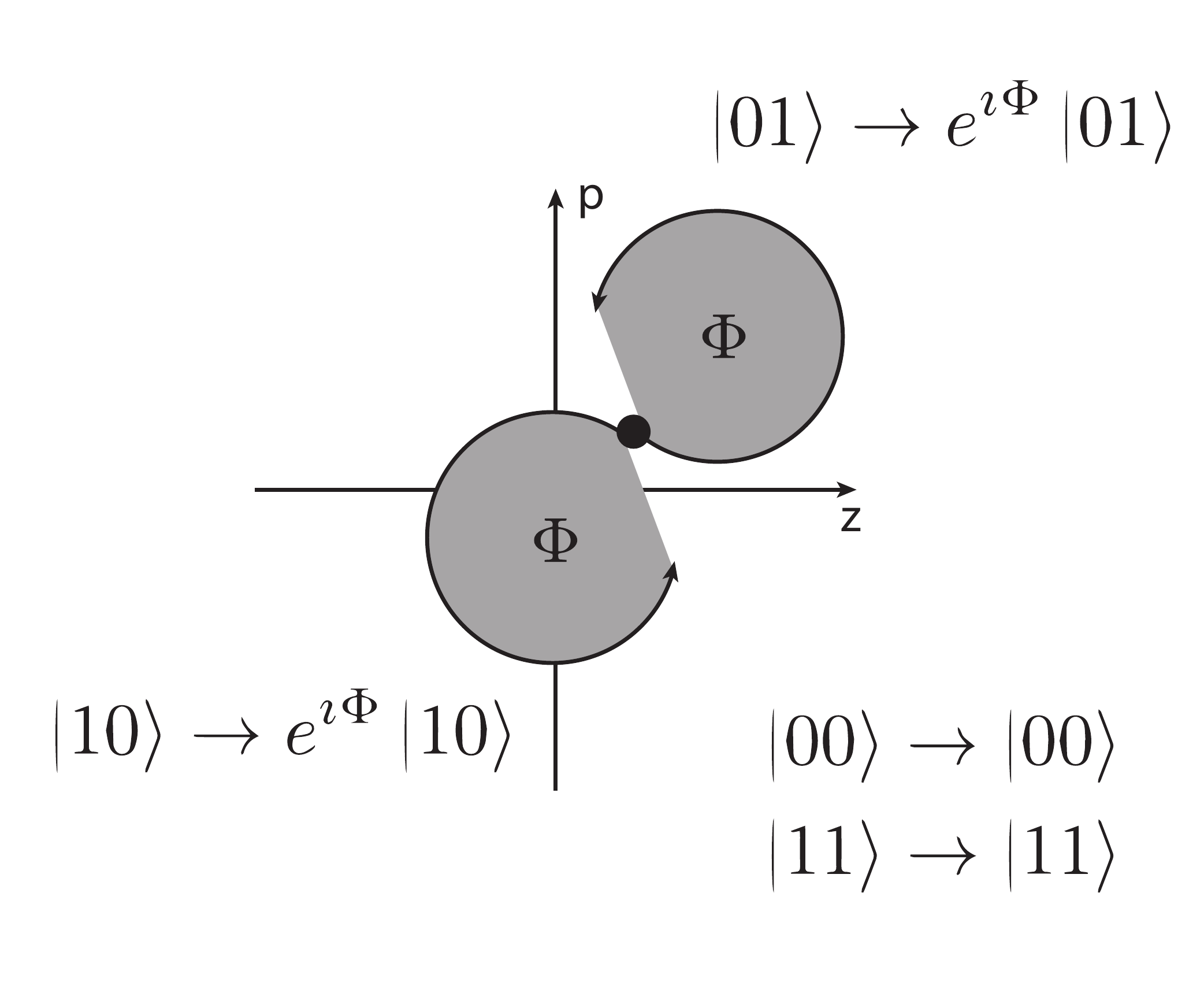}
    \caption{Phase space trajectories of two-ion states during a  geometric phase  gate. Spin-dependent forces drive particular ion states ($|10\rangle$ and  $|01\rangle$) along closed paths in phase space, imparting a geometric phase $\Phi$ set by the enclosed area. The $|00\rangle$ and $|11\rangle$ states do not couple to the control field and therefore accrue no geometric phase. Figure adapted from Ref. \cite{LeibfriedDidiGate2003}. The M\o{}lmer-S\o{}rensen gate proceeds in a similar fashion but applies phases to two-ion states in a rotated spin basis.}
    \label{fig:phasegates}
\end{figure}

A third type of two-qubit gate for ions is Leibfried's geometric-phase gate~\cite{LeibfriedDidiGate2003}. This gate uses a pair of detuned laser beams to generate a state-dependent force which traces a closed path in phase space, as shown in Fig.~\ref{fig:phasegates}. While this gate, too, utilizes the shared motion of the ions to generate coupling between them and is insensitive to the initial ion motional state, it differs from the MS gate in that it does not involve transitions between the $\one$ and $\zero$ qubit states. The geometric-phase gate was the first to achieve high fidelity, of $97 \%$ in $40$~$\mu$s \cite{LeibfriedDidiGate2003}, and it has since been used to demonstrate one of the highest-fidelity 2-qubit gates ($99.9 \%$ in $100$~$\mu$s) on trapped ions \cite{Ballance2QubitHyperfineGate2016}.  Despite these results, this type of geometric-phase gate has the drawback of not being applicable to FOFI qubits~\cite{LangerThesis2005} (MS gates do not have this limitation).  A variant type of geometric-phase gate \cite{BermudezGate2012} uses a strong microwave carrier to create a dressed state basis that is insensitive to environmental fluctuations and then traces a closed path in phase space with a state-dependent force from a single-sideband drive; so far this gate has been demonstrated with $97 \%$ fidelity in $^9$Be$^+$ ions with gate times of 250~$\mu$s \cite{TanDressedState2013}.

As mentioned previously, performing two-qubit gates with microwaves requires some method of coupling microwave fields to the shared ion motional mode, typically a magnetic field gradient. AC magnetic field gradients were used to enable MS-style entangling operations in $^{25}$Mg$^+$ with fidelity of $76 \%$ \cite{ospelkaus2011microwave} with duration of several hundred microseconds. Two-ion gates using static magnetic field gradients have also been demonstrated, initially achieving fidelity of $64 \%$ in 8 ms \cite{KhromovaGradientGate2012}. Recently improved versions of these gates have achieved fidelities of $99.7 \%$ for an AC gradient \cite{HartyNearFieldMicrowaves2016} and $98.5 \%$ for a static gradient \cite{WeidtMicrowaveGates2017}; both experiments required millisecond timescales for their two-qubit gates, however.

Since the errors in two-qubit gates are at least an order of magnitude higher than in single-qubit gates, reducing two-qubit gate errors in ions has been an active area of research for some time. Numerous sources of error currently limit the achievable fidelities in two-qubit ion gates. Some of the most prevalent include drifts of the control laser frequency and amplitude, magnetic field drifts in the laboratory (for magnetic-sensitive qubits), and off-resonant scattering. Furthermore, even though the most commonly used gates (MS and geometric-phase) are insensitive to the initial motional state of the ion, motional decoherence during the gate is also a significant source of error. Speeding up the gate to achieve entanglement faster than sources of decoherence can influence the ions is one way by which better performance might be achieved. (And faster gates are in general desired for higher computation speed as long as high fidelity can be maintained.)  Recently, fast two-qubit geometric-phase gates \cite{SteanePulseGates2014} achieved $99.8 \%$ fidelity using hyperfine qubits in just 1.6~$\mu$s \cite{SchaferFastIonGates2018}, although hundreds of milliwatts of power in focused Raman beams were required. This work also demonstrated the fastest-achieved two-qubit gate in ions, of less than 500 ns, but the achieved fidelity for such a high speed was only $60 \%$.  Fast, high-fidelity geometric-phase gates are enabled by shaping the envelopes of the gate-laser pulses.  However, this technique is not as effective for MS gates due to the presence of a so-called ``carrier" term in the MS interaction Hamiltonian, which stems from the fact that the MS gate involves transitions between qubit levels (as mentioned above).  As a result, speeding up high-fidelity MS gates may not be as straightforward, though a technique to mitigate the effect of the carrier term has been suggested \cite{MehtaSPIE2019}.

A different way to possibly realize very fast gates is to use a sequence of ultrafast pulses from a mode-locked laser to generate spin-dependent impulses and trace an entangling path in phase space. This method was demonstrated in \cite{WongCamposUltrafast2017}, but the overall gate speed achieved was only in the range $2$ to $20$~$\mu$s with fidelity no higher than $76 \%$. Once gate speeds become faster than the megahertz-scale trap frequencies, the control fields begin to excite multiple motional modes at once and the gate also begins to become dependent on the (typically uncontrolled) absolute optical phase. These factors have so far limited the fidelity of the fastest gates, illustrating the challenges involved in increasing gate speeds while maintaining high fidelity.

In addition to performing the gates faster, another common method to mitigate the effects of laboratory noise sources with slow time constants is to use spin-echo techniques. A single spin-echo pulse, occurring at a time in the gate when the ion spin and motion were unentangled, was used to achieve high two-qubit gate fidelities ($99.8 \%$ in $27$~$\mu$s~ \cite{BallanceHybridLogic2015}. While such dynamical decoupling pulses cannot be straightforwardly applied at times when the ion internal and motional states are entangled, new methods of dynamical decoupling pulse sequences during gate operation are a promising area of research~\cite{manovitz2017fast} and will be discussed further in Sec.~\ref{DFS}.

    \subsubsection{Gate Characterization: Tomography, Benchmarking, and Calibration}

Most of the experimental results so far have been given in terms of the fidelity, the squared overlap between the goal state $\psi_g$ and the experimentally-obtained state $\psi_e$: $F = | \langle \psi_g | \psi_e \rangle |^2$ for pure states. Fidelity is useful in that it is fairly easy to compute (assuming the goal state is known and that measurement errors can be neglected) and reduces all error sources to a single number. At the same time, precisely because it gives only a single number as output, fidelity is not a complete description of the quantum operation it describes. Knowing the fidelity gives little information as to the sources of imperfection in a gate or experiment, and it considers systematic or coherent errors on the same footing as stochastic errors which change from one experimental iteration to the next. This latter point means that fidelity can potentially give an over-optimistic assessment of a gate's performance. If the same gate is repeatedly applied to the qubit, stochastic errors will partially cancel out (in the same manner as a random walk), but systematic errors will add coherently and cause the fidelity to degrade much more quickly. For these reasons, methods to characterize the errors in a quantum process more accurately have long been an area of interest. This area has become known as Quantum Characterization, Verification, and Validation (QCVV).

The method of quantum process tomography (QPT) was proposed as a way to fully characterize a quantum process \cite{ChuangNielsenQPT1997, PoyatosCompleteChar1997}. QPT is able to determine the effects of a ``black box'' quantum operation on $N$ qubits (e.g., an $N$-qubit gate) by characterizing the gate's operation on $4^N$ input states---for a single qubit, the input states used could be $\zero$, $\one$, $(\zero + \one) / \sqrt{2}$ and $(\zero + i \one) / \sqrt{2}$ (though other sets of input states could be used). State tomography is performed on each output and matrix methods (described in \cite{ChuangNielsenQPT1997}) can be used to extract the $16^N - 4^N$ independent gate parameters. A clear issue with quantum process tomography is the exponentially-scaling number of input states needed to characterize the gate (only 4 for a single-qubit gate, but 16 for a 2-qubit gate, 64 for a three-qubit gate, etc.) and the sheer number of parameters extracted---12 to characterize a single-qubit gate, 240 for a 2-qubit gate, and 4032 parameters for a 3-qubit operation. Interpreting these parameters in a sensible way is clearly nontrivial. A second and equally daunting aspect of QPT is that perfect state preparation and measurement are assumed, and the process parameters output by the method can be significantly inaccurate in the case of realistic state preparation and measurement (SPaM) errors \cite{WeinsteinQPT2004}. Despite these limitations, QPT has been used to characterize different schemes for implementing entangling gates in ions~\cite{RiebeProcessTomography2005}.

Randomized benchmarking (RB) for gate characterization was developed as a way to deal with the issue of SPaM errors in qubits \cite{EmersonRB2005,KnillRB2008}. The most straightforward way to circumvent the issue of imperfect SPaM is to perform the gate not once, but a large and variable number of times (thus amplifying the effect of gate errors compared to SPaM errors), and determine the overall success probability as a function of the number of gates. RB essentially implements a random series of rotations about different axes, with the final rotation chosen such that the end result is the identity gate. (The gate series to be used are typically calculated ahead of time.) As such, in the absence of gate errors, the ion should end up in a measurement eigenstate. Since the final measurement result should be known based on the gates implemented, any results deviating from that answer can be classified as errors. Typically the operations to be used are $\pi/2$ rotations about different axes, but extensions to the original proposals have proposed different sets of gates to be used \cite{DankertUnitaryRB2009,Carignan-DugasDihedralRB2015,CrossDihedralRB2016} and methods of RB for multiple qubits \cite{MagesanScalableRB2011, MagesanCharacterize2012}. RB was first used to demonstrate average error probabilities of less than $0.5 \%$ in $\pi/2$ pulses in a $^9$Be$^+$ qubit \cite{KnillRB2008}, and has since become a standard gate characterization tool for QC research. While RB is a good way of extracting stochastic errors that determine gate fidelity, it performs less well in the case of correlated errors---which may be either cancelled out or amplified depending upon the exact random gate series applied---so RB thus provides little information about the magnitude of such correlated errors \cite{BallCorrNoiseRB2016,MavadiaExperimental2018}. Although it is a good method of decoupling SPaM errors from gate errors, randomized benchmarking does not fully characterize the gate errors that are present.

Gate set tomography (GST) represents one recent method developed to more fully characterize qubit gate errors ~\cite{BlumeKohoutGST2013}. In GST, a set of gate sequences that have been optimized to amplify all possible errors in a given gate set is applied repeatedly to the qubit. Amplification of the errors is achieved by applying each different sequence an exponentially-increasing number of times, measuring the resulting qubit state outcomes as a function of applied number of sequences, then fitting the resulting data to a noise model that identifies the different errors present in the gate set. One benefit of GST, when compared to RB, is that the process can bound the diamond norm error~\cite{blume2017demonstration}, which is the error metric by which rigorous fault-tolerant error correction bounds are set. GST has been used to demonstrate single-qubit gate operations in a trapped ion below a rigorous threshold for fault-tolerant operation \cite{blume2017demonstration}. Gate set tomography does suffer from a few significant drawbacks, however. First, a large number of different measurements are required to fully characterize even a single gate. Second, the algorithm to reconstruct gate errors can underestimate the diamond norm if additional gate sequences are not used to restrict gauge freedoms \cite{MavadiaExperimental2018}. Third, GST provides a rigorous bound on the diamond norm only if its assumptions are satisfied. These assumptions include gate operations being stationary and Markovian (i.e., memoryless). GST can thus account for both stochastic errors and calibration errors. But essentially all laboratory experiments also suffer from non-Markovian errors, such as slow drifts in magnetic fields, laser frequency, etc. While the GST algorithm can identify violations of Markovianity, it is not clear if its rigorous bounds survive in the presence of non-Markovian noise, and if this is the case, it is not obvious that the experimental overhead is justified.

A third and more recent technique for characterizing gate errors is robust phase estimation (RPE) \cite{kimmel2015robust}. RPE is a technique for extracting the systematic errors present in a set of gates. Like GST, RPE is robust against modest SPaM errors. Unlike GST, RPE presents a Heisenberg-limited method to extract a limited set of systematic gate errors which can be corrected (calibrated) by the experimentalist. For single-qubit rotations, the parameters of interest include the phase, rotation angle, and Bloch-sphere axis about which the qubit state is rotated. RPE was performed on a trapped $^{171}$Yb$^+$ ion and compared with GST \cite{rudinger2017experimental}, where it was shown that RPE could achieve accurate parameter calibrations at the $10^{-4}$ level with as few as 176 measurements, roughly two orders of magnitude fewer than required by GST for similar accuracy. RPE is specifically designed to allow experimenters to determine those systematic errors which can be improved, and thus does not provide information about stochastic errors. Some other gate-characterization method, such as RB or GST, must be used once the calibration procedure is finished to determine overall gate performance. An additional complication is that systematic errors may drift over the course of an experiment; however, RPE is efficient enough that periodic parameter re-calibrations may be possible with the technique.

Many experiments in the area of QCVV have been performed on trapped ions, in part due to the fact that ion gate errors are already small enough that characterizing them accurately is worthwhile. Furthermore, ions' identicality helps ensure that, once errors in a system have been characterized, that characterization remains accurate and useful for the experimenter over long times. At the same time, the generally slow speeds of trapped-ion gates mean that more complete methods of characterization---particularly GST or QPT---can introduce punishing experimental overheads. Finding an efficient method of QCVV for larger numbers of ions remains an active area of research.

    \subsubsection{Crosstalk}
As the number of trapped ions increases, so too does the difficulty in addressing only the desired qubits without crosstalk. In the absence of field inhomogeneities, the resonance frequencies for each ion are the same, and therefore transitions in each ion will be driven according to the beam intensity at each ion location. A high numerical aperture (NA) lens can be used to focus a control laser beam very tightly to a waist significantly smaller than the ion spacing. The location of the control beam can be switched to the desired ion using acousto or electro-optical deflection or MEMS devices \cite{KimMEMSMirror2007}. This technique has been used widely in linear chains of ions, and full control of chains of 20 ions has been demonstrated \cite{Friis20QubitEntanglement2018}. However, the number of ions that can be controlled in this way is limited by a number of technical complications, such as the lens size and the resolution of the beam deflector. Although the cm-scale wavelengths of microwaves make focusing more difficult, near-field microwaves generated by small trap features have been used to create ion-specific coupling \cite{PhysRevA.95.022337} by cancellation of the microwave field at the location of ions not chosen to undergo the gate.

A number of different strategies have been employed to mitigate the effects of crosstalk, especially those due to beams that address multiple ions at once. For example, ions can be shelved in states outside of the $|0\rangle$ and  $|1\rangle$  subspace that do not interact with a particular control field~\cite{RoosControl2004}.  Alternatively, composite pulse sequences have been implemented that correct for the errors introduced by unwanted ion addressing~\cite{herold2016universal}. In other experiments, the ion energy level degeneracy is intentionally broken in order to spectrally resolve transitions between different qubits. This method has been applied by shifting an ion off of the RF trapping null to introduce an excess micromotion sideband in only that ion~\cite{LeibfriedMicromotionAddress1999, AkermanUniversal2015} or by using a magnetic field gradient to shift energy levels via the Zeeman effect~\cite{MintertMicrowaveScheme2001, Timoney2011, PiltzLowCrosstalk2014}. The magnetic field gradient technique is particularly useful for driving qubit-specific gates using far-field microwaves that cannot be focused tightly, as discussed in Sec.~\ref{Gatetypes}.

The crosstalk reduction methods discussed thus far have treated the ions as stationary particles within their trapping potential. One final method for reducing crosstalk is to move ions in a 1-D chain \cite{NIST:ion_transport:2002,WarringIndividualAddress2013} or within a 2-D trap \cite{KielpinskiQCArchitecture2002} by varying the DC potentials. Bringing ions together into a single well can allow multi-qubit gates to be carried out, while separating them allows for single-qubit gates (and readout) to be implemented with low crosstalk. This approach nevertheless introduces additional complications. While ions have been shuttled over trap surfaces reliably and repeatably, the ion's motional state tends to heat up during the shuttling, and particularly during separation/joining of ion chains. Ion heating from shuttling may be mitigated to some degree by use of specially-tailored electrode-voltage waveforms, as will be discussed later in Sec.~\ref{2DArrays}.  It can also be mitigated through the inclusion of a sympathetic cooling species, as described further in Sec.~\ref{DualSpecies}. However, this cooling, as well as the ion shuttling operations themselves, will introduce latency into any QC algorithm.

\subsection{State Detection}
\label{Detection}
\begin{figure}
    \centering
    \includegraphics[width=\columnwidth]{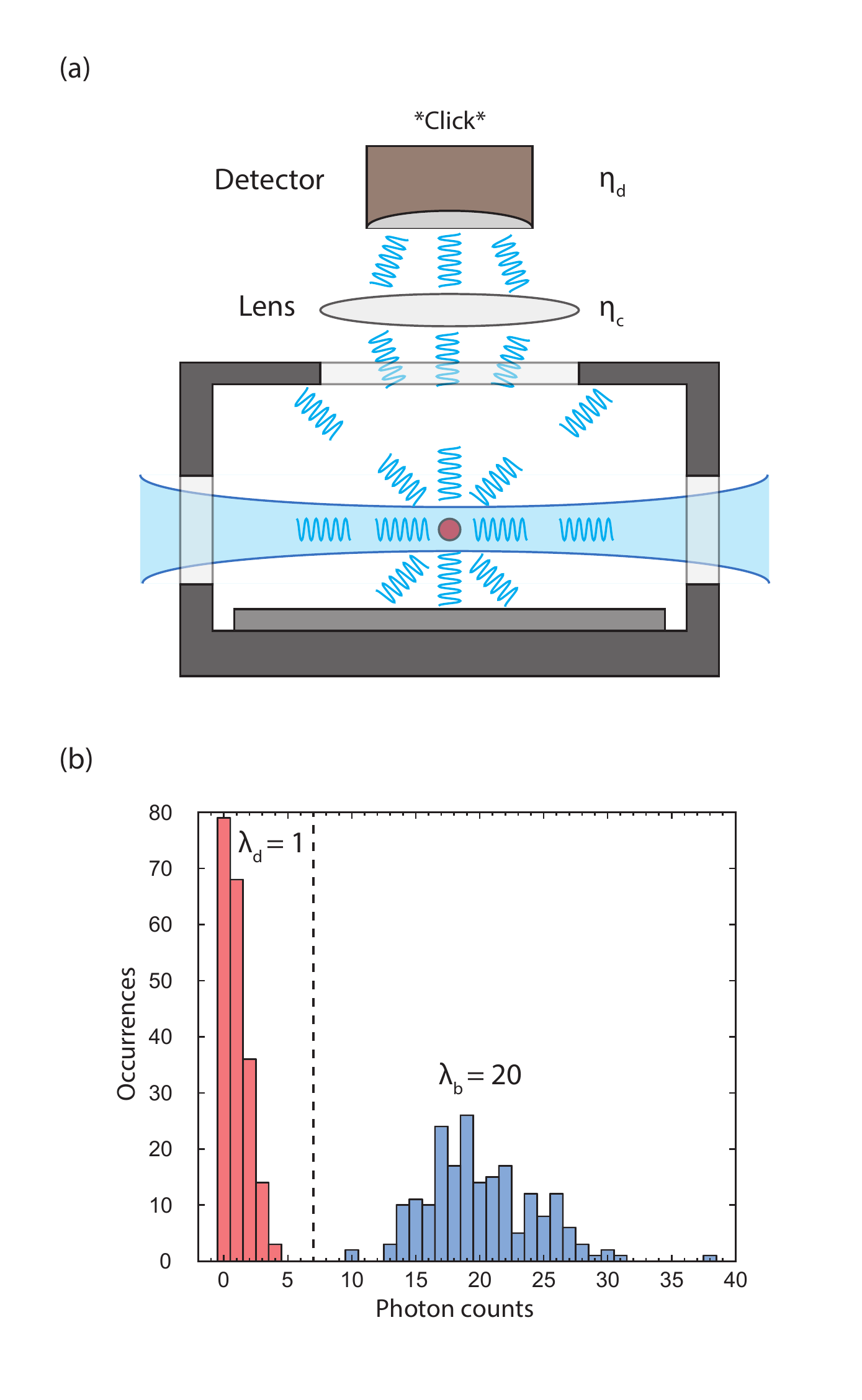}
    \caption{Trapped-Ion State Readout. (a) Schematic drawing of fluorescence detection of a trapped ion. The ion scatters many photons from a resonant laser beam that are collected by a large-NA lens with efficiency $\eta_{c}$. The collected photons are imaged on a detector, which registers photon counts, shown here as clicks, with its own efficiency $\eta_{d}$. (b) Simulated photon collection histograms for state detection of a trapped ion. The bright (dark) state photon counts are taken from a Poissonian distribution with mean $\lambda_{b}=20$ ($\lambda_{d}=1)$. The dotted vertical line shows a threshold value of 7 photons, demonstrating high accuracy determination of the ion state.}
    \label{fig:detection}
\end{figure}

Determination of the ion state needs to be accurate, fast, and, ideally, extensible to many ions. The primary mechanism used to date relies on state-dependent fluorescence~\cite{PhysRevLett.56.2797,PhysRevLett.57.1699,PhysRevLett.57.1696}. During measurement, a trapped ion is projected into either a so-called bright state that scatters many photons when illuminated with a detection laser or a so-called dark state that scatters very few photons. The scattered photons can be collected with a high-NA lens and detected with a high-efficiency detector, and the resulting photon counts can then be analyzed to infer the ion's state. This procedure is shown schematically in Fig.~\ref{fig:detection}.



For a sufficiently long illumination time, generally on the order of a few hundred microseconds to several milliseconds, a threshold value of photon counts can be established that differentiates between the bright and dark states with high accuracy. The photon arrivals for both bright and dark states typically follow a Poissonian distribution with the mean number of received bright-state photons $\lambda_b$ and dark-state photons $\lambda_d$ obeying $\lambda_b \gg \lambda_d$. Both $\lambda_b$ and $\lambda_d$ will generally be proportional to the measurement time, $t$; due to the exponentially-decreasing overlap between the distributions, very high fidelity can be achieved in reasonable measurement times. Using Poissonian statistics, one can calculate the necessary mean number of bright state photons to achieve an arbitrary detection fidelity. For example, if the mean $\lambda_d = 1$ and the mean $\lambda_b = 20$, a threshold can be set at 7 collected photons: if the ion is in the dark state, the probability is $99.99 \%$ that 6 or fewer photons will be collected, while for a bright-state ion there is a $99.97 \%$ chance that 7 or more will be collected. Simulated photon collection histograms for these parameters are given in Fig.~\ref{fig:detection}b, and experimental data using this technique can be found in, for example, Refs.~\cite{itano1988precise,NIST:HifiMicrogate:12}.  The rate of fluorescence scattering is limited to no more than half the linewidth of the excited state by saturation and is typically on the order of $10$'s of MHz, while photon detection probabilities (accounting for lens NA and detector efficiency) are on the order of $1 \%$, which gives rise to typical count rates of 100~kHz. Thus, as long as the dark-count rate is less than 5~kHz, 200~$\mu$s detection time should suffice to achieve the extremely high detection fidelities discussed above.

State-dependent fluorescence measurement can be applied in the case of hyperfine or Zeeman qubits as well. However, off-resonant scattering of the state-detection beam presents a possible error source, as the bright and dark states are separated by frequencies orders of magnitude smaller than the hundreds of terahertz splittings available in optical qubits. In $^{171}\mathrm{Yb}^{+}$, the ground state hyperfine splitting is relatively large ($12.6$ GHz) and the $|0\rangle$ and  $|1\rangle$ qubit states can be used as the bright and dark states for detection, though off-resonant scattering can still be a limiting source of detection error \cite{noek2013high}. It is important to note, however, that more efficient readout (achievable by, for instance, increased photon collection and detection efficiency) can mitigate the error arising from off-resonant scattering since fewer scattering events are required to determine the qubit state~\cite{CrainSNSPDdetect2019}. Time-resolved photon detection can also be employed to mitigate these errors \cite{WolkTimeResolvedDetect2015}.  In $^{43}\mathrm{Ca}^{+}$ and $^{9}\mathrm{Be}^{+}$, the FOFI qubit states are not well-suited for state-dependent fluorescence measurements.  Hence, the population in at least one qubit state is transferred with additional control pulses to a different state for the purposes of readout. Polarization selection rules can be exploited to dramatically change the photon scattering rates for these states used for readout, and the techniques of state-dependent fluorescence measurement can be applied directly. Despite the additional experimental complexity required to implement readout in these qubits, readout fidelities in excess of $99.9 \%$ have been achieved \cite{HartyHighFidelityIons2014}.

The fluorescence measurement technique has been extended to measure the states of several ions simultaneously by measuring total photon counts, with the total counts proportional to the number of ions in the bright state. Setting threshold values that accurately determine the number of bright and dark ions generally becomes more difficult for larger numbers of ions, since two bright ions emit only twice as many scattered photons as a single bright ion (whereas one bright ion can easily emit more than an order of magnitude more counts than a dark ion, as described previously). Additionally, the illumination time, which generally increases the number of photons measured per bright state ion, cannot be extended without bound. Besides slowing the experimental cycle time, longer readout times increase the probability of possible state misidentification due to effects such as spontaneous decay of the excited state or depumping from the cycling transition.

Another way to extend fluorescence detection to multiple ions is to spatially discriminate between the photons scattered by different ions. This method tends to work better for larger numbers of ions, even in the case of finite crosstalk, as the Poissonian distributions start to overlap for large ion numbers when total fluorescence is used. A magnifying imaging system has been used to focus the fluorescence from different ions in a chain onto different photomultiplier tube (PMT) detectors in a multichannel PMT array \cite{Debnath5QubitComp2016}, achieving roughly $99 \%$ fidelity per ion (limited by crosstalk between the different PMT channels). A similar idea involves mapping different ions to different pixels on an electron-multiplying CCD (EMCCD) array and collecting their individual fluorescence counts \cite{Zhang53IonSim2017}, which has achieved similar readout fidelity of $\sim 99\%$ per ion in times of $300 \, \mu$s for a chain of 53 ions. Even better EMCCD readout for a chain of four ions was demonstrated in \cite{BurrellMultiqubit2010}, with $99.99 \%$ fidelity in $\sim 400 \, \mu$s, although the ion-ion spacing of $14 \, \mu$m in this experiment was greater than in the 53-ion chain (reducing ion-to-ion crosstalk during readout).

More sophisticated readout schemes have also been developed with the goal of reducing the state readout time without sacrificing accuracy. By tracking the arrival times of photons scattered by the ions, adaptive techniques based on maximum likelihood analysis have been employed to achieve average readout times of hundreds of microseconds with accuracies of $99.9$--$99.99 \%$~\cite{MyersonReadoutIons2008,noek2013high,HartyHighFidelityIons2014}, with the possibility of even faster ($\sim10 \; \mu$s) readout with reduced fidelities around $99 \%$. Time-resolved state readout has also been applied to chains consisting of multiple trapped ions using a CCD camera~\cite{BurrellMultiqubit2010}.

An intriguing possibility for ion state detection is (nearly) background-free detection. Most of the background in typical ion detection schemes comes not from detector dark counts but rather from stray detection laser photons which scatter off of metal surfaces in the experiment. If fluorescence counts could be collected at a different wavelength from the laser used for excitation, the excitation laser photons could be filtered out and eliminated. The internal level structure of trapped ions could allow, for example, a two-photon excitation which then decays at a third wavelength which is detected. A first step in this direction was implemented via strong excitation (using 250 mW of laser power) on the quadrupole transition at 729 nm in $^{40}$Ca$^+$, followed by additional laser excitation at 854 nm to the $4P_{3/2}$ state, which allowed background-free collection of the 393-nm photons emitted as the ion returns to the ground state \cite{HendricksIonBlaster2008}. This was followed by experiments which achieved higher detection rates by using a more tightly-focused 729-nm beam \cite{LindenfelserBackgroundFree2017},  or using multi-stage excitation to the same $4P_{3/2}$ state in $^{40}$Ca$^+$ by first-stage excitation to the $P_{1/2}$ state via 397-nm light and applying multiple IR beams at 850 nm, 854 nm and/or 866 nm \cite{LinkeBackgroundFree2012}. In all three of these experiments, however, the $3D_{5/2}$ state---which is the $\zero$ qubit state in $^{40}$Ca$^+$---formed an intermediate state in the excitation chain. As a result, although photon count rates of up to 30 kHz were achieved with background rates less than 1 Hz, the achieved detection did not distinguish between the $\zero$ and $\one$ states and thus was not truly a state detection method. A way around this problem was demonstrated in 2013, also in $^{40}$Ca$^+$~\cite{FeiBackgroundFree2013}: high-intensity driving of the 732-nm quadrupole transition from the $\one$ state to the $3D_{3/2}$ state, followed by 866-nm excitation to the $4P_{1/2}$ state in order to achieve background-free detection of 397-nm decay photons while also protecting the $3D_{5/2}$ $\zero$ state. Although this demonstration achieved count rates of only a few per ms, and did not actually demonstrate state-sensitive detection due to lack of a 729-nm laser to excite to $\zero$, this result points the way to background-free and state-sensitive detection.

While background-free detection can reduce or eliminate the effects of scattered excitation photons on the detection fidelity, those photons might introduce an additional deleterious experimental effect in a large-scale trapped-ion quantum computer, namely, fluorescence-induced decoherence. Photons from the excitation beam itself, or those scattered off of one ion in the experiment, might hit another experimental ion. These photons---which scatter off of the $\one$ state and not the $\zero$ state---effectively constitute a projective measurement and can thus destroy the quantum information contained in the second ion. This has not been a limitation in many experiments to date that have performed all experimental operations with detection light off, then measured all of the ions at the end of the experimental sequence. In a quantum computer with error correction, however, it will be necessary to perform measurement steps during the experimental sequence as part of the error-correcting code, while nearby ions are still in superposition states which contain quantum information. Stray detection photons thus pose a risk to the operation of large-scale quantum information processors.

A solution to this second problem was worked out in 2005 \cite{SchmidtQuantumLogicSpectroscopy}: Use quantum logic to transfer state population information to a second auxiliary species and detect the state of the auxiliary species that will emit far-detuned light. Like entangling two-qubit operations, this technique first transfers information from the ``logic'' ion to a mode of the shared ion motion, then transfers that motional excitation to an internal excitation of the auxiliary ion. Readout on the auxiliary ion can then be performed without the danger of introducing decoherence to the logical ion. While adding a secondary ion species to the experiment may seem like significant overhead, it is likely that a second ion species will be needed anyway for large-scale quantum computation to perform sympathetic cooling (see Sec.~\ref{DualSpecies}). Quantum logic spectroscopy was originally demonstrated with $^{27}$Al$^+$ as the logic ion and $^9$Be$^+$ as the readout ion, and was originally implemented because the $^{27}$Al$^+$ possesses excellent optical atomic clock properties but lacks an accessible readout/cooling transition. Since then, quantum logic spectroscopy has been used to achieve multi-species readout fidelity as high as $96 \%$ in single-shot experiments \cite{BruzewiczQLAR2017}, and has been used to read out the parity of a two-qubit entangled state \cite{NegnevitskyMultiReadout2018}. Fidelity of $99.94 \%$ was achieved in an adaptive-readout experiment where the quantum-information transfer to the readout ion could be repeated many times during one measurement \cite{HumeAdaptiveDetection2007}, but the resulting long total readout times of $> 10$ ms represent a drawback, and the method took advantage of a unique level structure in $^{27}$Al$^+$ that is not available in most ions. While single-shot demonstrations of quantum logic spectroscopy have not yet surpassed $99\%$ fidelity, work in this important area is ongoing.

An additional method to mitigate undesired absorption of scattered measurement photons is to implement internal shelving operations before measurement in order to make some qubit ions transparent to detection photons \cite{RoosControl2004} while others are read out. While this method is intriguing, its ultimate achievable fidelity has not been explored, as maximizing the readout fidelity was not the goal of the original demonstration.

All trapped-ion-qubit state detection schemes benefit from high detection efficiency of photons emitted from the ions.  The total photon detection efficiency is given by the product of $\eta_c$, the fraction of scattered photons which are collected by the collection optics, and $\eta_d$, the probability a detector registers the event of an impinging photon. The value $\eta_c$ depends on the NA of the optics.  The total solid angle $\Omega_{SA}$ over which photons can be collected is given by $\Omega_{SA}/(4 \pi) = (\textrm{NA})^2/4$.  While high-NA optics are available (e.g., high-end microscope objectives), they either have very small working distances (100-$\mu$m-scale), which could interfere with ion trapping, or are very large in size (many inches). In practice, most collection lenses are located outside of the system vacuum chamber. As a result, most groups achieve values of $\Omega_{SA}/4\pi$ in the range of 0.01--0.1 (1--10 \%), with the highest values corresponding to 0.6 NA collection optics~\cite{noek2013high}.

The value $\eta_d$ is limited by the quantum efficiencies for the commonly-used single-photon detectors for trapped ions, which require the ability to detect blue-to-UV photons with reasonable efficiency.  The detector most typically used is the PMT, though EMCCD cameras and avalanche photodiodes (APDs) have also been commonly employed.  They all have values of $\eta_d$ in the range of 0.2-0.4 at the photon wavelengths of interest.  Recently, superconducting nanowire single photon detectors (SNSPDs) have been developed to detect UV photons emitted from ions with efficiencies as high as $\sim$0.8 \cite{SlichterSNSPD2017, WollmanSNSPD2017}.  and used to detect $^{171}$Yb$^+$ ions with 99.9\% fidelity in $11 \, \mu$s \cite{CrainSNSPDdetect2019}, although these devices operate only at cryogenic temperatures. Overall, the best total photon detection efficiencies have been around $2$--$4 \%$ \cite{noek2013high,CrainSNSPDdetect2019}.  It is an active area of research to improve these efficiencies, as will be discussed in later sections of this review.

\subsection{Quantum Control Demonstrations:  Quantum Computing Algorithms and Primitives}
\label{Algorithms}

Using the above-described methods for trapped-ion quantum control, significant progress has been made in demonstrating small-scale quantum algorithms and primitives, themselves elements that will play a part in more complex algorithms or quantum-error-correction schemes.  Here we highlight some of the key milestones among these experimental demonstrations.

A quantum algorithm is a procedure---typically written down as a set of gates and measurement operations on a qubit register---that solves some problem. In some cases, an algorithm can be further decomposed (either fully or partially) into primitives, repeated subroutines within the algorithm. To date, few quantum algorithms have been demonstrated, and those with only a few qubits each. A few key primitives have also been demonstrated. At the same time, trapped ion systems have generally achieved fidelities in these demonstrations as high as, or higher than, the overall fidelities achieved with other qubit modalities.

The first quantum algorithm to be implemented in a trapped-ion system was the Deutsch-Jozsa algorithm \cite{DeutschJozsaAlgorithm}. This algorithm determines whether a black-box oracle which takes an $N$-bit input and outputs a single-bit output is either constant (always outputting the same bit) or balanced (outputting a 1 or 0 depending on the input, with exactly half of inputs producing a 0 and the other half producing 1). While of little practical value, the algorithm is of note in that a classical computer requires multiple queries to the black box to determine whether the output is constant or balanced, whereas an $N+1$-qubit quantum computer can determine the answer with only a single query. In 2003, using the electronic and motional states of a trapped ion as two input qubits, the Deutsch-Jozsa algorithm was demonstrated for $N=1$ \cite{GuldeDeutschJozsa2003}. Although this experiment was a landmark in trapped-ion QC, only a single trapped ion was used and thus it represented only a very modest level of quantum control.

Quantum teleportation, first proposed in 1993 \cite{BennettTeleportation1993}, is a universal QC primitive that can enable any quantum computation if certain other resources are available \cite{GottesmanChuang1999}. Teleportation of an ion's quantum state was demonstrated in 2004 by two groups \cite{BarrettTeleportation2004, RiebeTeleportation2004} with average fidelities ranging from $73 \%$ to $78 \%$. These experiments demonstrated many of the same operations needed to operate a quantum computer: generation of an initial entangled state of the ions, separation of ions while maintaining entanglement, within-experiment measurements, and conditional operations based on those measurements. Furthermore, a total of three ions were simultaneously controlled during the experiments.

In 2005, the semiclassical quantum Fourier transform was demonstrated in a linear chain of three Be$^+$ trapped-ion qubits \cite{ChiaveriniQFT2005}. The experiment was noteworthy not only in that multiple ions were used in the calculation, but that the quantum Fourier transform is itself the key primitive used as the final step in Shor's factoring algorithm, a quantum algorithm which can perform useful tasks. Despite these important steps forward, what was demonstrated was in fact a semiclassical version of the QFT which did not require entangling gates between the ions. The squared statistical overlap---a rough proxy for fidelity in the multi-qubit experiment---was at least $87\%$ for all of the different input states to the Fourier transform algorithm.

Later in 2005, the first full quantum algorithm utilizing entanglement between multiple trapped-ion qubits was successfully demonstrated~\cite{BrickmanGroverSearch2005}. This demonstration of Grover's search algorithm in a two-ion system achieved successful ``search'' for the marked element with probability $60 \%$, in excess of the maximal classically-obtainable probability of $50 \%$.

In 2011, the first trapped-ion algorithm for universal, digital quantum simulation \cite{UniversalQSimLloyd1996} was used to simulate 2-D Ising interactions in chains of up to six ions \cite{LanyonDigSim2011}.  With as many as 100 gates performed, the experiment represented a major advance in the qubit number and complexity of quantum algorithms demonstrated, although the fidelities for the largest ion chains used did not exceed $77\%$.

In 2016, an implementation of Shor's algorithm on a five-ion system was used to factor the number 15 with success probability of $99 \%$ \cite{MonzScalableShor2016}. This experiment was a demonstration of what is arguably the most well-known quantum algorithm and it employed novel techniques to utilize qubits more efficiently than had been done previously.

Since 2016, three- and five-qubit fully-connected trapped-ion quantum computers have been built \cite{Debnath5QubitComp2016} and used to perform a variety of algorithms including Deutsch-Jozsa and Bernstein-Vazirani \cite{Debnath5QubitComp2016}, Grover search \cite{FiggattGroverSearch2017}, and a four-qubit error detection code \cite{LinkeErrorDetection2017}. Quantum chemistry calculations of the ground-state energies of $\textrm{H}_2$ and LiH molecules have also been performed in a few-ion system via the Variational Quantum Eigensolver (VQE) algorithm \cite{HempelVQE2018}, with a recent similar calculation for H$_{2}$O having been performed using up to 11 ions~\cite{NamWaterMolecule2019}. The variety of algorithms successfully demonstrated is a testament to the versatility of the programmable quantum computer, which allows arbitrary gates to be implemented between any two qubits in the module. Even without error correction, overall success rates in the $90$--$95\%$ rate were achieved for these algorithms. The five-qubit ion-trap quantum computer's performance in running the Bernstein-Vazirani algorithm and Hidden Shift algorithm was also compared against the five superconducting qubit Quantum Experience computer \cite{IBMQuantumExperience} made available online by IBM. The overall success probabilities of the ion-trap simulator, in the $85$--$90 \%$ range, were higher than those in the superconducting circuit device; however, the overall algorithm execution time was much faster in the superconducting device due to its sub-microsecond gate times. The superior success probabilities of the trapped-ion system were in part due to the fact that, unlike the superconducting system, the trapped-ion system was fully connected: gates could be performed directly between any two arbitrary ions. This not only highlights the utility of a high degree of connectivity in a QC system but also poses the question of whether trapped ions will be able to maintain the same degree of connectivity as the number of ions increases. Perhaps the most important takeaway from the demonstration is that the current state-of-the-art for fully-programmable quantum computers is about the same for superconducting and trapped-ion qubits: approximately 10 qubits, individual two-qubit gates in the $99\%$ range, and the ability to execute simple algorithms with reasonable overall chances for success.

Quantum emulation, or analog quantum simulation, represents a different approach to QC with trapped ions. Trapped-ion systems designated as emulators, rather than employing a universal gate set, engineer a  Hamiltonian for a system of ions which can be mapped to some other many-body system. To date, most trapped-ion quantum emulators have been based on the Ising Hamiltonian, which includes a single-spin dependent energy term (effective magnetic field) and spin-spin interactions. These interactions have been implemented by addressing the transverse modes in a linear chain of trapped ions \cite{KimTransverse2009} or in a 2-D system of ions confined in a Penning trap \cite{Britton2012}. Entanglement in these quantum simulators has been demonstrated for chains of a few ions \cite{KimFrustration2010}, whereas frustration in the ground state has been demonstrated in chains of 16 ions \cite{IslamIsingMagnetism2013}. Trapped-ion quantum emulators have allowed the study of dynamical effects, such as the spread of entanglement in chains of 15 ions following a sudden perturbation \cite{JurcevicQuasiparticle2014} or a system of 10 ions in both the transverse Ising and related XY model Hamiltonians \cite{JurcevicQuasiparticle2014,RichermeNonlocal2014}.  Additionally, simulations of the effect of disorder on energy transport have been performed in chains of 10 ions, where the disorder could be controlled very precisely by programming it in an ion-site-resolved manner \cite{SmithQTransport2016, MaierQTRansport2019}.  Recent results have also extended the number of ions that have been involved in the calculations, such as studying the dynamics of a sudden quench in a 53-qubit system in the transverse-field Ising model with long-range interactions \cite{Zhang53IonSim2017} and the demonstration of collective entanglement in a 2-D system of more than 200 ions via measurement of spin squeezing \cite{BohnetSpinDynamics2016} (although the ions in that 2-D system were not individually addressable or imaged for individual readout). Entangled states of 20 trapped-ion qubits in a linear chain were also recently demonstrated \cite{Friis20QubitEntanglement2018}, though the generated state was not fully entangled (i.e., it could not be verified that each ion was entangled with every other ion). Despite some limitations, this impressive set of recent experiments showcases the ability of trapped ions to perform calculations with many qubits. Demonstrations of this nature suggests that trapped ion systems may be able to perform simulations of other quantum systems with 50 to 100 qubits in the near future, and may yield interesting results for problems such as interacting spin systems and quantum glasses.

\section{Methodologies for Practical Trapped-Ion Quantum Computing}
\label{PracticalQC}
In order to build a practical quantum computer, one that provides an advantage over what classical computers can deliver, systems of trapped ions must be scaled up to sizes much greater than currently exist.  As discussed in the introduction, this involves more than just increasing the number of ions that can be simultaneously trapped. Scaling up also requires the means to control and measure a large number of ion qubits, while maintaining the high performance achieved in the few-ion proof-of-principle systems.  This includes developing the ability to manage the finite errors that arise in quantum gates due to noise, decoherence, and control imperfections, and to keep them from cascading as the number of operations required to implement practical quantum algorithms grows.  In the next few sections we will discuss methods that are being explored to address the challenge of building systems of trapped ions at greater scale and complexity.

\subsection{Architectures and Techniques for Scaling to Larger Numbers of Ions}
\label{ScalingMethods}
As mentioned in Sec.~\ref{ProsandCons}, ions have a great advantage over other qubit modalities with regard to increasing their number: they are fundamentally identical.  This reduces concern about the reproducibility of ion qubits and simplifies and reduces the amount of required system calibration.  However, as we will discuss below, there are still many other concerns that arise when considering how to increase the number of ions.
    \subsubsection{Linear Arrays}
    \label{LinearArrays}
Due to the relative simplicity of working with single, linear ion traps, most few-ion QC demonstrations have been performed with the ions trapped in one dimensional (1D) arrays.  Many of the advantages of working with 1D arrays are retained as the number of ions is increased and, as a result, the field has primarily pushed in the direction of making larger 1D arrays. As discussed in Sec.~\ref{Algorithms}, experiments have produced entangled states of 14 ions \cite{Monz14IonEntanglement2011}, Shor's algorithm has been implemented to factor 15 using five ions \cite{MonzScalableShor2016}, and a five-ion-qubit quantum processor has been realized, demonstrating a number of quantum algorithms \cite{Debnath5QubitComp2016,FiggattGroverSearch2017}; this work has been extended recently to the demonstration of a chemical simulation using up to 11 ions~\cite{NamWaterMolecule2019}.  In addition, quantum simulations involving up to 53 ions held in a 1D array have been performed \cite{Zhang53IonSim2017}.

Despite their simplicity, however, single linear arrays of ions encounter significant limitations as the number of trapped ions is increased.  This is primarily due to the fact that the speed of two-qubit gates between ions in a chain generally decreases as the total number of ions grows \cite{MonroeModularArch2014}.  This results from smaller Lamb-Dicke parameters (and corresponding weaker coupling to the ion motion) that arise due to the larger-mass chains, as well as the reduction of ion-ion coupling strength as the distance, $s$, between ions in the chain grows (scaling as $1/s^{\alpha}$ with $\alpha$ ranging from 1--3) \cite{LeungArbitraryEntangling2018}.  In addition, performing high-fidelity two-qubit gates becomes more challenging in large chains due to the increased susceptibility to unwanted spectral crosstalk between the large number of collective normal modes that are used to mediate the two-qubit interaction, as well as to an increased susceptibility of these modes to be heated by noise in the system if the gates take longer.  Recently, techniques have been proposed to mitigate some of these problems by employing pulsed control of ions' spin-motion coupling to enact high-fidelity gates between arbitrary pairs of ions in a 50-ion linear chain \cite{ZhuArbitrarySpeed2006,LeungArbitraryEntangling2018}.  While this result is highly promising and improves the feasibility of working with modestly-sized linear arrays of ions, it involves more complex control than is typically employed for multi-qubit gates, has yet to be demonstrated experimentally, and is unlikely to be an efficient way to scale to a significantly larger number of qubits.

One promising way around these issues is to break a single long ion chain into smaller, modular pieces, with each piece being a manageable size, such that high-fidelity, high-speed operations can be performed within each module.  The challenge, then, becomes how to best move quantum information between modules.

    \subsubsection{Two-Dimensional Arrays and the Quantum CCD Architecture}
    \label{2DArrays}
\begin{figure*}[t b h]
\includegraphics[width=2.0\columnwidth]{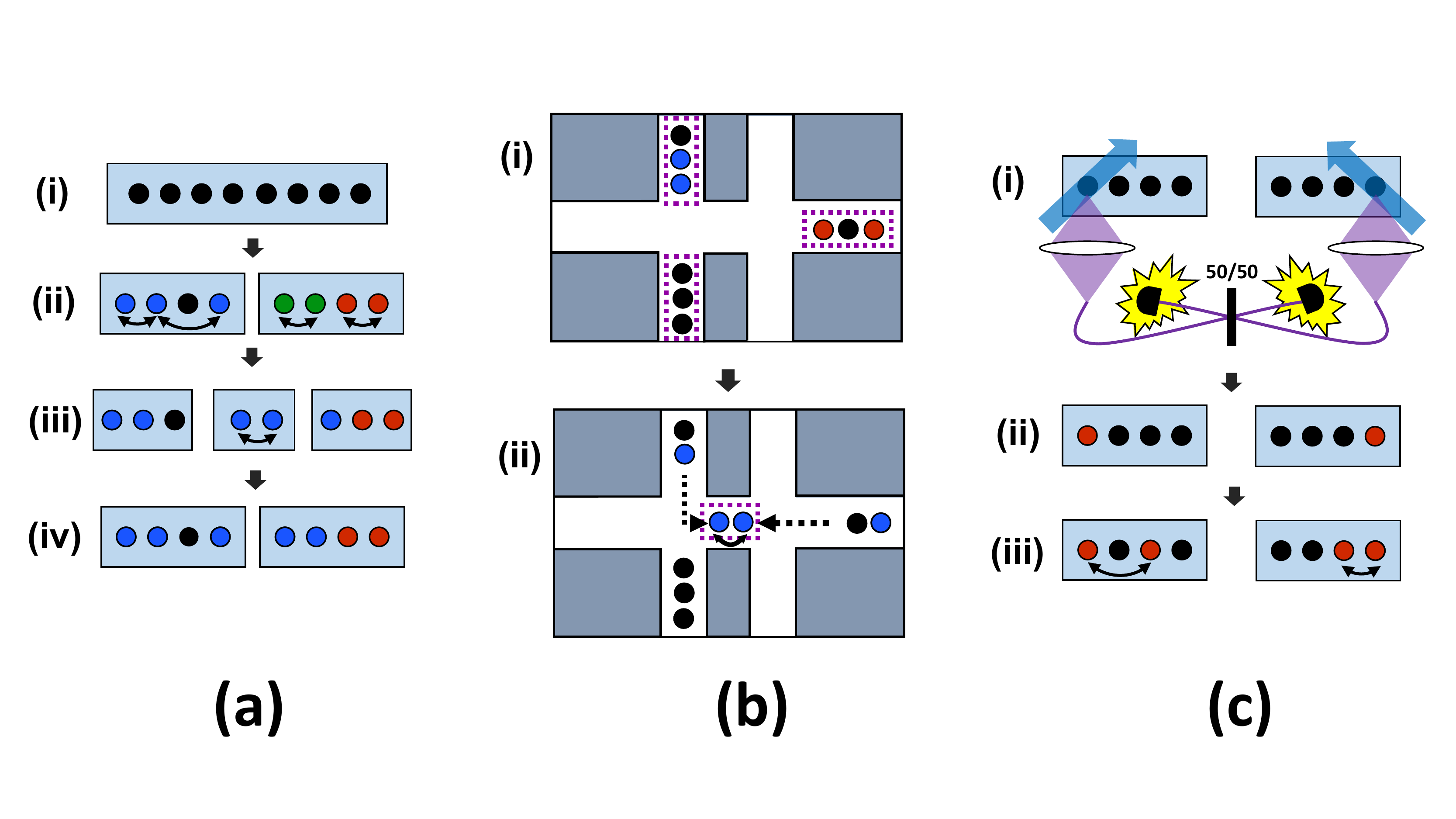}
\caption{Architectures for practical trapped-ion systems.  Circles represent ion qubits, and entanglement between qubits is indicated by like fill color.  Black fill indicates no entanglement.  Curved lines with arrows indicate an entangling gate has occurred.  (a) Linear Array: (i) A 1D chain of ions can be straightforward to produce, but performing entangling gates in such chains with high fidelity may prove more difficult as the chain size grows. (ii) One strategy is to split the chain into modules (blue boxes) to perform high-fidelity entangling gates. (iii) To entangle ions between modules, the modules can be reconfigured such that high-fidelity entangling gates can be performed between ions of formerly-different modules; (iv) the modules can then be subsequently returned to their original state.  (b) 2D Array and the QCCD: (i) Modules each consisting of a few ion qubits are arranged in a 2D array.  Ions, which may have been previously entangled, reside in so-called ``memory" zones (indicated by the dotted box), which can be optimized to allow for long qubit coherence time. (ii) To entangle qubits between modules, the ions are shuttled (indicated by dashed lines with arrows) through X-junctions to an ``interaction region" (indicated by a dotted box) where entangling gates are performed. Ions can then be shuttled back to return the modules to their original state, and the process can then be repeated to entangle ions in other modules.  (c) Photonic Interconnects: (i) Ions in each module start off un-entangled with one another.  A dedicated communication ion in each module is then excited by a laser (blue arrows) which results in the emission of photons (purple) that are entangled with the resulting internal ion state.  The photons are collected with optics (light gray) and interfered on a 50/50 beamsplitter (BS).  (ii) Simultaneous detection of photons by single-photon detectors (black hemispheres), positioned at the output ports of the BS, heralds entanglement between the communication ions. (iii) Intra-module entangling gates can then be performed to increase the number of entangled qubits across modules.  We note that, while the photonic interconnect architecture is shown here for linear ion arrays, it can also be readily applied to a 2D array.}
\label{fig:architectures}
\end{figure*}

Perhaps the most natural way to move quantum information from one location to another in trapped-ion systems is to move the ions themselves.  Indeed, the ability to do this, by varying the voltages applied to the ion trap electrodes to alter the trapping potential, is a real feature of working with trapped-ion qubits.  The conceptually simplest scheme one could imagine for this is to continue to work with 1D arrays, but to dynamically reconfigure the 1D array into smaller modules, as shown schematically in Figure \ref{fig:architectures}a.  In this scheme, quantum information is moved between two modules by first taking a subset of ions from each and creating a new module within which quantum information can be distributed.  Then, the modules are returned to their original configuration, now with the quantum information distributed between them. Repeating the operation for additional modules allows entanglement to be more widely distributed, in principle among an arbitrary number of modules.

From an implementation point of view, this requires, in principle, only minimal translation of the ions.  It does, however, require splitting and joining of ion chains, and such splitting and joining has been demonstrated in 1D ion arrays \cite{BowlerDiabaticTransportSplit2012,RusterSplitJoin2014}. Additionally, it requires some kind of swap operation between qubits in neighboring modules.  Such a swap can be accomplished via two techniques: either by performing a two-qubit SWAP gate (which interchanges the full quantum state between qubits and which can be composed of three CNOT gates) or by physically swapping the ions' positions.  However, in both cases, the number of swap operations required to move quantum information from one end of a 1D array to the other will scale linearly with the number of modules (and the number of qubits in the array), and thus challenges exist for doing this quickly and with low error.   For the technique based on SWAP gates, the gate introduces an error and incurs a cost in time per operation.  The technique based on ion position re-ordering has been demonstrated with high swap fidelity by rotating the ion crystal for two- and three-ion chains \cite{KaufmannChainReorder2017}. However, like the SWAP gate technique, it is time consuming, since it involves a number of split, join, and rotation operations.   As a result, the 1D array architecture can limit the distance over which quantum information can be distributed.

Extending the architecture to two-dimensional (2D) arrays has been proposed as a way to overcome this limitation \cite{KielpinskiQCArchitecture2002}.  In this scheme, ion qubits can be stored in modules that are distributed in a 2D plane and, in addition to ion-chain splitting and joining, ion transport is now relied upon to move quantum information around (see Figure \ref{fig:architectures}b).  Because of the introduction of the second dimension, this can be done, in principle, between any qubits in the plane without incurring the significant time and error costs of swapping quantum information via entangling gates or chain re-ordering operations.   While it is true that the time required to move information in the 2D architecture scales linearly with the distance over which the information is to be distributed (and over which the ions must be transported), the number of required split and join operations is independent of distance.  And since ion transport is typically much faster than a two-ion gate or a re-ordering operation (and lower error than the gate), the 2D architecture has the potential to offer a faster and higher-fidelity method to distribute quantum information over a many-ion array.

Beyond distribution of quantum information, 2D arrays offer an additional architectural benefit. As discussed in Ref.~\cite{KielpinskiQCArchitecture2002}, different regions in the array can be used for different functions.  For example, there can be ``memory" regions that are spatially separated from potential sources of decoherence, an ``interaction" region where control fields are present and gates are performed, a ``measurement" region where the ion qubit states are measured, and a ``loading" region where neutral atoms are initially ionized and trapped in the array \cite{LekitscheMicrowaveBlueprint2017}.  Spatially separating the functionality of different regions allows one to tailor and optimize the trap design for that particular function. The name for this architecture, coined in Ref.~\cite{KielpinskiQCArchitecture2002} and used widely in the field, is the Quantum Charge-Coupled Device (QCCD).

The flexibility offered by 2D arrays implies a few considerations, however.  First, as mentioned above, the speed of quantum information distribution is now set by the speed of ion transport.  To this end, there has been considerable work in the field done to develop methods for fast ion transport \cite{BowlerDiabaticTransportSplit2012,WaltherDiabaticTransport2012,LuFastShuttling2018}.  The difficulty of this problem and the speed limits that arise come from the desire to not heat or excite the motion of the ions as they are transported.  This desire stems from the fact that the fidelity of multi-qubit gates is significantly degraded for higher-temperature ion chains \cite{KirchmairHotGates2009}.  As we discuss in Sec.~\ref{DualSpecies}, ion motion can be cooled to mitigate this problem, but this cooling also comes with a time cost.  As a result, methods to improve the speed of ion transport focus on doing so with as low ion-motional excitation as possible.  In the so-called ``adiabatic" transport schemes \cite{FurstTransportAnalysis2014}, the speed limit is set, roughly, by the inverse of the ion trap secular frequency.  This can be understood as follows: in order to not excite the ion to higher motional states, the change in trap potential should be slow compared to the trap period to ensure that the potential has no near-resonant Fourier components that could drive transitions between motional levels.  Other, so-called ``diabatic" schemes \cite{BowlerDiabaticTransportSplit2012,WaltherDiabaticTransport2012,LuFastShuttling2018} can go faster than the adiabatic ones (without incurring much additional heating) by timing the transport precisely such that the total transport time is an integer number of trap oscillation periods.  In Ref.~\cite{BowlerDiabaticTransportSplit2012}, an ion transport speed of ${\sim}50$~m/s was demonstrated with only 0.1 quanta of motional excitation (the increase of the average population of the quantum harmonic oscillator defined by the ion trap); to give this speed some perspective, in 100 $\mu$s (a typical ion two-qubit gate time), the ion could in principle be moved 5 mm, a very large distance compared to the typical few-micron spacing between ions in the same chain.

A second issue that arises from the 2D array architecture is the increased complexity of the ion traps. In order to transport ions around on a 2D plane, a 2D arrangement of segmented trap electrodes is needed.  Additionally, transport in the array that enables arbitrary ion chain re-ordering, as well as high connectivity between ions (i.e., any ion can be transported to couple to multiple other ions) will require junctions in the paths over which the ions travel.  Junctions are locations in the trap where linear regions meet and form, for instance, ``T" or ``Y" shapes (i.e., 3-way junction) or an ``X" shape (4-way junction). Not only will a 2D arrangement of electrodes and the presence of junctions increase the overall number of required electrodes, it is difficult to imagine implementing such an arrangement entirely via mechanical assembly of the electrode structures, as is traditionally done with 1D linear ion traps.  Initial demonstrations of traps capable of implementing 2D transport relied on a combination of microfabrication, laser machining, and mechanical assembly to realize T \cite{HensingerTJunction2006} and X \cite{blakestad_junction_2009} junctions.  Such demonstrations, where the traps had just a single junction, can be considered as conceptual building blocks of a larger 2D array. These experiments paved the way for more complex designs by, for instance, exploring the challenges associated with increased sensitivity to noise and motional heating arising from the presence of gradients, or ``bumps" in the RF pseudopotential at the junction \cite{PhysRevA.78.063410}.   However, given the still somewhat-complex  fabrication and assembly processes of these traps, this work did not offer a clear technological path towards realizing large 2D arrays.

Fortunately, as mentioned in Sec.~\ref{MicroTraps}, microfabricated surface-electrode ion traps are a natural solution to this problem.  Indeed, it was this concern which was the prime motivation for developing the surface-electrode trap technology. In the last few years, transport in more complex 2D array building blocks has been demonstrated with surface electrode traps, including Y \cite{amini_racetrack, sandia_junction_2011} and X \cite{WrightXJunction2013} junctions.  In addition, 2D ion arrays have been demonstrated in hexagonal \cite{Sterling2014}, triangular \cite{Mielenz2016}, and square \cite{BruzewiczArrayLoading2016} lattice configurations, though transport has so far not been implemented due to a lack of junctions and ``streets" (the paths for ions to travel between lattice sites).  While surface electrode trap technology represents a very promising approach, it nonetheless remains an outstanding goal in the field to realize a 2D lattice array trap that includes junctions and streets to enable general transport of ions between sites.

    \subsubsection{Photonic Interconnects}
    \label{PhotonInter}
Another method for distributing quantum information between modules utilizes photons \cite{MonroeModularArch2014} to entangle ions located in separate regions.  This method, often referred to as remote entanglement (RE), works in the following way and is depicted schematically in Fig.~\ref{fig:architectures}c.  One ion (referred to as a ``communication ion") in each of two spatially-separated modules to be connected is excited by a fast laser pulse to a short-lived state and, upon rapid decay of this excited state, emits a single-photon.  The excited state is judiciously chosen such that when the photon is emitted, radiative selection rules ensure that it is maximally entangled with the electronic state of the ion qubit.  Different ion species and ion qubit levels can be used to produce single-photon qubits in either a frequency \cite{Moehring2007} or polarization \cite{BlinovIonPhotonEntangle2004} basis, where the photonic qubit basis states (or ``optical modes") have distinguishable frequencies or polarizations, respectively.  The photons from each ion are then collected and modematched using optics and interfered with each other on a 50/50 beam splitter.  A single-photon detector is placed at each of the output ports of the beam splitter and simultaneous detection of a photon by each of these detectors generates and heralds a maximally entangled state of the two ion qubits.  This occurs when the emitted photons are in different optical modes, which happens on average half of the time.  When the modes are the same (the other half of the time), the indistinguishability of photons ensures that they exit the beam splitter along the same path and no two-detector coincidence is observed (the Hong-Ou-Mandel effect \cite{HongOuMandel1987}); in this case, the RE generation is unsuccessful, but it can be retried repeatedly until success is heralded.  The rate of entanglement generation is then given by $\Gamma_{RE}=\gamma_{rep}(\eta_{c}\eta_{d})^2/2$, where $\gamma_{rep}$ is the attempt or repetition rate, $\eta_{c}$ is the photon collection efficiency (or the ratio of total photons collected in a single optical mode to the total photons emitted by one ion), and $\eta_{d}$ is the quantum efficiency of the detector.  This assumes unity probability that the ions are excited by the preparation pulse, which is typically (approximately) the case.

The repetition rate $\gamma_{rep}$ is limited fundamentally by the ion excited state lifetime (typically ${\sim}10$~ns) but more practically by the time it takes to re-initialize the ion state via Doppler cooling and optical pumping before each attempt ($\sim$1~$\mu$s) \cite{HuculModEntanglement2014}.

As mentioned previously, the best overall photon collection efficiencies achieved in trapped ion systems have been in the range of $0.02$--$0.04$. For purposes of RE, though, there is an additional requirement of optical mode-matching on the 50/50 beamsplitter, which necessitates coupling the collected photons from the two systems into a single optical mode. This is usually achieved by using the collection optics to couple light from each ion into a single-mode fiber with a typical efficiency of $\sim$0.1 \cite{HuculModEntanglement2014}.  As a result, the highest values of $\eta_{c}$ that have been achieved are $\sim$0.005.

Using the standard methods described above, the highest-achieved rate for trapped-ion RE generation is ${\sim}5$~Hz~\cite{HuculModEntanglement2014}.  Clearly, it is desirable to increase this rate, as it sets a practical limit to the operation speed of the quantum computer.  One strategy for doing this is to spatially-multiplex the RE process, that is, to use a larger subset of the ions in each module in parallel to make more attempts at RE in a given time.  Since the entanglement is heralded, the multiple pairs of entangled ions can be utilized as long as one can keep track of which photons come from which ion pairs.  However, this strategy for increasing the RE rate comes at a cost in both ion number, as well as in the complexity of collecting, routing, and interfering photons from multiple ions.  In order to minimize this overhead, it would thus be advantageous if the per-ion-pair RE rate was improved.  Given the fundamentally limited values of $\gamma_{rep}$ and the fairly high $\eta_{d}$ values already achieved,  it seems the biggest opportunity for this improvement is through increase of $\eta_{c}$.

One possible path to improved photon collection efficiency is to couple the ion to a high-finesse optical cavity.  The cavity can be designed to alter the ion's photon emission pattern from one that is largely isotropic to one that leads to significantly-enhanced emission into the cavity mode.  As a result, the collection efficiency need not be limited by the NA of the collection optics.  Much work has been done towards demonstrations of ion-cavity coupling \cite{MundtCavity2002, KellerCavity2004, HerskindCavity2009, StuteCavity2012, SteinerCavity2013, StuteCavity2013, TakahashiCavity2018, BallanceCavity2017, BegleyCavity2016}.  In general, the size of the collection enhancement increases with increased cavity finesse and decreased cavity mode volume.  Achieving very high finesse, which is often practically limited by the reflectivity of the cavity mirrors, has thus far proven difficult in general, and especially for blue-to-UV wavelengths.  Indeed, nearly all demonstrations of ion-cavity coupling have been implemented using infrared photons even though the ion-level transitions that produce these photons are not the rapid ones that are typically used for high-fidelity readout or high-rate remote entanglement.  In Ref.~\cite{BallanceCavity2017}, coupling of a Yb$^+$ ion to a UV cavity via 369-nm photons was demonstrated, but the cavity finesse degraded significantly over the course of a few months once it was placed in the UHV environment required by the trapped ion.  Decreasing the cavity mode volume to increase the collection efficiency, which involves moving the cavity mirrors closer to the ion, has also proven difficult.  This is because the mirrors are coated with dielectrics in order to achieve high reflectivity and they typically charge up to a degree that prevents ion trapping for very small mirror separations. Despite these difficulties, research aimed at demonstrating increased ion-cavity coupling strength is likely to continue given its potential impact on both photonic interconnects as well as ion-state measurement.

If RE generation rates are improved, these photonic interconnects offer a powerful means for distribution of quantum information.  It not only allows for connection between modules or zones of a single ion trap, but even between ion trap modules located in separate vacuum chambers \cite{MaunzRemoteInterference2007, Moehring2007}.  Furthermore, the connection speed is, in a practical sense, independent of the physical distance between modules, even over the scale of an entire laboratory, since the information travels at light speed.  While the discussion of photonic interconnects has thus far focused on connection between just two modules, the architecture is extensible to multiple modules.  To do so, and realize high connectivity between many modules, a many-port optical switch may be required, as is suggested and described in Ref.~\cite{MonroeModularArch2014}.

\subsection{Error Reduction and Mitigation}
\label{ErrorReducandMit}
As quantum computers, and the algorithms that are run on them, grow in size and complexity, errors that occur throughout computations will have to be managed.  These errors result from two general mechanisms.  The first is decoherence, which arises from undesired coupling of the qubit to its environment.  For ions, examples of this undesired coupling are spontaneous emission or fluctuating fields that shift the qubit energy levels or heat ion motion.  The second error mechanism arises from imperfect control fields.  These imperfections could take the form of miscalibrated or noisy control-field amplitude, frequency or polarization; they typically result in quantum gate errors, though we note that noisy control fields can also lead to decoherence.

There are two main strategies that can be employed to manage these errors: reduce the rate at which they occur or detect and correct them.  In the next few sections, we will discuss progress with trapped ions on these fronts.

\subsubsection{Decoherence-Free Subspaces and Composite-Pulse Control}
\label{DFS}
Perhaps the most obvious way to reduce errors is to reduce the sources of decoherence and control-field imperfections.  Indeed, all groups working with trapped ions take great pains to do this, but this approach can only be taken so far; it is not reasonable to expect that these sources can be eliminated completely.  A complementary strategy is to make the system somehow less sensitive to error sources.

One promising approach for doing this is using a so-called decoherence-free subspace (DFS) of qubits \cite{LidarDFS1998, ZanardiDFS1997, DuanDFS1998}, whereby a ``virtual" qubit \cite{JonesQCArchitecture2012} is formed by combining multiple ``physical" qubits to create a subspace of states that are insensitive to certain decoherence sources.  The general idea of this approach is based on the fact that many sources of decoherence affect qubits in close physical proximity equally.  An example of this for ions would be a spatially homogeneous (on the scale of the inter-ion spacing), but temporally fluctuating, magnetic or optical field that shifts the splitting between the $\zero$ and $\one$ ion-qubit levels.  In this case, a DFS can be created using two physical qubits to encode one virtual qubit in the entangled states of the form $\alpha\zero\one+\beta\one\zero$.  Such states are relatively immune to dephasing since a change in the $\zero\rightarrow\one$ splitting results in the same phase accrual for both virtual qubit basis states.

A first implementation of this technique with trapped ions used the hyperfine ground states of two $^9$Be$^+$ ions to encode the DFS \cite{KielpinskiDFS1013}.  In this work, it was shown that the coherence time of the DFS, as compared with a single-qubit state, could be improved by a factor of $\sim$ 50 in the case where noise was purposely applied to the system via an off-resonant laser with intensity fluctuations (which induced time-varying AC-Stark shifts).  In addition, it was shown that the coherence time could be extended by a factor of $\sim$3.5 for ambient noise, demonstrating that a large fraction of this ambient noise was common-mode to the two ions and that it could be mitigated via a DFS. Subsequently, it was shown that the DFS could be used to extend the coherence time of the Be$^+$ hyperfine-qubit system to greater than 7 seconds \cite{langer2005long}. A two-ion DFS was later investigated using Zeeman qubits in $^{40}$Ca$^+$, where it was demonstrated that entanglement could be used as a resource for extending quantum coherence up to 20 seconds \cite{HaeffnerRobustEntanglement2005}.  This technique was even extended using eight $^{40}$Ca$^+$ ions in a DFS state based on optical qubits, demonstrating a coherence time of $\sim$320 ms, limited by the lifetime of the metastable excited qubit level.

Beyond showing that a DFS can improve the coherence time, or memory, of trapped-ion qubits, it was crucially shown theoretically that universal quantum computation could be performed on DFS qubits in trapped-ion systems \cite{AolitaDFS2007, CenDFS2006}.  Subsequently, experiments were performed demonstrating the realization of a universal set of quantum gates acting on $^{40}$Ca$^+$ ions in a DFS \cite{MonzDFSGates2009}.

While the DFS shows great potential for reducing errors, it has only been used in ion experiments in a limited fashion.  This seems to be primarily because most research groups work with just a few ions, making even a modest factor of two in qubit number overhead somewhat prohibitive.  It is possible that as systems with larger numbers of ion qubits become more common, work in the area of DFSs will become more prevalent.

A second approach for reducing errors, arising from both imperfect control and decoherence, is to use composite control pulse sequences \cite{MerrillCompPulseReview2014, KabytayevCompPulses2014}.  The general idea of this approach is to use multiple control pulses, separated in time, to decrease the sensitivity of the qubit to noise in the control fields or the environment.   The simplest example of this is spin echo~\cite{PhysRev.80.580}, which involves applying a $\pi$-pulse in such a way as to refocus a qubit that has undergone dephasing. Extending the technique to multiple pulses allows one to tailor the refocusing to counteract noise at particular frequencies; such a generalization is often referred to as  dynamical decoupling \cite{ViolaDynamicDecoup1998}, and it can be thought of as an active method to spectrally filter noise.  Such techniques have been developed and implemented with trapped ions to extend their coherence time \cite{BiercukDynamDecoup2009, SzwerDynamDecoup2011,Wang10MinuteCoherence2017}, as well as to demonstrate tuneable noise filters \cite{BiercukDynamDecoup2009}.

In addition to using dynamical decoupling to reduce quantum memory errors, it has also been shown that it can be used during quantum gates to improve their fidelity by counteracting dephasing during the gate time \cite{WestDynamDecoupGates2010, BermudezGate2012}. Such techniques have been employed with trapped ions to realize two-qubit gates \cite{TanDressedState2013,BallanceHybridLogic2015,manovitz2017fast}, with the current best 99.9\% two-qubit gate fidelity in trapped ions enabled by such a scheme \cite{Ballance2QubitHyperfineGate2016}.

The error in gates arising from imperfect control can also generally be reduced through the use of composite pulse sequences, such as BB1 or CORPSE/SCROFULOUS \cite{WimperisBB11994,CumminsPulseSequence2003}. Such pulse sequences essentially render a gate operation resilient to a particular type of error (such as control-field frequency-detuning or amplitude errors), at the expense of a longer sequence. These control-error-compensation sequences have been implemented for hyperfine qubits driven by microwaves \cite{TimoneyErrorResistant2008} and Raman gates \cite{MountErrorCompensation2015}, and have typically reduced single-qubit gate errors by a factor of 3 to 4. In Ref. \cite{MountErrorCompensation2015}, this allowed errors below $4 \times 10^{-4}$ to be achieved. Higher-order pulse sequences can provide even greater error suppression, but at some point the increase in overall gate length is not justified by the error reduction.  In addition to gate errors, composite pulse sequences have also been developed and demonstrated to reduce crosstalk errors arising in optical-qubit control \cite{MerrillCompCrossTalk2014, LowMethodology2016, McConnellHeisenberg2017}.

    \subsubsection{Error Correction}
    \label{ErrorCorrection}
Useful quantum algorithms that exceed what can be done classically often require an impressive number of gate operations: for example, with its gate depth polynomial in the input size, Shor's algorithm requires on the order of $10^{10}$ operations to factor a $1024$-bit number \cite{BeckmanEfficient1996}. Despite efforts and great progress to reduce errors from decoherence and imperfect control, the errors have not been lowered to a level that is sufficient to successfully perform an algorithm with this number of operations with a usefully-high fidelity.  As a result, while the reduction of errors is necessary, a second strategy that is focused on detecting and actively correcting errors is also required.  This error detection and correction is particularly challenging in quantum systems because the measurement of quantum superposition states collapses them.  Copying the states to create redundancy, a technique often employed in classical error correction, is not available to quantum error correction due to the well-known No-Cloning theorem \cite{WoottersNoCloning1982,DieksNoCloning1982}.  Despite these challenges, quantum error correction is indeed possible, but it comes at significant cost in required numbers of qubits and operations.  A discussion of how quantum error correction works and the concepts of quantum fault tolerance are beyond the scope of this article. (We refer the interested reader to an excellent introduction of the topic \cite{GottesmanQEC2009}, as it is general to all qubit modalities.)  However, we note a few key elements that are germane to trapped ions.

First, one must define a logical qubit. In the case of quantum error correction, this is a space of multi-qubit, entangled states that are contained in a number of physical qubits.  This logical-qubit state space is designed to allow for correction of a particular set of errors that can occur on the physical qubits.   It has been shown that this can be done for an arbitrary error, in principle, with a code that uses as few as 5 physical qubits \cite{KnillQEC1997} to correct one error on the logical qubit.  As noted below, however, the overhead is usually significantly higher.  Some codes require coupling of only nearest-neighbor qubits to one another \cite{BravyiKitaevSurfaceCode1998, FowlerSurfaceCode2012}, while others may benefit from longer-range couplings \cite{SvoreFTThresholds2006}.  The former are usually preferred from a practical implementation standpoint, though clearly ions are also well-suited to the latter due to their ability to be transported and to the availability of long-distance photonic interconnects.

Quantum error correction cannot result in a net improvement if the initial gate error rates are too high.  For a given error correction code, which in addition to the logical qubit states includes the set of quantum operations required to detect errors and to perform logical operations on the encoded qubit, a threshold can be defined as the maximum physical-qubit error rate that, if present, will result in the logical qubit error rate being reduced below that of the physical qubits.  (Strictly speaking, this is the definition of a ``pseudo-threshold", but for the sake of this review we will use the term ``threshold" more loosely; a discussion on the difference is given in Ref.~\cite{SvoreFT2006}.)  For example, a code known as the surface code \cite{RaussendorfSurfaceCode2007} has a particularly high threshold of ${\sim}0.01$, though it requires a higher minimum number of physical qubits than the ideal 5 per logical qubit, which hints at the tradeoffs that must be made when choosing a particular code.  The overhead to achieve a given logical error rate is strongly dependent on the physical qubit error rate, with dramatic increases in overhead required if physical qubit errors are close to the threshold.  As a result, reducing physical qubit errors leads to a reduction in required resources for QC.  This is a chief reason why trapped ions, with their high operation fidelities, are a particularly good qubit modality.

As noted above, in order to correct errors, they need to be detected, and this is generally accomplished in quantum error correction via the use of so-called ancilla qubits (or ancillas).  The ancillas are first entangled with the qubits that store the quantum data (data qubits) and are then measured.  The process of managing quantum errors basically consists of repeated cycles of ancilla-data qubit entanglement, ancilla measurement (error detection), and feedback to the data qubits (error correction).  As a result, the practice of error correction not only requires many additional qubits, but is very measurement intensive.  Unlike quantum algorithms that do not consider error correction, where the entire quantum register is ``read out" or measured at the end of the computation, the inclusion of error correction requires that ancillas are measured throughout the computation and that the data qubits are not unintentionally measured in the process (which would lead to un-correctable errors in the corresponding round of error correction).  It turns out that this is not so easy to achieve in practice, especially when all the physical qubits are identical, as in the case of trapped ions.  This is because it takes many photon absorption and re-emission events to determine the state of an ancilla ion qubit (due to limited photon collection and detection efficiency), yet the absorption of a single photon by a data ion effectively measures it.  This therefore places very stringent requirements on system crosstalk \cite{BruzewiczQLAR2017}, though, as discussed in Sec.~\ref{Detection} and further in Sec.~\ref{DualSpecies}, this problem can be greatly mitigated through the use of a second ion species.

Despite the importance of quantum error correction, the large number of required qubits and quantum operations have resulted in just a few experimental implementations to date.  The first such example of this in a trapped-ion system demonstrated correction of single spin-flip errors using three physical qubits to encode one logical qubit~\cite{ChiaveriniQEC2004}.  While this was an important first step, the three qubit code used is known to be insufficient to correct arbitrary quantum errors, and indeed, phase-flip errors could not be simultaneously managed.  Additionally, only a single error correction cycle was implemented.  A similar three-qubit code was later used to correct phase-flip errors (and not spin-flip errors), this time in a repetitive fashion \cite{SchindlerQEC2004}.  More recently, error detection of both spin and and phase flip errors in a seven-ion logical qubit was demonstrated and sequences of gate operations were performed on the encoded qubit, though error correction was not implemented \cite{NiggQED2014}.  In addition, error detection of both spin and and phase flip errors in a four-ion logical qubit was demonstrated using one additional ancilla ion to determine which error had occurred \cite{LinkeErrorDetection2017}.  Importantly, this detection was implemented in a fault-tolerant manner, such that a single physical qubit error could not lead to an undetectable logical qubit error.  However, the ultilized code was such that errors could not be uniquely identified, and therefore could not be corrected.

Due to  its seeming necessity, work aimed at experimental demonstrations of fault-tolerant quantum error detection and correction of arbitrary errors, that is, the realization of a logical qubit that has smaller errors than the physical qubits of which it is composed, is a highly active area of current research in the trapped-ion QC field \cite{BermudezAssessing2017, TroutSurfaceCodeSim2018}.

   \subsubsection{Dual-Species Ion Systems}
   \label{DualSpecies}
While the identical nature of trapped-ion qubits is a major net advantage of these systems, there are a few instances where it proves to be a hindrance.  In Secs.~\ref{Detection} and \ref{ErrorCorrection}, we already discussed how measurement of ancilla ions can cause undesired measurement, and therefore decoherence, of other identical data ions.  Also, as discussed in Sec.~\ref{PhotonInter}, establishing a photonic interconnect via remote entanglement involves the scattering of multiple photons by communication ions.  However, just like in the case of measurement, absorption of these photons by data ions is a decoherence process and will occur with significant probability when the communication and data ions are identical.

Another issue that arises is related to laser cooling of ions. As discussed in Sec.~\ref{Multiqubitgates}, the fidelity of trapped-ion two-qubit gates can be limited by excitation or heating of the ion motional modes that are used.  This heating can arise from multiple sources, including technical noise and anomalous heating, as described in Sec.~\ref{AnomHeating}, or due to non-idealities in the control of ion transport or ion module splitting and joining.  Whatever the cause of the motional excitation, it will in general occur during a quantum algorithm, and therefore must be cooled during the algorithm as well.  However, laser cooling involves the scattering of many photons, an intrinsically decohering process, and thus cannot be performed directly on data qubits without destroying the data.  A solution to this problem is to use a separate ion (or ions), coupled to the data ions via the motional modes, to cool the data ion.  This technique, known as sympathetic cooling, is effective since the decoherence-inducing cooling lasers, in principle, only interact with the spin of the sympathetic cooling ion, which does not store any necessary quantum information.  In practice, however, a problem remains due to the identical nature of ions: much like in the case of qubit measurement, there is a high probability that a stray cooling photon will be absorbed by, and thus decohere, the data qubit.  It is worth noting that this cannot be managed solely by reducing control cross-talk.  The sympathetic cooling process requires close proximity between data and cooling ions and, since the cooling process results in fairly isotropic re-emission of photons by the cooling ion, there is in many cases an unacceptably large probability that the data ion will absorb a re-emitted photon \cite{BruzewiczQLAR2017}.

There is fortunately a simultaneous solution to the problems associated with the need for in-algorithm qubit measurement, remote-entanglement generation, and laser cooling: using two species of ions, one for the data ions and the other for the ancilla, communication, and sympathetic cooling ions.  In all cases, a second ion species mitigates the problem because photons scattered by one species are far-detuned from any transition in the other, and are thus absorbed by it with very low probability.

There are a few different operations that are required in a dual-species ion chain, which depend on the particular application.  For measurement and communication applications, the quantum information needs to be swapped from the data qubits to the ancilla or communication ions, which involves an interspecies quantum operation.  Such interspecies operations require control fields that address transitions in both ion species.  In the case of measurement, since only the quantum state populations, and not the coherences, are of interest, this information swap does not necessitate an interspecies two-qubit entangling gate (e.g. an MS-like gate); rather, a simpler transfer scheme such as that used for quantum logic spectroscopy may be all that is needed~\cite{SchmidtQuantumLogicSpectroscopy, BruzewiczQLAR2017}.  That said, these simpler schemes often require high-fidelity ground-motional state preparation in order to achieve high population-swap fidelity, which in some sense transfers the burden to sympathetic cooling.  As a result, an MS-like gate, with its relatively higher insensitivity to motional state populations, may in some cases be more desirable despite its higher degree of difficulty .

For inter-module connections via photonic links, an interspecies two-qubit entangling gate is required since the goal is to generate entanglement between modules, and this can only be done by swapping the entanglement from communication qubits to data qubits.  In addition, the establishment of remote entanglement requires the scattering of multiple photons which can induce recoil heating of the chain; therefore, sympathetic cooling may be required in this application.

In the above applications, it is important to consider the choice of the motional mode used for sympathetic cooling and inter-species operations.  The chosen mode must have substantial Lamb-Dicke parameters for all the control fields that need to be coupled to the ion motion so that operations can be performed at high speed without requiring high control-field intensity.  It is also important to consider the masses of the ions in the dual species chain. It is generally desirable to use two species of ions that have similar mass (e.g., a mass ratio, $\mu$=$m_1/m_2\sim$1) \cite{WubennaSympCooling2012, HomeMixedSpecies2013}.  This is because, for few-ion chains, as $\mu$ tends away from 1, the components of the motional-mode eigenvectors tend to small values for one ion species or the other.  And, as discussed above, Lamb-Dicke parameters (which are proportional to the eigenvector components) need to be substantial for both ions. Values of $\mu$ closest to 1 can be achieved using different isotopes of the same element, though the isotope shifts are typically not large enough to prevent decoherence at a level required for QC. Instead, choosing ions of different elements is widely considered to be the best path forward.

Given the recognized importance of dual-species ion systems for scalable architectures, much work has been done to perform key operations in small dual-species ion chains \cite{HomeMixedSpecies2013}.  The first demonstrations of dual-species sympathetic cooling were done in Penning traps, where $^{24}$Mg$^+$ was used to cool other Mg$^+$ isotopes \cite{DrullingerSympCooling1980} and where $^9$Be$^+$ was used to cool $^{198}$Hg$^+$ \cite{LarsonSympCooling1986} and multiple isotopes of Cd$^+$ \cite{ImajoSympCooling1996}; it was later demonstrated in a Paul trap with two different isotopes of Cd$^+$ \cite{BlinovSympCooling2002}.  Sympathetic cooling to near the motional ground state, as is likely required for high-fidelity quantum operations has since been done in $^{24}$Mg$^+$-$^9$Be$^+$ \cite{BarrettSympCooling2003}, $^{27}$Al$^+$-$^9$Be$^+$ \cite{SchmidtQuantumLogicSpectroscopy}, $^{27}$Al$^+$-$^{25}$Mg$^+$ \cite{ChouAlClockcomp2010},  $^{43}$Ca$^+$-$^{40}$Ca$^+$ \cite{HomeCompleteMethods2009}, $^{40}$Ca$^+$-$^{88}$Sr$^+$ \cite{BruzewiczQLAR2017}, and $^{171}$Yb$^+$-$^{138}$Ba$^+$ \cite{InlekMultiNode2017} two-ion chains, as well as in a $^{9}$Be$^+$-$^{40}$Ca$^+$-$^{9}$Be$^+$ three-ion chain \cite{NegnevitskyMultiReadout2018} and a $^{9}$Be$^+$-$^{24}$Mg$^+$-$^{24}$Mg$^+$-$^{9}$Be$^+$ 4-ion chain \cite{JostEntangledOsc2009}.

As discussed in Sec.~\ref{Detection}, the first dual-ion-species quantum operations were aimed at improving the accuracy of optical atomic clocks, where sympathetic cooling, state preparation, and state measurement were performed using a two-ion $^{27}$Al$^+$-$^9$Be$^+$ chain \cite{SchmidtQuantumLogicSpectroscopy}.  Similar techniques using dual-species to assist quantum state measurement were demonstrated in two-ion $^{40}$Ca$^+$-$^{88}$Sr$^+$ \cite{BruzewiczQLAR2017} and $^{171}$Yb$^+$-$^{138}$Ba$^+$ \cite{InlekMultiNode2017} chains.  In a $^{43}$Ca$^+$-$^{40}$Ca$^+$ \cite{HomeSympCoolMem2009} two-ion chain, it was explicitly verified that the coherence of the data ion could be maintained during sympathetic cooling.  In addition, a coherence time of 10~min was demonstrated in a $^{171}$Yb$^+$ ion qubit, a measurement that was enabled by sympathetically cooling the Yb$^+$ ion with a $^{138}$Ba$^+$ ion~\cite{wang2017single}. Quantum operations in larger dual-species ion chains were performed using the four ion $^{9}$Be$^+$-$^{24}$Mg$^+$-$^{24}$Mg$^+$-$^{9}$Be$^+$ configuration, where single and two-qubit gates were implemented between the Be$^+$ data ions in the presence of the Mg$^+$ cooling ions to demonstrate a gate set capable of universal QC \cite{HannekeProgrammable2009}.  This chain was further utilized to demonstrate key primitives of the QCCD architecture by splitting and joining the chain, transporting ions between different trapping zones, and performing sympathetic cooling following transport before implementing two qubit gates between Be$^+$ qubits \cite{HomeCompleteMethods2009}.

Two-qubit gates between different ion species were first demonstrated in the two-ion $^{43}$Ca$^+$-$^{40}$Ca$^+$ \cite{BallanceHybridLogic2015} and $^{25}$Mg$^+$-$^9$Be$^+$ \cite{TanMultiElement2015} chains.  In Ref.~\cite{BallanceHybridLogic2015}, a maximally-entangled Bell-state was generated with 99.8(6) \% fidelity, which was used to show a violation of Bell's inequality.  In Ref.~\cite{TanMultiElement2015}, an MS gate was implemented to generate a Bell state with 97.9(1) \% fidelity, which was also then used to show a violation of Bell's inequality.  In this work, the CNOT gate, consisting of the MS gate and additional single-qubit gates, and the SWAP gate were demonstrated.  In particular, it was shown that the dual species gates could be used to better swap the qubit populations between ions, as compared to the technique of quantum logic spectroscopy, in the case where the ion chain has significant motional-state excitation.  In a $^{9}$Be$^+$-$^{40}$Ca$^+$-$^{9}$Be$^+$ chain, a primitive of error correction was demonstrated using the Ca$^+$ ion as an ancilla, which utilized multiple MS gates between Be$^+$ and Ca$^+$ \cite{NegnevitskyMultiReadout2018}.  Three-qubit, dual-species Greenberger-Horne-Zeilinger (GHZ) states were also generated in this work with a fidelity of 93.8(5) percent.  A two-qubit MS gate in a  $^{171}$Yb$^+$-$^{138}$Ba$^+$ chain was realized with a fidelity of 60\%, limited by excessive motional heating \cite{InlekMultiNode2017}.  However, in this experiment, the first dual-species architectural primitive of a photonic interconnect was demonstrated, whereby entanglement between the Ba$^+$ communication qubit and a photon polarization qubit was generated with a fidelity greater than 86\%.

In general, experiments with dual-species ion chains are more challenging than those using a single species due to the increased number of required lasers and complexity of optical, as well as motional, control.  As a result, they have only been performed in a few groups and the best demonstrated gate fidelities are not yet at the level of what has been achieved using only one species.  Given the significant role dual-species chains are likely to play in QC, the number of groups working with them is growing, which will likely lead to improvements in demonstrated performance.  In addition, work is now being done to implement quantum operations on molecular ions in the presence of atomic ion ancillas \cite{ChouMolecularIon2017, HudsonMolecularQC2018}, which may lead to new techniques for trapped-ion QC.

\section{Integrated Technology for Control of Trapped Ions}
\label{ScalableHardware}
In order to build practical trapped-ion-based quantum computers, development of hardware technology for controlling and measuring large numbers of ions with low error will likely be required.  Through such development, there is great potential for improved control and measurement of even small systems, but it will be imperative to explore any tradeoffs that exist between scalability and performance in systems of any size.  In the following sections we will discuss progress in the field toward realizing hardware with the potential for improved scalability in trapped-ion QC.

    \subsection{Chip-Scale Ion Traps}
\begin{figure}
    \centering
    \includegraphics[width=0.8\columnwidth]{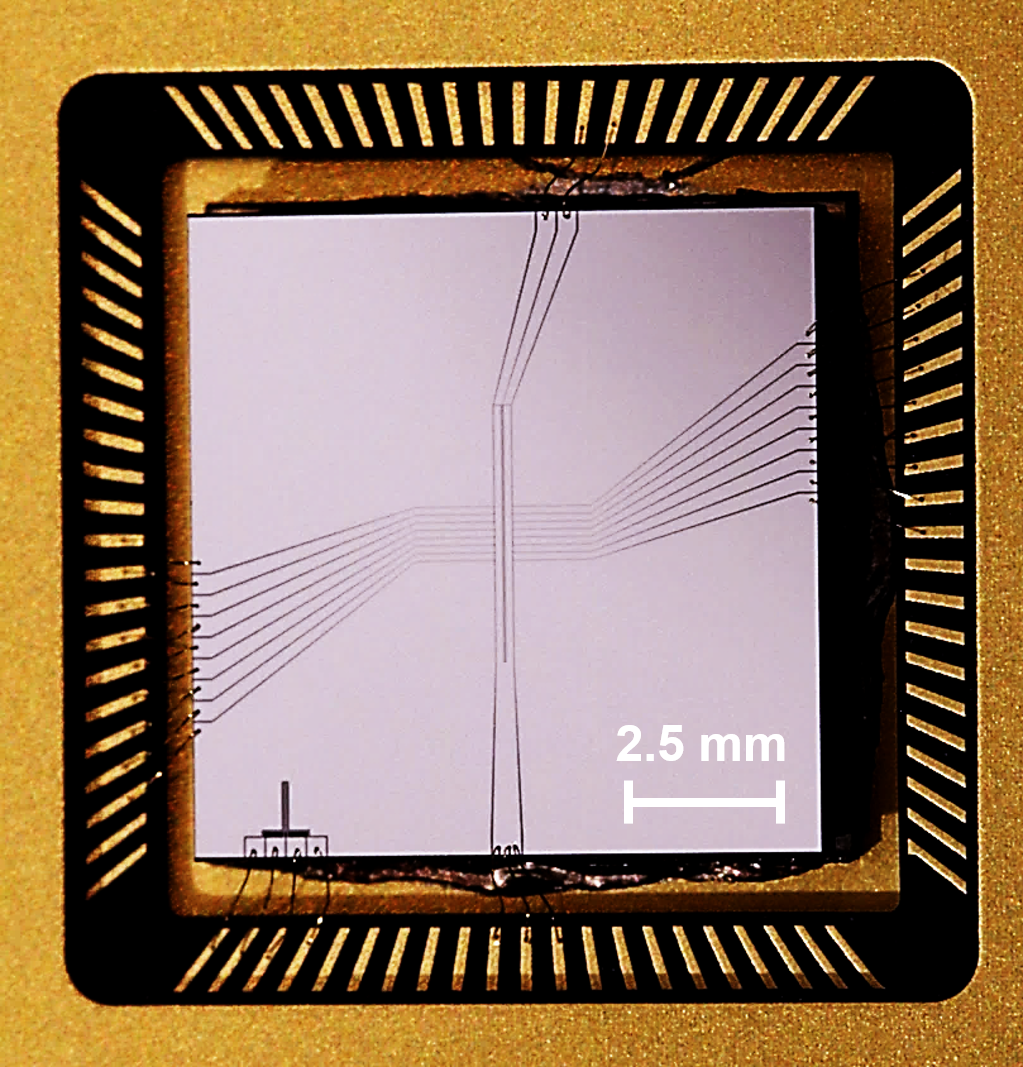}
    \caption{Photograph of a surface-electrode ion trap chip.  The 1-cm-square trap (gray) is mounted in a ceramic pin grid array (gold and black) to which it is connected electrically via wirebonds.  The trap, designed and fabricated at Lincoln Laboratory, consists of a sapphire substrate upon which a 1-$\mu$m-thick layer of aluminum metal is deposited, and patterned via optical lithography, to define the trap electrodes.  This particular trap is designed to confine ions in a linear array ${\sim}50$~$\mu$m above the surface of the chip.}
    \label{fig:chiptrap}
\end{figure}

The surface electrode trap technology will likely be required as the need arises for an increased number of ion trap electrodes and an increased electrode-configuration complexity (see Secs.~\ref{MicroTraps} and \ref{2DArrays}).  A typical surface-electrode ion trap consists of a substrate, or chip, composed of material such as sapphire, quartz, or silicon, with metal electrodes patterned on its surface (above which the ion is trapped).  These electrodes are typically formed via a few-micron-thick metal deposition, followed by optical lithography and chemical etch to define the electrode pattern; electroplating techniques are often used, as well.  Such fabrication methods offer great flexibility in design, allowing for arbitrary electrode shapes and patterns, as well as for multiple metal layers (separated by insulating layers) that are useful for wiring and routing of electrical signals to the electrodes.  In addition, these methods deliver extremely tight dimensional tolerances, providing sub-micron-feature resolution.  Since the positions of the ions are set by these dimensions, the ion location above a chip-scale trap is thus very precisely determined.  As a general rule of thumb, the scale of the ion-to-electrode-surface distance is set by the lateral dimensions of the electrodes, and so shrinking the lateral scale necessitates moving the ion closer to the trap surface.  Because of current issues associated with anomalous motional heating (see Sec.~\ref{AnomHeating}), it has not thus far been desirable to make traps with ion-to-surface distances closer than a few tens of microns.  As a result, the lateral scale of a typical zone in a surface electrode trap, which is comprised of $\sim$ 5-10 electrodes, is a few hundred microns.  However, as the trap fabrication resolution is far from this limit, the technology is amenable to significant reduction in size if ion heating can be reduced or tolerated.  A photograph of a typical surface electrode trap chip is shown in Fig.~\ref{fig:chiptrap}.

Wafer-scale fabrication can now be performed on wafers as large as 12" in diameter so that, in principle, very large trap arrays could be realized on one substrate.  While this offers a potentially powerful way to scale up, from a modularity point of view, this may or may not be the most practical way forward.  Rather, another advantage of surface electrode traps is that they can be made modular by physically tiling together many smaller chips.  As discussed in Ref.~\cite{LekitscheMicrowaveBlueprint2017}, chip-to-chip electrical connections are not required since ions are sensitive only to the electric fields produced above the electrodes, and ions can therefore be transported over gaps between the chips if these chips are aligned precisely.  Modularity can also be realized using chip-scale traps in architectures that utilize photonic interconnects.  Indeed, a hybrid architecture can be imagined in which modestly-sized chips are used to implement QCCD-based operations and chip-to-chip coupling is enabled via remote entanglement \cite{MonroeKimScaling2013}

Perhaps the most transformative advantage of chip-scale traps is that they provide a format for integration of potentially scalable ion control and measurement technology, as well as integration of classical signal processing and computing technology.  The entire half-space beneath the trap electrodes can be utilized to this end, and in the next few sections, we will discuss progress and the potential on this front.

    \subsection{Integrated Photonics for Light Delivery}
    \label{IntPhot}
In order to control and measure trapped ions, a number of lasers with different wavelengths (typically around five) are required.  This number gets multiplied by roughly two when working with dual-ion-species systems.  In addition, these lasers need to be sent to each location where an ion resides, or where a quantum operation is to be performed.  As a result, the number of laser beams that must be delivered to precise locations in an ion trap array grows as the array size grows.  Current methods to address individual ions with lasers typically employ free-space optics such as mirrors, acousto-optic modulators (AOMs), and lenses, located outside the ion trap vacuum system, to steer and tightly focus the light through vacuum chamber windows and to the desired positions.  AOMs are also used as  high-extinction and high-speed optical switches, and as precise tuners of the optical laser frequency and phase.  Steering and switching laser beams with these optics to address a small number of ions held in a linear array, with the beams traveling orthogonally to the linear trap axis, can be and, indeed have been, used effectively.  However, it is hard to conceive of a way to use these techniques to address ions trapped in a 2D array with low crosstalk, especially when using surface-electrode traps.  This is because laser beams delivered by free-space optics are typically propagated parallel to and across the chip surface, and therefore cannot be generally used to address an ion at the center of the array without hitting ions at the edge. (Such parallel propagation is implemented to avoid striking the chip, which would create light scatter and which could generate photoelectrons that can result in surface charging and reduced trap performance \cite{wang2011laser}.)

The tight focusing of laser beams serves two important functions: first, it reduces crosstalk (in principle) because the laser field is more confined.  Second, it creates a higher-intensity laser-beam spot; since higher intensities lead to faster quantum operations in ions, this results in higher-speed performance for a fixed laser power, or lower required laser power for a fixed operation speed.  However, focusing of beams comes at a cost.  The tighter the beam focus, the more susceptible the system becomes to beam-pointing error (i.e., beam-position jitter or  misalignment).  In addition, a tighter beam focus at one position leads to an increased beam divergence away from the focal point because of the fundamental nature of diffraction.  Such beam divergence is likely to increase crosstalk (a problem that beam-focusing was supposed to help solve), especially in 2D ion array architectures.  Furthermore, for chip-based ion traps, the divergence limits the lateral size of the chip in the direction of beam propagation, since the beams must be prevented from striking the chip surface \cite{BrownCodesign2016, GuiseBallGrid2015}.

\begin{figure*}[t b h]
\includegraphics[width=2.0\columnwidth]{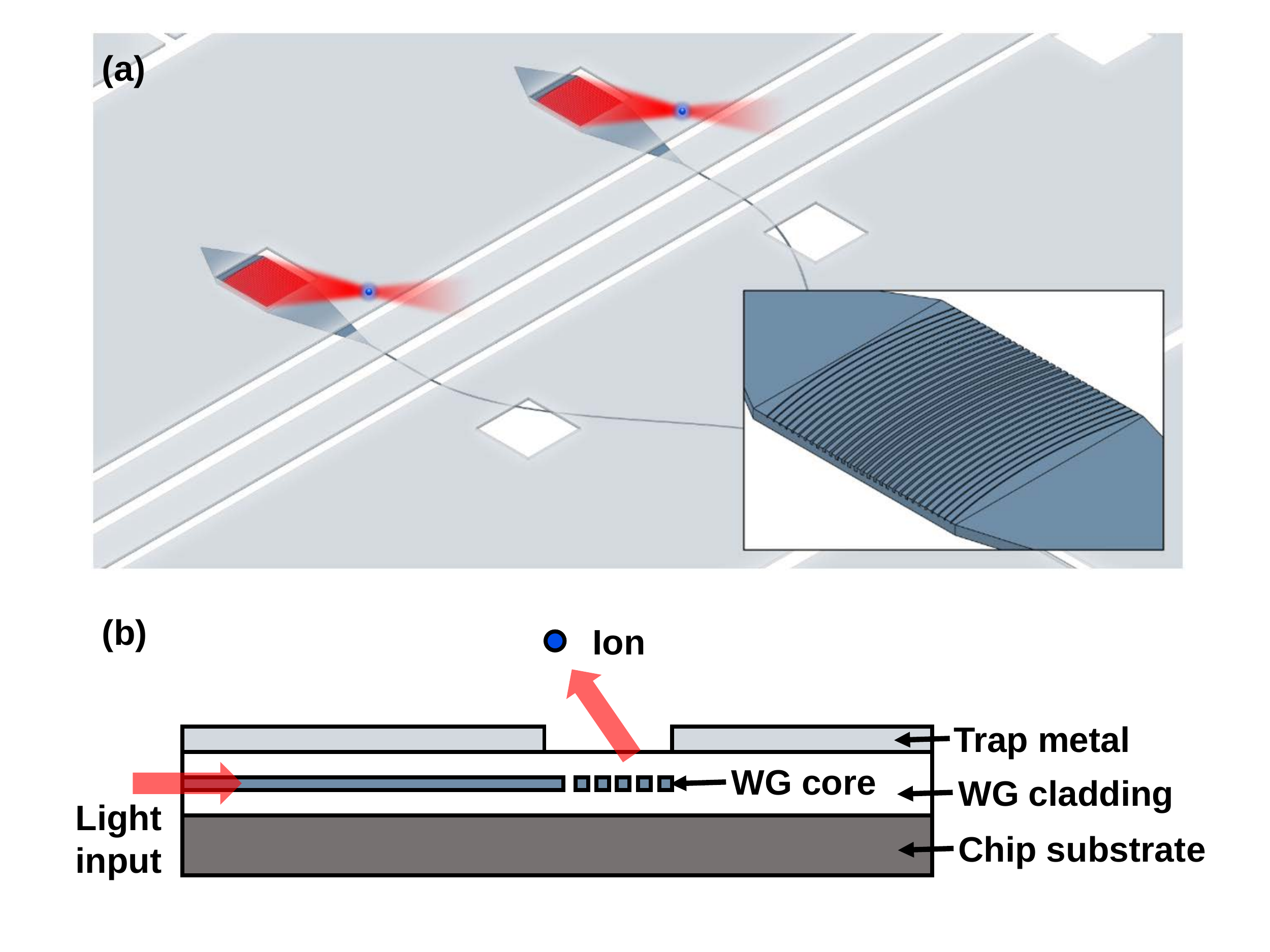}
\caption{Integrated photonics for light delivery.  (a) Cartoon of integrated photonic waveguides and grating couplers delivering light to two locations above a surface electrode ion trap.  The waveguides and gratings consist of a patterned high-optical-index core (blue-grey) surrounded by a lower-index cladding (white).  Square windows are opened in the trap metal (light grey) to allow the light (red) emitted from the grating coupler to reach the ion (blue circle).  The inset shows a zoomed in view of a grating coupler. (b) Schematic cross section of an ion trap with integrated photonics.  Layers, including the chip substrate (dark grey) are shown, with thicknesses not to scale.}
\label{fig:intphot}
\end{figure*}

The use of integrated photonics has been suggested as a means to address the challenges of precisely delivering a large number of tightly-focused laser beams to a 2D ion array with low crosstalk \cite{KielpinskiIntPhotonArch2016, MehtaIntegrated2016,MehtaThesis2017}, and many decades of work in the area of silicon integrated photonics~\cite{ThomsonSiPhotonicsReview2016} can potentially be leveraged to this end.  Different integrated photonic components can be considered, but a crucial one is the optical waveguide, which is essentially an optical fiber fabricated into the chip substrate.  As shown in Fig.~\ref{fig:intphot}, waveguides consist of a core material, patterned on a chip, that is surrounded by other, cladding materials that have a lower optical index than the core.  As in fiber-optics, the optical mode will be confined to the higher index core, such that light can be routed on chip following the patterned core path.  Such patterning is typically done with the same micro/nano-fabrication techniques used to fabricate surface electrode traps.   The flexibility this offers is profound.  The width and thickness of the waveguides can be chosen to ensure single spatial mode propagation and polarization maintaining capabilities, and by varying the waveguide cross-sectional geometry, the optical modes can be shaped. And not only can light be routed and shaped in complex fashion around the chip, but waveguides also allow for splitting of the light into multiple paths (i.e., on-chip beam splitters) to enable fan-out.

As discussed in Sec.~\ref{MicroTraps}, surface-electrode traps are well-suited to host integrated photonic components.  The necessary materials can be deposited and patterned on the trap substrate, and then the trap electrode metal can be deposited and patterned on top of the photonics layers.  The light can thus be routed below the ions and be brought to a location in the center of the array without passing through ions on the edges.

Since the ions are trapped above the surface of the chip, the light from the waveguides must be directed vertically into free space.  This could be done, in principle, with integrated turning mirrors located at the ends of the waveguides \cite{TangTurningmirror2002}.  However, light exiting waveguides into free space diverges rapidly, and so the mirrors would need to be curved to refocus the light, which requires complex fabrication techniques that are not common to typical wafer-scale processes.  An alternative approach, that may be easier from a fabrication perspective, is the use of diffractive vertical grating couplers (see Fig.~\ref{fig:intphot}) \cite{MehtaIntegrated2016,MehtaThesis2017}.

Grating couplers are made by periodic variation of the optical index of the waveguide material~\cite{HeitmannGratingCoupler1981}.  This is typically accomplished by patterning the variation in the core material at the end of the waveguide along the propagation direction.  Light incident on the grating coupler will be diffracted out of the chip plane at an angle set by the grating period, the particular optical indices of core and cladding material, and the wavelength of the light. Holes in the metal electrodes are patterned to allow light to pass through so that it can reach the ion location. By appropriately curving the diffraction grating teeth and by varying the grating period along the length of the coupler, focusing can be achieved in the two directions transverse to the light propagation.  The size of the focused spot is, as expected, limited by diffraction; however, since the grating couplers can be placed very close to the ion (e.g., tens of microns away, limited by the ion-to-trap-surface distance) grating couplers need only be tens of microns in lateral scale to achieve few-micron spot sizes, for visible light, at the ion locations. The very tight spot sizes allow light intensities at the ion location to be comparable to those achievable via free-space beams, despite coupling and waveguide losses. Importantly, as shown in Fig.~\ref{fig:intphot}, the grating coupler, like the waveguides, is essentially a planar device, making fabrication straightforward using standard lithographic techniques.

Light can be delivered to the chip from the side via coupling into waveguides that run to the chip's edge, or via input grating couplers anywhere on the chip (i.e., using the grating couplers in reverse to how their operation is described above). This input coupling can be done from free space, but can also be achieved with optical fiber butt-coupled directly to the chip since the integrated optics can be designed to match the optical mode of the fiber.  With waveguides and grating couplers, and fiber-optic inputs, it is thus possible, in principle, to deliver high-intensity light to ions in a 2D surface-electrode trap array with low crosstalk and with no beams travelling in free space until they exit the grating couplers only a few tens of microns away from the ions.  This highlights another potential advantage of integrated photonics. The highly-stable beam paths afforded by this technology, and their lithographic registration to the ion trap electrodes (and thus the ions themselves), suggests that they may provide improved control of the laser beam position and phase with respect to the ions, as compared with that provided by free-space optics.

Integrated photonics for delivery of light to trapped ions is just beginning to be explored experimentally.  Early, motivating work, demonstrating coupling of light to ions via a fiber attached to an ion trap \cite{KimIntFiberTrap2011} showed the potential benefits of eliminating free-space optics, but did not chart a clear path to scalability.   Recently however, silicon nitride (SiN) waveguides and grating couplers, clad by quartz and silicon dioxide (SiO$_2$), were integrated into a surface-electrode trap and were used to deliver 674-nm light to Sr$^+$ ions~\cite{MehtaIntegrated2016}.  In this work, on-chip routing and splitting of light was demonstrated, as was high-fidelity, low-crosstalk trapped-ion quantum control.  However, many challenges remain. Chief among them are delivering multiple wavelengths of light to the ions and to demonstrate 2D array control.  Multi-wavelength delivery is particularly challenging because individual integrated photonics components typically work over a narrow optical bandwidth. Waveguides optimized to guide a single spatial mode at one wavelength are not optimized for others (and can even be multi-mode) and the angle of emission out of a grating coupler is very wavelength-dependent.  As a result, it is likely that different and distinct waveguide and grating coupler devices will be required for each wavelength of light needed.

Another challenge to widespread use of integrated photonics is optical loss.  The waveguide materials must be highly transmissive over the wide range of wavelengths needed for full trapped-ion control and readout.  This is not so easily achieved, especially for the near-UV wavelengths that many commonly-used ions require.  The reason for this is two-fold.  First, there are few materials that are highly transparent in the UV. (Indeed, silicon, the most mature material system for integrated photonics, is completely opaque over the entire UV-to-visible spectrum.)  Second, for a given waveguide roughness that results from non-idealities of fabrication, Rayleigh scattering induces loss that scales poorly with decreasing wavelength $\lambda$ (as 1/$\lambda^4$). As a result, low-loss integrated photonic devices must be developed using materials that can be fabricated with low roughness.  This work is ongoing, as devices made of materials transparent in the UV-to-visible wavelength range, such as SiN \cite{SoraceAgaskarSPIE2018}, gallium nitride (GaN) \cite{XiongGaN2011}, aluminum nitride (AlN) \cite{XiongAlN2012, LuAlN2018, LiuUVAlN2018}, lithium niobate (LiNbO$_3$) \cite{ZhangLiNbO32017}, and alumina (Al$_2$O$_3$) \cite{WestAl2O32018}, are being explored.

It should be noted that even for highly-transmisive, low-roughness materials, integrated waveguides will always introduce more loss as compared with free space optics; such loss includes input and output coupling efficiency to and from waveguides.  However, grating coupler efficiency can be made high \cite{NotarosHighEffGratings2016, MehtaGrating2017}, and optical loss can be made low \cite{WestAl2O32018}, such that the tight focusing afforded by this technology may win back in intensity what is lost in power.  That said, the lost optical power will be going somewhere.  It will generally not all be absorbed, but rather, some will be scattered, and this scattered light could lead to undesirable crosstalk.  The trap metal placed over the photonics should help to block much of this light from reaching ion locations, but it remains an open challenge to show that scattered light will not be a problem for scalability.

In addition to light delivery, the beams generally need to be switched on and off, and integrated optical modulators may provide this functionality, as suggested in Ref.~\cite{KielpinskiIntPhotonArch2016}.  Most demonstrations of integrated modulators have focused on performance at IR wavelengths; however, a number of the modulator material systems that have been explored are transmissive in the UV-to-visible wavelength range, for instance, SiN \cite{YegnanarayananSiNModulators2018, MohantyBlueSiNModulators2018}, InGaN/GaN \cite{FengGaNmodulator2018}, AlN \cite{XiongAlN2012, ZhuAlNMod2016}, and LiNbO$_3$ \cite{MehtaLiNbO3VisMod2017, WangLoncarModulators2018, DesiatovVisLiNbO3Mod2019}.  Whatever the most promising materials may be, we anticipate that this switch technology will be under widespread and rapid development over the next few years given the crucial role switching plays in trapped-ion control.  It is, however, important to note that integrated switches may not be required.  Since trapped ions can be moved, high-extinction switching may be achieved by simply displacing the ions just a few microns outside of the laser beam focus.

\subsection{Integrated Optics and Detectors for Light Collection and Measurement}
As the size of ion arrays grows, it will become increasingly necessary to develop methods and technology to collect and detect photons emitted by a large number of individual ions.  This must be done with high efficiency since the ability to accurately measure the state of an ion is largely determined by the number of photons that can be collected from the ion and detected during the measurement time.  In addition, as discussed in Sec.~\ref{PhotonInter}, the rate of remote entanglement generation for photonic interconnects increases with increased photon collection and detection rates.

Optics and single-photon detectors integrated into surface electrode ion traps offer a potentially powerful means to achieve this.  In principle, one could imagine having one detector positioned below each ion to be measured \cite{LekitscheMicrowaveBlueprint2017, MehtaThesis2017}. Since integrated detectors would be located very close to the ions, they could have a compact form factor, and yet their active area could still collect photons from a large solid angle.  Incorporation of integrated optics placed between the ion and the detector could further improve the collection efficiency, as well as provide spatial filtering to help prevent stray light, or light from neighboring ions, from reaching the detector.  Integrated collection optics could also be used to couple photons emitted by ions into single-mode integrated photonic waveguides. These waveguides could route photons as desired.  For instance, on-chip remote entanglement could be generated by interfering photons emitted by remotely-located ions using a waveguide beamsplitter and by detecting the interference with integrated detectors.  Alternatively, photons from waveguides could be coupled off-chip to fiber for longer-distance remote entanglement. Integrated photon collection, routing, and detection could also potentially enable performing RE with significantly larger numbers of ions.  Not only would this offer a path towards scalability, but it could also be used to increase the rate of RE via spatial multiplexing.  In addition, an integrated approach to RE may offer improved fidelity due to superior optical mode matching that can likely be achieved by integrated photonics and fiber optics as compared with free-space optics.

Initial work aimed at integrating collection optics into ion traps involved incorporating a multimode fiber into a surface-electrode ion trap \cite{VanDevenderFiberTrap2010}.  The fiber, placed below the trap, collected 280-nm photons emitted by a Mg$^+$ ion through a 50-$\mu$m hole in the trap and delivered them to an off-chip PMT for detection; the collection NA was 0.37.  In a separate experiment, a five-element lens array was integrated below a slotted region of a surface-electrode trap \cite{ClarkIntegrated5Collection2014}.  The array had each of its elements coupled to fiber and collected 397-nm light from Ca$^+$ ions with 0.37 NA.  Metallic, spherical micro-mirrors have been monolithically integrated into a surface electrode trap~\cite{merrill2011demonstration} and were were used to collect 397-nm photons from Ca$^+$ ions with a NA of 0.63.  Metallic diffractive mirrors were also monolithically integrated into similar traps and used to collect 370-nm photons from Yb$^+$ ions with a 0.68 effective NA \cite{Ghadimi2017}.

For photon detectors, a first proof-of-principle experiment demonstrated the integration of a commercially-obtained photodiode beneath a surface electrode trap having transparent electrodes made of indium-tin-oxide (ITO), and fluorescence from a trapped Sr$^+$ ion was detected~\cite{EltonyITOTrap2013}. Cryogenic SNSPDs have been monolithically integrated into surface electrode ion traps where high detection efficiency and low dark counts were demonstrated, as was compatibility with the RF fields required for ion trapping~\cite{SlichterSNSPD2017}.  Avalanche photodiodes (APDs) have also been monolithically integrated into a surface electrode ion trap fabricated in a CMOS foundry~\cite{MehtaThesis2017}, and stable trapping of Sr$^+$ ions has been demonstrated in this trap.  However, a demonstration of using integrated SNSPDs or APDs to measure fluorescence from ions has yet to be reported.  These two detector technologies are likely to be complementary.  While both have been demonstrated with performance compatible with high-fidelity ion-state measurement, SNSPDs generally have higher detection efficiency and lower dark counts than APDs; however, they require special processing for fabrication and must be operated at cryogenic temperatures.  APDs, on the other hand, can be made in robust CMOS foundry processes~\cite{FieldAPD2010} and can be operated at room temperature, so they may be a more practical choice for ion QC applications.

Whatever the technology, future demonstrations will likely be focused on  integration of detectors and collection optics to achieve high-speed, high-fidelity measurement in a way that is not susceptible to stray or scattered light, and that delivers an advantage over conventional collection and detection techniques.

    \subsection{Integrated Electronics}
As discussed previously, we anticipate that the need for ion-motion control is likely to increase in the near future, as more research is focused on demonstrating modularity in ion chains.  Such control is enabled by varying voltages on the trap electrodes, and these voltages are typically generated by a significant number (approximately one per electrode) of digital-to-analog converters (DACs).  These DACs are typically housed in electronics racks remotely-located from the ion trap, with signals delivered to the electrodes by a large array of wires that must pass through vacuum feedthroughs.  While this approach works well for few-electrode traps, it is likely to become increasingly unmanageable as the trap complexity grows.  Furthermore, the long signal paths necessitated by the remote location of the DACs are susceptible to noise; this is typically mitigated by incorporating electronic filters, but such filtering reduces the voltage switching speed and is thus not ideal if ion transport time is at a premium.

It has been suggested that integration of these electronics may potentially provide a great benefit \cite{GuiseInVacElec2014, mehta_cmos_2014, LekitscheMicrowaveBlueprint2017, StuartDACTRap2018}.  Work in this direction has demonstrated 80 commercially-available DAC channels connected, in vacuum, to a surface-electrode ion trap via traces on a printed circuit board \cite{GuiseInVacElec2014}, thus eliminating the need for cumbersome and noise-susceptible wiring.  Trapping and transport of Ca$^+$ ions were demonstrated in this system.  In addition, chip-scale traps with monolithically-integrated trench capacitors for on-chip electrical filtering \cite{Allcock2012, GuiseBallGrid2015} and through-substrate vias for connection of electrical signals to the back-side of the trap chip \cite{GuiseBallGrid2015} have been demonstrated.

The potential for monolithic integration of electronics was first shown explicitly in Ref.~\cite{mehta_cmos_2014}, where an ion trap was fabricated in a commercial CMOS foundry and stable loading and trapping of Sr$^+$ ions was demonstrated.  This opened the door to taking advantage of the enormous capabilities of CMOS electronics for ion traps.  Building on this result, recent work demonstrated monolithic integration of 16 DAC channels into a surface electrode ion trap fabricated in a 180-nm CMOS foundry process, where Ca$^+$ ions were trapped and transported robustly \cite{StuartDACTRap2018}.  Additionally, the DAC noise was characterized and it was shown that it could be dynamically filtered using active on-chip electronic switches.

While integrated electronics show great promise, there still remains much work to do to show that they do not introduce significant deleterious effects.  For instance, it must be shown that on-chip power dissipation can be managed, and that the currents flowing in the circuits do not generate fluctuating magnetic fields that cause decoherence of ion qubits. That said, the demonstration of functional CMOS DACs integrated into ion traps potentially paves the way for integrated electronic devices beyond DACs.  For instance, circuits could be incorporated to shape and count pulses from integrated photon detectors~\cite{MehtaThesis2017}, and even on-chip analog and digital processing is possible.  Such processing could, for example, reduce latency in error correction feedback. In addition to the active devices afforded by CMOS electronics, it is possible to use the many available wiring layers to route voltage signals around the trap \cite{mehta_cmos_2014, StuartDACTRap2018}, or even to provide complex arrays of current lines to generate magnetic fields for quantum gates and local definition or shimming of the ion-spin quantization axis.

\section{Outlook}
\label{Outlook}
Quantum computing with trapped ions has progressed significantly over the last couple of decades, yielding many exciting results.  Despite significant outstanding challenges, we believe these results demonstrate that there is great potential for building a practically useful quantum computer consisting of ion qubits.  However, there is still much science and engineering to be done in order to determine how to realize this potential.  It is of course difficult to predict where the next few decades will lead, but in these final sections, we speculate, based on the current status of the field, as to what types of trapped-ion systems and control techniques might be used, and what experiments might be done, to help us best make this determination.

\subsection{The NISQ Regime}

As has been seen in the previous sections, the requirements for a fully fault-tolerant trapped-ion quantum computer capable of executing complex algorithms are daunting. It is likely to be some years before a truly large-scale quantum processor becomes available. In the meantime, it is natural to ask, are there interesting things to do with the sorts of quantum computers that are likely to be available in the very near future?

Preskill coined the term Noisy Intermediate-Scale Quantum (NISQ) to describe the sorts of quantum processors that seem imediately realizable \cite{PreskillNISQ2018}. The defining characteristics of NISQ-era processors include qubit numbers on the order of 100 and---given the huge overhead that would otherwise be required---the absence of full schemes for quantum error correction. The hundred-qubit scale sets a limit on what sort of problems these devices may address, while the lack of error correction points towards a need to reduce or mitigate errors or to find ways to limit their effect on the computation without introducing prohibitive overhead.

One promising application during the NISQ era is that of quantum emulation. In particular, hundred-qubit quantum computers may be able to perform bespoke quantum simulations of other systems---for example, to analyze the behavior of solid state systems or determine chemical structures---and may be able to surpass the performance of classical computers in doing these calculations. Fifty qubits has been identified as a threshold \cite{LundBosonSampling2017} beyond which existing classical computers may be unable to accurately simulate the behavior of a quantum system. In this respect, the recently-demonstrated control over chains of more than 50 ions \cite{Zhang53IonSim2017} suggests that trapped-ion systems may be able to solve some problems that are intractable for classical computers within the next few years.

One further intriguing application of trapped-ion systems is that of quantum sensing. Trapped ions hold promise as sensors of time (i.e. optical clocks), electric fields, and magnetic fields (see~\cite{DegenQuantumSensingRMP2017} and references therein). The techniques of QC---namely, the generation of highly-entangled states of multiple ions---may enable trapped-ion based quantum sensors with very high performance or unique capabilities. Ions encoded in decoherence-free subspaces may be able to use their impressive coherence times to enhance sensitivity \cite{SchmidtKalerEntangledSensor2012}, while quantum error-correcting codes may enable improved sensor precision or reduce susceptibility to environmental noise \cite{OzeriHeisenberg2013, DurImprovedmetrology2014}. A full exploration of this topic is beyond the scope of this review, but a recent review paper on quantum sensing \cite{DegenQuantumSensingRMP2017} discusses trapped ions as quantum sensors and the possible uses of entanglement-enhanced states for sensing. Of particular note is a recent proposal to use dissipatively-engineered error-correcting codes to mitigate noise-induced decoherence in a trapped-ion magnetic-field sensor \cite{ReiterDissipativeSensing2017}, although this idea has not yet been demonstrated. The use of highly-entangled multi-ion states for improved sensing is another area that may yield fruit in the near term.

\subsection{Further Considerations}

Now that we have reviewed potential methods for scaling ion systems in size, including the technologies that are being developed to enable such scaling, we can consider how the ion- and qubit-specific choices may impact these systems.  Though qubits have been encoded in several different combinations of states in many different ion species (and pairs of species) to demonstrate various aspects of the feasibility of ion-based QC, particular qubit states, atomic species, and gate methods will lead to particular implications for the performance of larger-scale processors.  Here we discuss these considerations and potential ramifications for future systems.

  \subsubsection{Choice of Ion Species}

Experimenters developing QC systems with trapped ions have several species to choose from (see Table~\ref{tab:ion_props} for properties of several of the ion species used most often in QC experiments).  Basic trapping and control have been demonstrated with almost all of the alkaline-earth and alkaline-earth-like ions, and those working in the fields of atomic clocks and frequency standards have considered and worked with these and a few more (e.g. Al$^{+}$, In$^{+}$, Lu$^{+}$). Yet, it is worth considering the potential benefits and drawbacks of using particular ion species as systems are scaled up in size and capability.  As laid out in Secs.~\ref{sec_trapped_ion_qubits} and~\ref{sec_ion_control}, there are many ion-qubit and quantum-logic-gate types, and as described in Sec.~\ref{ScalableHardware}, there are many technologies that need to be developed in order to build more scalable systems.  Different ion species present different tradeoffs among these options, due to their differences in mass, energy spectra, coupling-strength of states in the spectra to electromagnetic radiation, nuclear spin, and particular isotopic abundances.

Ion mass is of importance for several reasons, even if not considering the effect of the mass on the ion's electronic structure.  First, the RF pseudopotential in a Paul trap is mass dependent, and hence larger masses require larger voltages to achieve similar secular trap frequencies.  Hence there is a direct impact on achievable speed of operations if the potential that can be applied is limited by dielectric or vacuum breakdown, or by the in-chip integration of electronic technologies which are ultimately also limited by insulator breakdown and current carrying capacity in small devices.  Power dissipation, which scales linearly with the total trap capacitance (relevant as systems grow in size) and quadratically with the RF voltage amplitude, will also be higher for higher mass ions for the same potential.  Second, the force that can be applied to an ion via an optical-dipole force, as required for motion-based multi-qubit logic, generally goes down as the ion mass goes up, via the sideband Rabi frequency's dependence on the Lamb-Dicke parameter.  Hence two-qubit gates with heavier ions will be slower, assuming comparable optical power.  Third, as mentioned in Sec.~\ref{DualSpecies}, the choice of ion species for sympathetic cooling is heavily influenced by the ions' masses, since the efficiency of multi-mode energy-transfer (and therefore efficiency of cooling in general) is maximized for equal mass.  Since emitted photon re-scattering can reduce qubit coherence in ions with like spectra, equal masses are not viable (even with individual optical addressing), so two species of similar mass are preferred.  Finally, ions of light mass suffer more than heavier ones from photon scattering error during gates driven via optical fields~\cite{PhysRevA.75.042329_2007}.  This is because Rayleigh scattering (from the ions during the gate), which does not affect the electronic-state coherence, nonetheless imparts momentum to the ions, and therefore their coupled motion will deviate from the desired phase-space path, leading to phase error.

Another important factor to consider for scalable systems is the wavelength $\lambda$ of the light required for ion-qubit control.  While quantum logic may be performed with RF or microwave fields as described previously, some light will always be required for state preparation and readout, and likely also for ionization of the neutral precursor, as methods such as electron-impact ionization and direct ion production using laser ablation lead to unwanted side effects like prohibitive stray-field production and a lack of isotope selectivity.  Therefore, laser wavelengths for these operations should be in technologically attainable ranges, with sufficient power available, and if integrated technologies for light delivery and detection are pursued, the wavelengths must be compatible with the materials and structures employed.  In general, more optical power is available with narrower linewidths, and is more easily spectrally stabilized, for wavelengths in the red to IR part of the spectrum when compared to the blue to UV wavelength range (there are exceptions to this, notably harmonics of YAG lasers for which there is an industrial base, allowing significant power for pulsed Raman excitation where frequency precision is relaxed).  Moreover, integrated photonic delivery technologically favors the longer wavelength range due to the challenges of precisely producing sub-wavelength-scale structures and of finding and fabricating materials with low optical loss at short wavelengths.  Practical considerations of working with shorter wavelength light include the challenge of finding materials for efficient polarizing optics, lenses, and mirrors, as well as fibers for low-loss transmission.  Tolerances for general mode matching, including input and output coupling to waveguides and fibers and combining light from separate modes at a beamsplitter are also much more difficult to reach robustly as the wavelength decreases.  On the other hand, shorter wavelength light does present a few advantages for ion-based quantum logic.  The Lamb-Dicke parameter controlling motional coupling to light is inversely proportional to wavelength, meaning gates driven with shorter-wavelength light can be faster, all things being equal.  Additionally, detection of blue and UV light is in general more effective than for longer wavelength light due to to the higher energy of each photon, particularly in the case of photo-multiplier tubes and superconducting nanowire detectors.  While this is not strictly true in the case of detection based on semiconducting technologies, where detector design must be optimized for particular wavelength ranges, high detection efficiency is available in the blue and near-UV.


Beyond simple wavelength determination, the electronic structure of the ion has bearing on the required level of manipulation and the precision that can be attained.  For instance, the presence of metastable $D$ (or $F$) levels enables the use of optical qubits for quantum logic and can aid in low-error detection of non-optical qubits.  However, it also provides a scattering channel during Raman excitation~\cite{PhysRevA.75.042329_2007} and generally leads to the requirement of additional laser wavelengths for repumping or quenching of these levels when necessary.  Likewise, hyperfine structure can enable the use of long-coherence-time FOFI qubits and Raman-based logic, but the proliferation of levels in high-nuclear-spin isotopes leads to a challenge in efficient quantum control, coupled with more potential decay paths that can limit state preparation and detection fidelity~\cite{PhysRevA.79.020304}.  The extra levels present in such systems also provide additional paths for leakage errors, which can complicate quantum error correction, as mentioned in Sec.~\ref{ProsandCons}.

The nuclear spin of available isotopes is also a consideration; some species have no stable zero-nuclear-spin isotopes (most notably Be$^{+}$), and some species' stable non-zero-nuclear-spin isotopes have very large nuclear spin (e.g. $^{43}$Ca$^{+}$ and $^{87}$Sr$^{+}$), complicating quantum control and manipulation.  Isotopic abundances can also make efficient loading of the required species challenging ($^{43}$Ca$^{+}$ also falls into this category); enriched sources can be employed, though impurity loading will eventually still limit array uniformity.  Alternatively (or in conjunction), the use of remote pre-cooling of the neutral precursor, with multiple stages of isotope selectivity~\cite{sage2012loading,BruzewiczArrayLoading2016}, can improve array purity.  Very recently, there has been interest in long-lived radioactive isotopes, where the difficulties of dealing with an unstable atom are hopefully outweighed by the beneficial level structure.  In particular, $^{133}$Ba$^{+}$, with half-life of 10.5~y, is being investigated as a potential hyperfine qubit candidate, due to the favorable wavelengths of Ba$^{+}$ combined with the small nuclear spin of 1/2~\cite{PhysRevLett.119.100501}.  The results are promising, in particular the very high potential state-preparation-and-readout fidelity afforded by the nuclear spin and the presence of a long-lived $D$ state addressable using IR light. The combination of the nuclear spin and the long-lived $D$ state also allows for an optical qubit with $m=0$ clock states, which should make this qubit highly insensitive to magnetic fields.  At the same time, loading and manipulating large arrays of radioactive ions may present further challenges when compared with stable species.

\begin{table*}[t b h !]
\caption{Properties of ions of interest for QC.  First-order field-independent (FOFI) transitions that have been used are indicated in the last column with the required magnetic field; ``(Clock)'' indicates that the nominally zero-field clock states ($m_F=0$) are used.  The symbol $I$ is nuclear spin, and $\lambda_{1/2}$, $\lambda_{3/2}$, $\lambda_{D}$ refer to the wavelenths of the transitions from the ground state to the $P_{1/2}$, $P_{3/2}$, and (if present) $D_{5/2}$ levels with decay rates $\gamma_{D}$.  Qubit types that are typically encoded:  Z (Zeeman), H (hyperfine), F (Fine structure), and O (optical).  Gate types that are typically used:  R (Raman), O (Optical), M (Magnetic [AC or static gradient]).  *These isotopes have no good method of state discrimination; they have been used for sympathetic cooling, essentially as Zeeman qubits with no state detection.  $\dagger$This isotope of barium is radioactive, with a half life of 10.5~y. $\ddagger$Light at 532~nm (355~nm) from a doubled (tripled) YAG laser has been used to drive Raman transitions in Ba$^{+}$ (Yb$^{+}$).}
\begin{ruledtabular}
\begin{tabular}{ccccccccc}

Ion  & m  & I & $\lambda_{1/2}$,$\lambda_{3/2}$,$\lambda_{D}$  & $\gamma_{D}$ & $\omega_{\textrm{F}}/2\pi$  & Qubits  & FOFI, $B_0$ & Gates \\
  & (amu) &  &  (nm)  &  (s$^{-1}$)  & (THz)  &     & (mT) &  \\
\hline

\rule{0pt}{4ex}Be$^{+}$ &  9  & 3/2 &  313, 313, N.A.  & N.A. & 0.198  &  H  &  11.94 & R, M\\
\rule{0pt}{4ex}Mg$^{+}$ &  25   & 5/2 &  280, 280, N.A.  & N.A. & 2.75  &  H  &  10.9  & R, M\\
         &  24   &  0 & '' &  '' & ''  &  *  &   & R\\
\rule{0pt}{4ex}Ca$^{+}$ &  40   & 0 &  397, 393, 729  & 0.855 & 6.68  &  Z, F, O  &   & R, O\\
         &  43   & 7/2 & ''  & '' &  ''  &  H, O  &  14.61 &  R, M\\
\rule{0pt}{4ex}Sr$^{+}$ &  87 & 9/2 &  422, 408, 674  & 2.90  & 24.0  &   &  & \\
         &  88   &  0 & '' & '' & ''  &  O  &   &  O, R\\
\rule{0pt}{4ex}Cd$^{+}$ &  111  & 1/2 &  226, 214, N.A.  & N.A.  & 74.4  &  H  & (Clock) & R \\
         &  112,114  & 0 & ''  & '' & ''   &   *   &  & R \\
\rule{0pt}{4ex}Ba$^{+}$ &  133$\dagger$ & 1/2 &  493$\ddagger$, 455, 1762  & 0.0286 & 50.7  &  H, O  &  (Clock) &  \\
         &  137  &  3/2 & '' & '' & '' &  H, O  &  (Clock) & \\
         &  138  &  0 & ''  & ''  & '' & O  &   & R\\
\rule{0pt}{4ex}Yb$^{+}$ &  171  & 1/2 & 369$\ddagger$, 329, 411  & 139 & 99.8  &  H  &  (Clock) & R, M\\
         &  172, 174  & 0 & '' & '' & '' & O  &   & \\

\end{tabular}
\end{ruledtabular}

\label{tab:ion_props}
\end{table*}

\subsubsection{Choice of Qubit and Gate Type}

The two most popular qubit-gate pair choices in current experiments are hyperfine qubits manipulated using Raman gates and optical qubits manipulated using quadrupole transitions, in both cases using lasers for excitation.  While there is significant recent work based on utilizing magnetic field gradients, sometimes in combination with RF or microwave fields, to control (nominally hyperfine) qubits, taken together, the single and two-qubit gates with the best fidelity and at least reasonable speed are currently laser-based.  We will hence compare the two pairs mentioned above in terms of the tradeoff between speed, power, and error due to spontaneous emission, the main fundamental error source for these gates.  We note that both Raman and direct optical (quadrupole) gates may be applied to Zeeman qubits, e.g. two-qubit phase gates can be performed either using Raman coupling between the qubit states or by coupling one of the qubit states to the metastable level to impart a phase.  This is also true of the fine-structure qubit.  Therefore, this comparison is instructive when selecting among laser-based gates in general.

We will consider hyperfine qubits undergoing stimulated Raman transitions and subject to spontaneous scattering from an auxiliary level (almost always a rapidly decaying $P$ level) on the one hand, and optical qubits undergoing direct optical qubit excitation and subject to spontaneous decay from the metastable state (almost always a long-lived $D$ level) on the other.  The qubit types have significant tradeoffs for memory/storage separate from gates; these include the much longer decay times, and hence longer coherence times, available in the hyperfine qubits which can be contrasted with the more straightforward state-preparation and detection and more amenable laser wavelengths available in the optical qubits.  In terms of single and two-qubit gates, the direct optical gates can be performed using only a single laser beam, while the Raman gates require two or three laser fields and their relative interferometric stability.  There is a corollary to this, however, in that, for single-qubit gates, the two Raman beams may be arranged with parallel $k$ vectors such that the effective Lamb-Dicke parameter is essentially zero, allowing for effecting transitions that are insensitive to the level of ion-motional excitation.  Thus single-qubit gates performed in this way are much less reliant on cooling the ion motion to the ground state of the trap potential when compared with optical transitions.  Two-qubit gate operation requires imparting a force on the ions, so in this case, the Raman beams must have a difference-$k$ vector with an appreciable component along the displacement direction of the mode(s) of interest.

The total power required for a single-qubit Raman excitation with Rabi frequency $\Omega_{\textrm{R}}$, while achieving a certain error probability $\epsilon_R$ due to Raman scattering to the ground $S$ manifold, assuming two equivalent beams of waist size $w_0$, is~\cite{PhysRevA.75.042329_2007}

\begin{equation}
P_{\textrm{Raman,1Q}}=\frac{4\pi}{3} \hbar\, c\, w_{0}^{2}\, \frac{k_{3/2}^{3}}{\epsilon_R}\, \Omega_{\textrm{R}}=\frac{2\pi^{2}}{3} \hbar\, c\, w_{0}^{2}\, \frac{k_{3/2}^{3}}{\epsilon_R}\, t^{-1}_{\textrm{gate}}.
\label{p_raman}
\end{equation}


\noindent Here $k_{3/2}$ is the wavevector magnitude of the $S_{1/2}$-to-$P_{3/2}$ transition, and we have optimized the Raman beams' polarization.  By choosing a particular error probability, we can obtain the required power as a function of the gate speed (the second equation rewrites the power in terms of $\pi$-time $t_{\textrm{gate}}$).

A similar expression may be obtained for the power required for a single-qubit optical transition with Rabi frequency $\Omega_{\textrm{Q}}$ on a quadrupole transition with wavevector magnitude $k_{D}$ to a $D$ state with decay rate $\gamma_{D}$~\cite{James1998}:

\begin{equation}
P_{\textrm{Opt,1Q}}=\frac{1}{10} \hbar\, c\, w_{0}^{2}\, \frac{k_{D}^{3}}{\gamma_{D}}\, \Omega^{2}_{\textrm{Q}}=\frac{\pi^{2}}{40} \hbar\, c\, w_{0}^{2}\, \frac{k_{D}^{3}}{\gamma_{D}}\, t^{-2}_{\textrm{gate}}.
\label{p_optical}
\end{equation}

\begin{figure}[t b]
\includegraphics[width=1.0\columnwidth]{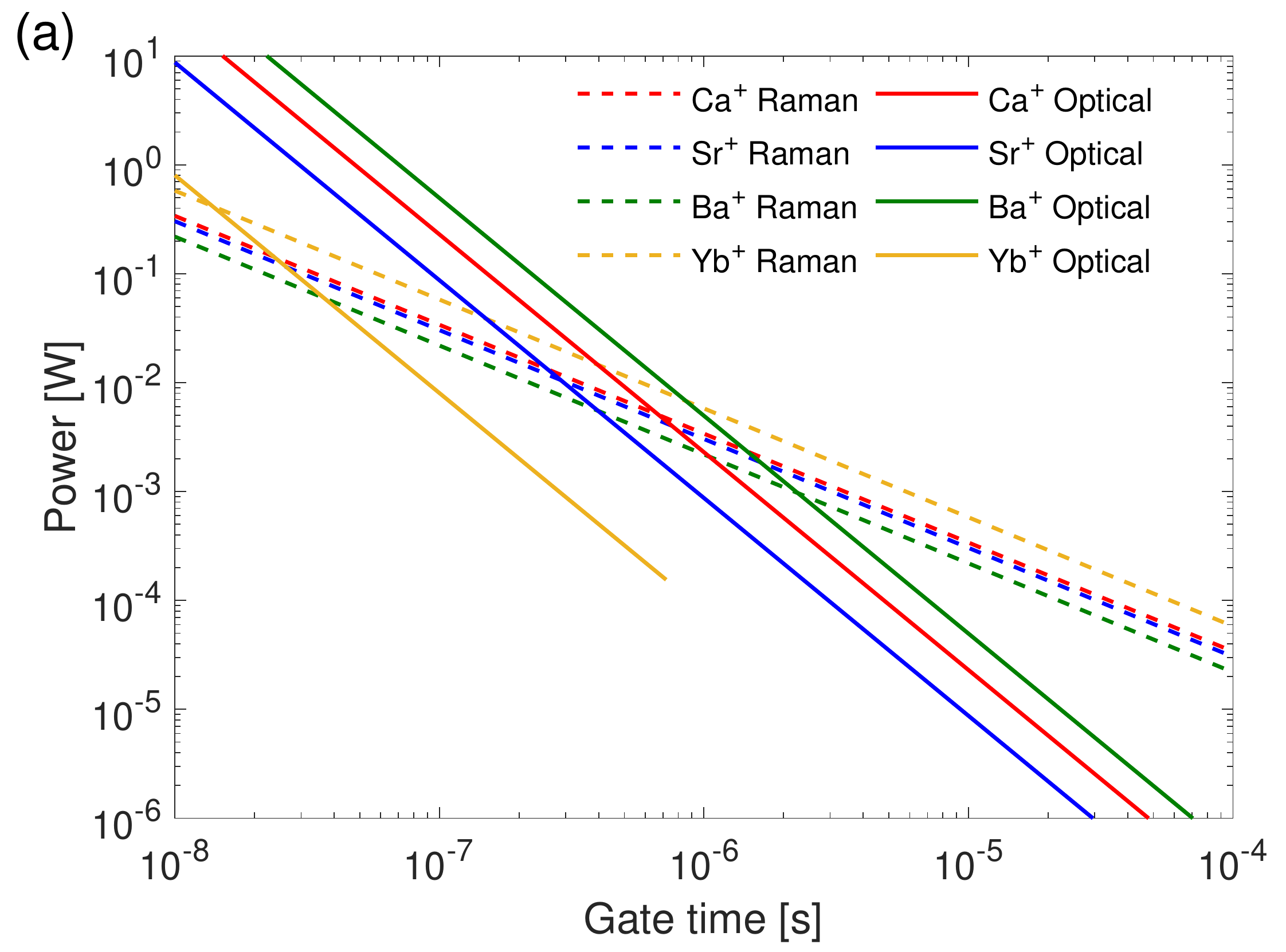}\\
\includegraphics[width=1.0\columnwidth]{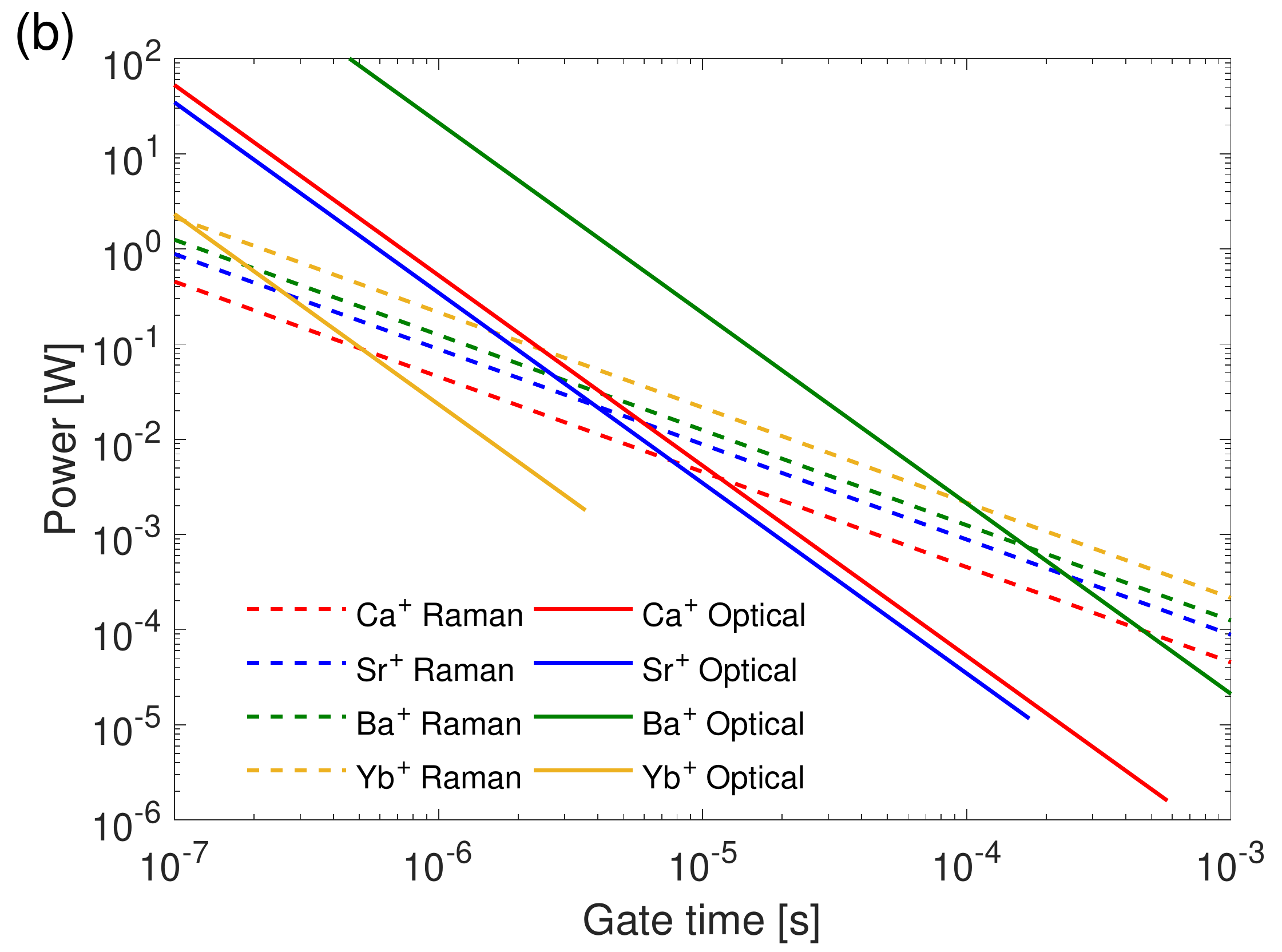}
\caption{Optical power required to drive optical and Raman gates as a function of gate time for several ion species of interest.  The power is the total assuming an equal split between two Raman beams, the optical gate is performed between the ground state and the $D_{5/2}$ state, and in both the optical and Raman case, a constant 20~$\mu$m beam waist is assumed.  (a) Single qubit gates with an error of $10^{-4}$ error or less. (b) Two-qubit gates with an error of $10^{-3}$ or less.  Lower error from spontaneous emission in the optical case requires going to a shorter gate time (by increasing the power), whereas scattering error can be reduced in the Raman case for the same gate time by increasing both detuning and power.}
\label{fig:gate_comparison}
\end{figure}

\noindent In this case, we have maximized the Rabi frequency with respect to the laser beam polarization and direction with respect to the quantizing magnetic field for a $\Delta m_{j} = 2$ transition, where $m_{j}$ denotes the Zeeman sublevel.  As can be seen, equations~\ref{p_raman} and~\ref{p_optical} are very similar.  In both cases, longer-wavelength transitions and smaller beams are strongly preferred, both considerations which suggest the use of integrated photonic technologies for scalable designs.  See Fig.~\ref{fig:gate_comparison}a for a comparison of single-qubit Raman and optical gates for several species of interest; here we plot the power required as a function of $\pi$-pulse gate time for a spontaneous scattering error of $10^{-4}$ or less. The main trends among species are due to the differences in the optical gate wavelengths and $D$~state decay times.  The main difference in the two expressions for required power when considering Raman and optical excitation is due to the Rabi frequency's dependence on the electric field $E$ in a Raman transition in the presence of fine structure [$\Omega \propto E_1 E_2\ \omega_{\textrm{F}} / (\Delta_R(\Delta_R-\omega_{\textrm{F}}))$ for Raman detuning $\Delta_R$ and fine-structure splitting $\omega_{F}$] versus a direct optical transition ($\Omega \propto E$).

This difference in scaling with the applied laser field intensity, in combination with the fact that the spontaneous emission error probability in the optical-gate case is inversely proportional to the gate speed, leads to a situation in which the gate-type with the lower required power (to get to a certain error probability) changes as a function of gate speed:  at longer gate times, the direct optical gate requires less power than the Raman gate, while the opposite is true for shorter gate times.  The cross-over gate-time depends not only on desired maximum error probability, but also on the level structure of the ion species used:

\begin{equation}
t_{\textrm{C-O,1Q}}=\frac{3}{80}\left( \frac{\lambda_{3/2}}{\lambda_{D}} \right)^{3} \frac{\epsilon_{R}}{\gamma_{D}}.
\end{equation}

\noindent where the wavelengths $\lambda_{3/2}$ and $\lambda_{D}$ correspond to the wavevectors defined above.  Values for $t_{\textrm{C-O,1Q}}$ are shown for various ions for an error probability per $\pi$-pulse of $10^{-4}$ or less in Table~\ref{tab:raman_optical}.


Two-qubit gates, which are generally slower than single-qubit gates due to the requirement to excite the motion of the ions (not just of the electrons), can be compared in a similar fashion.  The gates will operate at a Rabi frequency that is reduced by the Lamb-Dicke parameter $\eta$, meaning that the required power will be larger.  The required power for a Raman-driven two-qubit gate, with the same scattering error, will be increased by $1/\eta^2$ when compared to a single-qubit gate; since the two-qubit gate (with the same power and detuning) will be $\eta$ times slower (ignoring here multiple-phase-space-loop gates), one factor of $\eta$ (and increased detuning) is needed to achieve the same error in the longer gate duration.  Another is required to achieve an equivalent gate time $t_{\textrm{gate}}$.  The required power in this case is

\begin{equation}
P_{\textrm{Raman,2Q}}=\frac{16\pi^{2}}{3} c\, w_{0}^{2}\, \frac{k_{3/2}}{\epsilon_R}\, m\, \omega_{\textrm{T}}\, t^{-1}_{\textrm{gate}},
\end{equation}

\noindent where $m$ is the mass of one ion and $\omega_{\textrm{T}}$ is the the (angular) trap frequency.  Similarly, for direct optical gates, the power needed for a two-qubit gate will be increased by $1/\eta^{2}$, in this this case because the Rabi frequency is proportional to the optical electric-field amplitude:

 \begin{equation}
P_{\textrm{Opt,2Q}}=\frac{\pi^{2}}{10} c\, w_{0}^{2}\, \frac{k_{D}}{\gamma_{D}}\, m\, \omega_{\textrm{T}}\, t^{-2}_{\textrm{gate}}.
\end{equation}

\noindent Hence heavy ions will need proportionally more power for two-qubit gates, as expected by the requirement to excite the collective ion motion.  Longer-wavelength transitions are still favorable, though not to the degree to which they are for single-qubit gates.  In Fig.~\ref{fig:gate_comparison}b we plot the power required for a two-qubit gate as a function of gate time for a spontaneous scattering error of $10^{-3}$ or less, also for the species shown in Fig. \ref{fig:gate_comparison}a.
The cross-over gate time $t_{\textrm{C-O,2Q}}$ will then be increased by a factor of $\eta_{P}^{2}/(2 \eta_{D}^2)$ over $t_{\textrm{C-O,1Q}}$, the factor of $2$ due to the quadratic dependence of $P_{\textrm{Opt,2Q}}$ on the gate time; cross over times for two-qubit gates are also listed in Table~\ref{tab:raman_optical}.  The choice of gate type clearly depends heavily on the species of interest and the speed versus power trade-off.  We should reiterate that only spontaneous emission errors are considered here.  Recoil can also be a significant source of error for lighter ions participating in two-qubit gates~\cite{PhysRevA.75.042329_2007}, and so these errors must be considered when spontaneous emission rates are near $10^{-4}$ or below.

\begin{table*}[t b h !]
\caption{Cross-over gate durations $t_{\textrm{C-O}}$ and total required gate-beam power $P(t_{\textrm{C-O}})$ when the power needed for Raman and optical gates are equivalent (cf. Fig.\ref{fig:gate_comparison}); this assumes a 20~$\mu$m beam waist.  Single-qubit (two-qubit) gate times and powers are given for an error of $10^{-4}$ or less ($10^{-3}$ or less), considering only spontaneous emission errors.}
\begin{ruledtabular}
\begin{tabular}{cllll}

\multirow{2}{*}{Ion } & \multicolumn{2}{l}{Single-qubit gate ($\leq 10^{-4}$ error)} & \multicolumn{2}{l}{Two-qubit gate ($\leq 10^{-3}$ error)} \\
                     &   $t_{\textrm{C-O,1Q}}$ (ns)  &  $P(t_{\textrm{C-O,1Q}})$ (mW) &   $t_{\textrm{C-O,2Q}}$ ($\mu$s)  &  $P(t_{\textrm{C-O,2Q}})$ (mW) \\
\hline
\rule{0pt}{4ex} $^{43}$Ca$^{+}$  &  680 & $5.0$    &  12        &  $3.9$  \\
$^{87}$Sr$^{+}$  &  290                 &  $10$    &  3.9       &  $22$  \\
$^{137}$Ba$^{+}$ &  2300                &  $0.97$  &  170       &  $0.73$  \\
$^{171}$Yb$^{+}$ &  14                  &  $420$   &  $0.11$    &  $2000$  \\

\end{tabular}
\end{ruledtabular}

\label{tab:raman_optical}
\end{table*}

We additionally note that ion species with low-lying $D$ levels have a fundamental lower limit in achievable error for Raman gates (single- and two-qubit) due to scattering to the $D$ level (only decay to the $S$ is considered above) which does not go to zero in the limit of large Raman detuning.  For the ions considered in Fig.~\ref{fig:gate_comparison} and Table~\ref{tab:raman_optical}, this limit is in the neighborhood of $10^{-4}$ except for Yb$^{+}$, in which it is a couple orders of magnitude smaller~\cite{PhysRevA.75.042329_2007}. This is not the case for direct optical gates to the $D$ levels, where the error can always be decreased by reducing the gate time (by increasing the optical power).  Therefore, if very low ultimate error is a consideration, e.g. as is required for many currently studied error-correcting codes, one would employ optical gates in species with $D$ levels, or Raman gates in species without $D$ levels.

While the qubit-gate combinations considered here may ultimately be limited by photon scattering, magnetic-field-gradient gates avoid this error mechanism.  If the gate speeds using these techniques can be significantly increased, and if likely challenges associated with addressing and power dissipation can be overcome, magnetic-field-gradient gates have the potential to achieve even higher fidelities than the laser-based gates analyzed above.

\subsubsection{Choice of System Temperature}

While laser cooling is employed to reduce the kinetic energy of an atomic ion to equivalent temperatures of ${\sim}1$~mK via Doppler cooling and ${\sim}20$~$\mu$K via subsequent resolved-sideband cooling, the trap itself can remain at room temperature (or even above~\cite{UCB_heater_heating_rates_2018}) during QC operations; the internal electronic qubit is effectively isolated from this heat source.  There are, however, effects of trap-electrode temperature to consider for scalable systems.  For example, UHV pressures are required for long ion lifetimes, and the use of cryogenics to achieve low pressures has the added benefit that a wide range of materials may be used, since outgassing is exponentially suppressed at low temperatures.  On the other hand, cooling power is limited at very low temperatures due to the $T^{3}$ dependence of the heat capacity of most materials, resulting in challenges to power handling of dissipation from, e.g., integrated optical and electronic technologies.  The observed superlinear scaling of anomalous motional heating (AMH) with temperature would suggest lower logic errors can be attained by working at low temperature, but most of the gain is accrued by getting to the 50--100~K range \cite{PhysRevA.89.012318_2014}.  It therefore may be most prudent to work at an intermediate temperature of a few to several tens of kelvin where most of the molecular constituents of air are reduced to very low vapor pressure, but where sufficient cooling power is available.  One caveat to this is in the case of working with species subject to reaction with hydrogen, such as Be$^{+}$; longer ion lifetimes may be attainable at lower temperatures (the vapor pressure of hydrogen drops significantly below ${\sim}$20~K).  Another advantage of working at very low temperatures is the reduction in ion swapping events due to elastic collisions with background gas molecules as has been observed in room-temperature dual-species scenarios~\cite{BallanceHybridLogic2015}.

Aside from trap temperature, trap frequency and ion mass also impact AMH and can be examined to determine how best to minimize its deleterious effects for quantum logic.  The heating rate $\dot{\bar{n}}$ of a particular mode of vibration with harmonic-oscillator excitation $n$ is related to the electric-field noise spectral density $S_{E}(\omega)$ at a trap frequency $\omega$ ($\omega=2\pi\times f$, with frequency $f$ in hertz) by

\begin{equation}
\dot{\bar{n}}=\frac{q^2}{4 m \hbar \omega}S_{E}(\omega).
\label{eq_heating}
\end{equation}

\noindent Here $q$ and $m$ are the ion's charge and mass, respectively, and $\hbar$ is the reduced Planck constant.  The error in the most widely used ion two-qubit gates~\cite{PhysRevA.62.022311_2000,LeibfriedDidiGate2003} due to AMH is directly proportional to $\dot{\bar{n}}$ for heating rates slow compared to the gate speed~\cite{Ballance2QubitHyperfineGate2016}.  Experimentally, the scaling of $S_{E}(\omega)$ with frequency is typically measured to be in the range of $\omega^{-1}$ to $\omega^{-1.5}$~\cite{PhysRevLett.109.103001_2012,PhysRevB.89.245435_2014,PhysRevLett.120.023201,PhysRevA.97.020302_2018}, leading to a scaling of $\dot{\bar{n}}$, and hence the two-qubit gate error, of $\omega^{-2}$ to $\omega^{-2.5}$.  Therefore higher trap frequencies are paramount for countering gate errors when dominated by AMH.  This is true even for fixed laser (or microwave) gate-drive power:  the gate speed will be proportional to the sideband Rabi frequency, $\eta \Omega$ for $\Omega$ the carrier Rabi frequency and $\eta\propto\omega^{-1/2}$. Thus, even though there is a reduction in gate speed for higher trap frequency, the total error from AMH will go down as at least $\omega^{-1.5}$ due to the stronger dependence of $\dot{\bar{n}}$ on $\omega$.  Since the trap frequency goes as $\omega\propto m^{-1/2}$ (for both axial and radial modes, as trap size is reduced~\cite{NIST:SET:QIC:05}) and since $\dot{\bar{n}}\propto1/m$ (see Eq.~\ref{eq_heating}), the scaling of gate error in this case is very weakly dependent on ion mass for fixed trapping voltage.  In this limit of AMH-dominated error, as the highest obtainable trapping frequency is best, the applied voltage should be as high as possible---dielectric breakdown will likely be the ultimate limit for small structures, justifying the assumption of fixed voltage.  Though the mass dependence is weak, lighter ions also suffer recoil error if the logic is laser based~\cite{PhysRevA.75.042329_2007}, so there is a tradeoff to consider here, with dependence on the relative amount of error from these two sources.

\subsubsection{Implications}

These considerations, taken together, suggest particular scenarios for scaling systems of trapped ions.  If gate duration of approximately 10$^{-5}$~s is not a limitation to overall processor speed, lower optical power requirements can be obtained by using optical logic gates as opposed to stimulated Raman excitation.  If, however, much faster gates are required, with two-qubit gate durations at the microsecond scale or below, less power is required via Raman. It should be noted that extrapolations of spontaneous scattering error to significantly shorter gate durations may not be valid, since performing two-qubit gates much faster than the trap oscillation period can lead to nonidealities in standard gate operation due to operation outside the Lamb-Dicke regime, and spontaneous scattering may no longer be dominant.  Additionally, to keep scattering errors low in this regime, more power may be required than such extrapolations would suggest since one must obtain the required phase difference between the desired and other motional modes, all of which will be driven appreciably for very fast gate operation~\cite{SteanePulseGates2014}.

For architectures capitalizing on the potential scalability of integrated photonics approaches, optical gates have an additional advantage of being operated at generally longer wavelengths in the red and IR. Raman gates will likely require high optical powers to be delivered at blue and UV wavelengths where loss in optical waveguides is somewhat higher; this loss will likely always be worse due to the scaling of scattering loss with wavelength.  Hence, parallel operations over a large array will likely require lower total input optical power, and suffer less on-chip power dissipation, if gates are done optically.

In applications where low memory error is paramount, e.g. for distillation of high-fidelity entanglement from multiple copies of remotely-generated entangled pairs, hyperfine qubits, especially those that can allow for FOFI transitions, are the best option, with Zeeman qubits as a potential second choice if the system requirements allow for sufficient magnetic shielding.  Optical and fine-structure qubits will always be limited by metastable-state lifetimes, so may be less appropriate for cases that require long periods without error correction, e.g. NISQ quantum emulation.

With regards to hyperfine qubits, high ion mass leads to an increase in the required power for Raman gates.  And heavier ions require higher voltages on ion traps, a challenge to scalability.  However, the nuclear-spin-1/2 isotopes, with their more straightforward state preparation and repumping schemes, are present in the heavier species.  The lighter ions allow higher trap frequencies, and therefore may allow faster gate operation (if sufficient power is available). They are also easier to move with optical-dipole forces, and so require lower power for Raman excitation, but their wavelengths are far into the UV, potentially limiting applicability of standard integrated photonics technologies as may be desired in large arrays.  The high voltages and optical powers required with heavy ions and the UV wavelengths of the light ions suggest the use of medium-weight ions for portable applications, such as QC-based sensors, where a compromise on electrical and optical power (at more reasonable wavelengths, including red/IR-accessible optical qubits) is possible.

While very long coherence times are available using hyperfine qubits in non-zero nuclear-spin ions, fault-tolerant QC may be hindered by the leakage possible in these atomic systems due to the presence of the many hyperfine sublevels in the ground-state manifold.  A recent analysis shows that Zeeman qubits in zero nuclear-spin ions can lead to fewer required resources to reach a given logical qubit error, assuming magnetic field fluctuations can be reduced to a nominal level, due to these ions' relative resilience to leakage errors~\cite{PhysRevA.97.052301_2018}.

Magnetic gradient gates, either based on static or dynamic gradients, have not been analyzed in any detail here, as they have so far proven to be substantially slower than optical gates. Since microwaves are generally used for these gates, there are also concerns for larger systems related to crosstalk and required power. The lack of spontaneous scattering could potentially lead to higher-fidelity gates, but in order to make the gates faster, ions will likely need to be trapped closer to the gradient-producing structures (wires or permanent magnets); the scaling of AMH with ion-electrode distance would seem to preclude more than a modest increase in gradient via reducing this distance alone, however.  Operation at cryogenic temperatures, where AMH is significantly lower and wiring resistances---whether in normal metal or in a superconducting material---can be greatly reduced to minimize required microwave power, is a possible avenue for further development of this gate methodology. However, present capabilities are not commensurate with what is achievable optically.  Near-term scalable systems will likely be based on the Raman and direct-optical gates described above.

If there were to be a ``general purpose'' ion, it would probably be Ca$^{+}$.  It is widely used in experiments, QC-based and otherwise, all types of qubits and gates have been demonstrated using this species, and it has been used to demonstrate very high fidelity two-qubit gates, state-preparation, measurement, and very long coherence times in trapped-ion systems.  It has also been used to demonstrate many QC primitives and algorithms, as well as in quantum simulation investigations.  The wavelengths needed are relatively convenient, roughly spanning the visible spectrum, and there is a choice of having fully functional ion qubits with ($^{43}$Ca$^{+}$) or without ($^{40}$Ca$^{+}$) nuclear spin, each of which has optically addressable levels for shelving or quantum operations.  The ground state hyperfine splitting in $^{43}$Ca$^{+}$ is a manageable 3.2~GHz, large enough to provide spectroscopic addressability, but not so high that it cannot be easily spanned or that dealing with microwave transmission becomes prohibitive.  Moreover, its mass is very near the geometric mean of the lowest and highest masses of ions routinely used in QC experiments (namely $^{9}$Be$^{+}$ and $^{171}$Yb$^{+}$), so it plays well with other ions if dual-species operation is desired, i.e. for remote entanglement generation, syndrome readout, and sympathetic cooling in an ion register. Ca$^{+}$ is thus likely to be a convenient choice for an ion to build a system around, particularly if flexible operation is desired or if the ultimate QC-related application is unclear.

Considering dual-species operation in general, while Ca$^{+}$ would be a reasonable selection to pair with many ion species, pairs of ion species closer in mass are preferable if the main goal is for a sympathetic coolant and/or syndrome-extraction/entanglement-transfer ancilla.  Pairs that make sense in this regard, as well as when considering wavelength similarities (for ease of use), are Be$^{+}$/Mg$^{+}$, Ca$^{+}$/Sr$^{+}$, Sr$^{+}$/Ba$^{+}$, and Ba$^{+}$/Yb$^{+}$.  Three of these pairs are currently being pursued for QC in major efforts, some by multiple groups.  The light-ion duo, Be$^{+}$/Mg$^{+}$, has a compact (though rather UV) wavelength range, and if power is available, very high trap frequencies and fast operation should be possible~\cite{TanMultiElement2015}.  Rayleigh scattering imparting random momentum kicks may be a limit to ultimate fidelity. The Ca$^{+}$/Sr$^{+}$ pairing also has a good overlap in wavelength ranges and provides two very similar systems, with optical qubit and optical gate possibilities.  There is the potential in addition to use a single laser wavelength near 400~nm for interspecies Raman-based logic~\cite{BallanceThesis2014}, though scattering to the $D$ levels may ultimately limit Raman-gate fidelity.  The heavy-ion pairing Ba$^{+}$/Yb$^{+}$ is being pursued for remote-entanglement in a modular QC architecture~\cite{InlekMultiNode2017}, with the very favorable Ba$^{+}$ wavelengths serving to provide photons for entanglement generation.  Interspecies operations here require more power than with the other ions (see Fig.~\ref{fig:gate_comparison}), but these species benefit from the availability of high-power pulsed lasers at YAG harmonics for Raman gates.  The fourth pair of ion species mentioned above, Sr$^{+}$/Ba$^{+}$, while not being actively employed in many current experiments, would appear to have significant potential. With a very favorable mass ratio, no UV wavelength requirements, and the flexibility to have two optical qubits, or an optical and hyperfine pair, these species may be useful for on-chip applications requiring photon collection and/or transmission.  Moreover, the possibility of utilizing $^{133}$Ba$^{+}$ in this pair adds the potential benefits of a nuclear-spin 1/2 hyperfine structure.  These are obviously more combinations that can be pursued if a large wavelength range is tolerable, and this variety allows tailoring to paricular applications.

\subsection{Future Experiments to Enable Practical Trapped-Ion Quantum Computers}

We anticipate that a number of very useful research directions and experiments will be pursued to help enable and/or assess the prospects for practical QC with trapped ions.

\textit{Understanding and mitigating anomalous motional heating:} In many experiments, the fidelities of multi-ion-qubit gates are limited by motional heating.  While this is currently not true in all gate demonstrations (including those with the highest reported fidelities), we expect the precision of quantum control of trapped ions to continue to improve, and so at some point, motional heating will be the dominant limitation if it is not mitigated.  Likely key to this mitigation will be understanding the source of the electric field noise in ion traps.  We therefore believe that crucial experiments remain to be done to take the ``anomalous" out of anomalous heating.  In particular, it must be determined what ion trap materials and/or surface preparation techniques are best-suited for high-fidelity quantum control of ions, keeping in mind that such materials should be compatible with scaling systems to greater size.

\textit{Development of new techniques for robust ion control:} While ions have been controlled and measured with very high fidelity, achieved fidelities are not high enough to obviate the need for quantum error correction and the substantial overhead that comes with it.  Furthermore, the demonstrated high fidelity has been achieved only in few-ion systems, and it may worsen as the system is scaled up by orders of magnitude; at the very least, we should probably not expect it to improve without effort.  As a result, experiments must continue to focus on quantum control of trapped ions that not only improves fidelity, but that is likely to do so even in systems of considerably larger size.  One direction for this is to develop control methods and hardware that maintain their precision as they are scaled up.  Another direction is to assume that control imperfections are likely to be magnified as the system size grows, and to develop techniques that need not be as precise.  One promising approach based on the latter strategy, is that of dissipation engineering, whereby dissipation is used as a resource, rather than a hindrance, to generate quantum states with high fidelity in a manner that is relatively insensitive to the amplitudes and/or frequencies of the control fields \cite{PoyatosDE1996, LinDE2013}, as compared with strictly unitary operations.

\textit{Techniques for faster gates:} One of the biggest drawbacks to working with trapped-ion qubits is the speed with which gates can be performed; these gates, though high fidelity, are slow compared with some other qubit modalities.  All gates performed on ions to date have utilized the coupling of a control field to either an electric or magnetic multipole of the ion, and the gate speed is thus fundamentally limited by the strength of this coupling (for a fixed control-field intensity).  In the case of two qubit gates, which require imparting the momentum of control-field photons to an ion, the mass of the ion is also a limit to the speed.  A conceptually straightforward way to increase gate speed is to increase the control-field intensity, but when considering the goal of working with large numbers of ions, each likely requiring its own gate field, the prospect of increasing the per-ion field power is daunting.  One promising solution, as discussed in Sec.~\ref{IntPhot}, is to employ integrated photonics to enable tight focusing of gate lasers, and thus deliver high intensity light to ions with moderate power.  However, this approach should not be the only way forward that is considered.  Instead, the development and demonstration of fast, high-fidelity ion control techniques that do not depend on high control-field intensity should be an emphasis of research in the community.  It is particularly important that such work focus on two-qubit gates, which may set the speed limit to quantum processing with ions.  In doing so, it is likely that techniques will have to be developed that operate outside the Lamb-Dicke regime with highly-excited ion-motional states. In fact, work along this line has already begun \cite{SchaferFastIonGates2018}.  Considerations besides laser intensity that arise in this part of parameter space include off-resonant driving of motional modes other than the desired one, and in the case of M{\o}lmer-S{\o}rensen gates, off-resonant excitation of the carrier~\cite{SteanePulseGates2014}.

\textit{Noise Characterization:} As discussed in Sec.~\ref{ErrorReducandMit}, noise affecting qubit memory and control presently limits trapped-ion coherence times and gate fidelities.  Analyses of error mitigation protocols and error correcting codes always assume some model of the noise, which may not correspond to the actual noise present in trapped-ion systems.  However, the determined efficacy of a particular error-handling strategy typically depends sensitively on the details of the assumed noise.  It will therefore be of great importance to develop and implement efficient techniques to measure the types, magnitudes, and correlations of noise in trapped-ion systems, particularly for systems of intermediate scale (e.g. 100 ions or greater), where different noise sources are likely to predominate compared to few-ion systems.  Only then can we hope to have meaningful estimates of the performance of larger quantum computers in the presence of noise.

\textit{Demonstration and performance analysis of fault-tolerant error correction:}  In order to truly assess the prospects for building a practical trapped-ion quantum computer, a fault-tolerant logical qubit must be demonstrated.  That is, it must be explicitly verified that a number of physical ion qubits can be assembled and controlled in order to detect and correct any realistic errors that are likely to occur in a large quantum system, and this must be done in such a way that reduces the error rate of the logical qubit as compared with the physical qubits.  While such a demonstration would be heroic in its own right, an analysis of the performance of quantum error correction must also be undertaken in order to learn how it is likely to work in systems of larger size, including how a universal set of fault-tolerant quantum gates on logical qubits might be implemented.

\textit{Determining the benefits and limitations of integrated control and measurement hardware:}  As discussed in Sec.~\ref{ScalableHardware}, one of the most promising paths towards realizing a scalable trapped-ion quantum computer is to develop integrated ion-control and measurement technology.  In many cases, there is potential that this integration will not only lead to a capability to control large numbers of ions, but will also lead to improved performance in small systems.  At the same time, there is no assurance that this integration will not introduce new problems.  Experiments aimed at assessing the benefits and challenges associated with integration will therefore be important.  Simply, the integrated hardware must be designed, built, and tested in small trapped ion systems in order to determine its long-term potential.

\textit{Experiments to inform architectural analysis:} In a few places throughout this review, we have speculated on which architectures might be most promising for the implementation of practical trapped-ion quantum computers.  However, there is currently insufficient data to determine which architectural primitives (or combinations of primitives) are likely to be best.  The ideal details and parameters of these architectures also remain to be determined.  We believe that experiments aimed at making such determinations will be crucial to charting the direction of future trapped-ion system development.  For instance, it will be important to explore the tradeoffs for QC in linear chains of ions of varying length.  As the chains grow longer, two-qubit gates will likely get slower and have higher error; however, in order to process quantum information using a given number of qubits, a smaller number of split/join and transport operations will be required, as compared to shorter chains.  Since these transport operations take time and introduce motional excitation detrimental to gate performance, a practically-optimal linear chain (multi-qubit module) length may be discovered, and ought to inform the size of linear arrays that are developed for ion QC systems.  This is just one example, but there are many experiments that will be beneficial to explore the tradeoffs between time, fidelity, and resources (both in qubit number and gate operations) for a given set of architectural primitives.  These include comparing the performance of 1D and 2D ion arrays, as well as exploring the benefits and drawbacks of modular approaches, such as that based on photonic interconnects.

Due to the status of trapped ions as a leading qubit technology, trapped-ion experiments have already played a key role over the past two decades in advancing the field of QC and in highlighting challenges that must be overcome to achieve large-scale quantum information processing. While the preceding list of suggested experiments probing the long-term prospects for trapped-ion QC is surely not exhaustive, we believe it gives a flavor of what questions will be asked and investigated over the coming years. Trapped ions are likely to continue to be a powerful tool for exploring the capabilities and limitations of QC.

\begin{acknowledgments}
We thank Karan Mehta for his careful reading of the manuscript and for his comments and suggestions.  We also thank Wes Campbell and Eric Hudson for useful discussions.  This work is sponsored by the Assistant Secretary of Defense under Air Force Contract $\#$FA8721-05-C-0002. Opinions, interpretations, conclusions and recommendations are those of the authors and are not necessarily endorsed by the United States Government.
\end{acknowledgments}

\bibliography{APR_TrappedIonQC}

\end{document}